\documentclass[a4paper,11pt]{article}
\pdfoutput=1 

\usepackage{jheppub} 
\usepackage[bottom]{footmisc}
\usepackage{amssymb}
\usepackage{amsmath}
\usepackage{amsthm}
\usepackage[usenames,dvipsnames]{xcolor}
\usepackage{epsfig}
\usepackage{dcolumn}
\usepackage{tikz}
\usetikzlibrary{shapes.geometric, arrows}
\usepackage{upgreek}
\usepackage{setspace}
\usepackage{subfig}
\usepackage{enumitem}
\usepackage{array,multirow,bigdelim,arydshln}
\usepackage{appendix}
\usepackage{xparse}
\usepackage{nccmath}
\usepackage{stmaryrd}
\usepackage[T1]{fontenc} 
\usepackage{mathtools}
\usepackage{physics} 
\usepackage[export]{adjustbox}
\usepackage{multirow}
\usepackage{graphicx} 
\usepackage{float} 
\graphicspath{{./images/}}
\usepackage[nottoc]{tocbibind}
\usepackage{hyperref}
\usepackage[utf8]{inputenc}
\usepackage{CJK}
\usetikzlibrary{decorations.markings}
\usetikzlibrary{decorations.pathmorphing}
\usetikzlibrary{intersections}
\usetikzlibrary{calc}
\hypersetup{
	colorlinks,
	urlcolor=Maroon,
	linkcolor=Maroon,
	citecolor=Maroon
	}

\NewDocumentCommand{\binomial}{omm}
 {%
  \genfrac(){0pt}{}{#2}{#3}%
  \IfValueT{#1}{_{\!#1}}%
 }
\NewDocumentCommand{\eulerian}{omm}
 {%
  \genfrac<>{0pt}{}{#2}{#3}%
  \IfValueT{#1}{_{\!#1}}%
 }

\usepackage{latexsym}
\usepackage{tikz}

\theoremstyle{plain}

\theoremstyle{definition}

\def\oh{\mathcal{O}}
\def\ah{\mathcal{A}}

\def\rh{\mathcal{R}}

\def\fh{\mathcal{F}}

\def\ih{\mathcal{I}}
\def\vh{\mathcal{V}}

\def\bi{\begin{itemize}}
\def\ei{\end{itemize}}
\def\it{\textit}

\DeclareMathAlphabet{\mathbbold}{U}{bbold}{m}{n}

\def\bea#1\eea{\begin{eqnarray}#1\end{eqnarray}}
\def\be#1\ee{\begin{equation}#1\end{equation}}
\def\ba#1\ea{\begin{align}#1\end{align}}

\usepackage{amsmath}
\usepackage{multicol}
\usepackage{bbm}

\usepackage{amsthm}
\usepackage{mathrsfs}
\usepackage{upgreek}
\usepackage{amssymb}
\usepackage{bm}
\usepackage{setspace}
\usepackage{array,multirow,arydshln}
\usepackage{bigdelim}
\usepackage{scalerel}
\usepackage{diagbox}

\usepackage{tabularx}

\usepackage{tikz}

\usetikzlibrary{shapes.geometric,arrows,arrows.meta,decorations.pathmorphing,decorations.markings,patterns}

\def\<{\langle}
\def\>{\rangle}

\tikzset{
    vector/.style={
        decoration={snake, aspect=0.75, mirror, segment length=2mm},
        decorate
    },
    photon/.style={decorate, decoration={snake, amplitude=1pt, segment length=4pt}}
}

\usepackage[percent]{overpic}
\usepackage{multirow} 
\usepackage{slashed}

\title{Cuts and Contours}

\author[a]{Carolina Figueiredo,}
\author[b]{Marcos Skowronek}

\affiliation[a]{Jadwin Hall, Princeton University, Princeton, NJ 08540, USA}
\affiliation[b]{Department of Physics, Brown University, Providence, RI 02912, USA}

\emailAdd{cfigueiredo@princeton.edu}
\emailAdd{marcos\_skowronek\_santos@brown.edu}

\abstract{The traditional formulation of string amplitudes via worldsheet integrals provides a parametrization of the moduli space that fails to expose the complete singularity structure of the amplitudes. This problem is solved by the positive parametrization of string amplitudes given by \textit{surfaceology}. In this work, we use this formalism to study a number of properties of string amplitudes at tree-level and one-loop. We introduce several global prescriptions for an integration contour for which the integrals are finite everywhere in kinematic space. At tree-level, this is done in two ways: one directly implements the Feynman $i\varepsilon$ to analytically continue from Euclidean to Lorentzian worldsheets; the other is a generalization of the closed Pochhammer contour to arbitrary number of points. At loop-level, we present a systematic way of extracting cuts directly from the worldsheet integrand. This provides a powerful set of unitarity constraints, which we use to test the consistency of different ``stringy'' UV regularizations of field theory amplitudes. In addition, we identify the massive threshold expansion of the integrand, which allows us to reduce the problem to a finite set of Feynman integrals in Schwinger parametrization and provide a straightforward contour prescription reminiscent of its field-theory version.}

\begin{document}

\addtocontents{toc}{\protect\setcounter{tocdepth}{2}}
\maketitle

\section{Introduction}\label{sec:intro}

The study of scattering amplitudes in string theory has been an active research topic for decades. There have been numerous advances in studying their properties, such as their unitarity cuts \cite{Eberhardt:2022zay}, their high-energy behavior \cite{Gross:1987ar,Gross:1987kza,Banerjee:2024ibt}, their $\alpha'$-expansion \cite{DHoker:1988pdl,Schlotterer:2012zz,Stieberger:2016xhs,Mafra:2022wml} or their analytic continuation in terms of integration contours \cite{Witten:2013pra,Eberhardt:2023xck,Eberhardt:2024twy,Manschot:2024prc}. To understand many of these features, it is essential to have a strong grasp on how the amplitudes behave in the regions where singularities arise.

Despite this, there is an obvious limitation to the formalism in which string amplitudes are conventionally studied. The on-shell scattering amplitudes are written as integrals over the moduli space of punctured Riemann surfaces, parametrized in terms of the positions of vertex operators on the worldsheet, $z_i$ ($i=1,\ldots,n$), that enter an $n$-point amplitude (as well as the modular parameters for higher-genus surfaces). For example, the integral describing a scattering process of $n$ open strings at tree-level can be written as:
\begin{equation}
    \ah_n[12\cdots n] = (\alpha^\prime)^{n-3}\int_{z_1<\ldots<z_n} \frac{d^n z_i}{\text{SL}(2,\mathbb{R})}\ \prod_{i<j} \varphi(z_i) z_{i,j}^{2\alpha' p_i\cdot p_j},
\end{equation}
where $z_{i,j}=z_i-z_j$ and $\varphi(z_i)$ depends on the specific theory we consider. The momenta dependence enters through the Koba-Nielson (KN) factor, and the dot products are defined with mostly plus signature. For simplicity, from now on we will work in units where $\alpha^\prime =1$.

Crucially, the integrand is invariant under $\text{SL}(2,\mathbb{R})$ transformations, which means that one has to make a gauge-fixing choice to consider the appropriate number of degrees of freedom. Typically, this is done by fixing the positions of a number of $z_i$'s (e.g. three of them at tree level). After gauge-fixing, we are left with an integral that doesn't manifest all the singular loci. 
In order to access all singularities, one needs to perform so-called \emph{blowups} on the worldsheet \cite{mumford1969irreducibility}, which are local changes of variable that separate degenerate singular regions. 
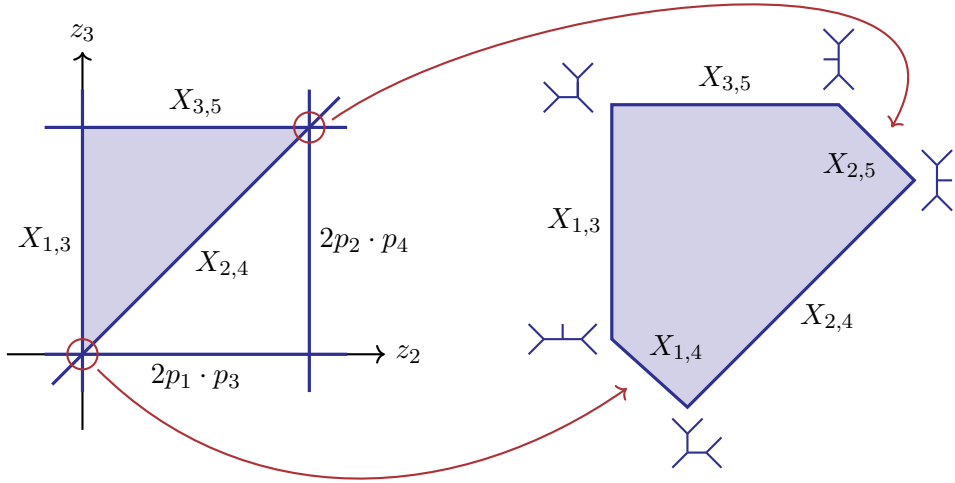
\begin{figure}[t]
    \centering
   \begin{tikzpicture}

\coordinate (T1) at (0,3);
\coordinate (T2) at (3,3);
\coordinate (T3) at (3,0);
\coordinate (C) at (0,0);

\fill[Blue!15] (C) -- (T2) -- (T1) --cycle;

\draw[->,thick] (-1,0) -- (4,0) node[right] {$z_2$}; 
\draw[->,thick] (0,-1) -- (0,4) node[above] {$z_3$}; 

\draw[Blue,very thick] (-0.5,3) -- (3.5,3); 
\draw[Blue,very thick] (0,-0.5) -- (0,3.5);
\draw[Blue,very thick] (3,-0.5) -- (3,3.5);
\draw[Blue,very thick] (-0.4,-0.4) -- (3.4,3.4);
\draw[Blue,very thick] (-0.5,0) -- (3.5,0);
\node[above] at (1.5,3) {$ X_{3,5}$};
\node[left] at (0,1.5) {$ X_{1,3}$};
\coordinate (MCT2) at ($(C)!0.5!(T2)$);
\node[below,xshift=10] at (MCT2) {$ X_{2,4}$};
\node[below] at (1.5,0) {$ 2 p_1 \cdot p_3$};
\node[right] at (3,1.5) {$ 2 p_2 \cdot p_4$};
\draw[Maroon, thick] (0,0) circle[radius=0.2];
\draw[Maroon, thick] (3,3) circle[radius=0.2];

\coordinate (A1) at (7,0.2);
\coordinate (A2) at (8,-0.7);
\coordinate (A3) at (11,2.3);
\coordinate (A4) at (10,3.3);
\coordinate (A5) at (7,3.3);
\coordinate (MA1A5) at ($(A1)!0.5!(A5)$);
\coordinate (MA1A2) at ($(A1)!0.5!(A2)$);
\coordinate (MA5A4) at ($(A5)!0.5!(A4)$);
\coordinate (MA3A4) at ($(A3)!0.5!(A4)$);
\coordinate (MA2A3) at ($(A3)!0.5!(A2)$);

\fill[Blue!15] (A1) -- (A2) -- (A3) -- (A4) -- (A5) -- cycle;
\draw[Blue,very thick] (A1) -- (A2) -- (A3) -- (A4) -- (A5) -- cycle;

\draw[->,Maroon,thick] (0.2,-.2) to[out=-45,in=-145] (7.2,-0.45);
\draw[->,Maroon,thick] (3.3,3.1) to[out=35,in=60] (10.7,3);

\node[left] at (MA1A5) {$ X_{1,3}$};
\node[above,xshift=10] at (MA1A2) {$ X_{1,4}$};
\node[above] at (MA5A4) {$ X_{3,5}$};
\node[below,xshift=-10] at (MA3A4) {$ X_{2,5}$};
\node[below,xshift=10] at (MA2A3) {$X_{2,4}$};

\draw[Blue,thick] (6.8,0.4) -- (6.6,0.2);
\draw[Blue,thick] (6.8,0) -- (6.6,0.2); 
\draw[Blue,thick] (6.6,0.2) -- (6.1,0.2);
\draw[Blue,thick] (6.35,0.2) -- (6.35,0.4);
\draw[Blue,thick] (5.9,0) -- (6.1,0.2);
\draw[Blue,thick] (5.9,0.4) -- (6.1,0.2);

\draw[Blue,thick] (6.55,3.4) -- (6.3,3.4);
\draw[Blue,thick] (6.55,3.4) -- (6.55,3.65);
\draw[Blue,thick] (6.55,3.65) -- (6.75,3.85);
\draw[Blue,thick] (6.55,3.65) -- (6.35,3.85);
\draw[Blue,thick] (6.55,3.4) -- (6.75,3.2);
\draw[Blue,thick] (6.55,3.4) -- (6.55,3.6);
\draw[Blue,thick] (6.1,3.2) -- (6.3,3.4);
\draw[Blue,thick] (6.1,3.6) -- (6.3,3.4);

\draw[Blue,thick] (9.8,3.5) -- (10.,3.7);
\draw[Blue,thick] (10.2,3.5) -- (10.,3.7);
\draw[Blue,thick] (10.,3.7) -- (10.,4.1);
\draw[Blue,thick] (10.,3.9) -- (9.8,3.9);
\draw[Blue,thick] (9.8,4.3) -- (10.,4.1);
\draw[Blue,thick] (10.2,4.3) -- (10.,4.1);

\draw[Blue,thick] (11.1,1.9) -- (11.3,2.1);
\draw[Blue,thick] (11.5,1.9) -- (11.3,2.1);
\draw[Blue,thick] (11.3,2.1) -- (11.3,2.5);
\draw[Blue,thick] (11.3,2.3) -- (11.5,2.3);
\draw[Blue,thick] (11.1,2.7) -- (11.3,2.5);
\draw[Blue,thick] (11.5,2.7) -- (11.3,2.5);

\draw[Blue,thick] (8,-1.3) -- (8.25,-1.3);
\draw[Blue,thick] (8,-1.3) -- (8,-1.05);
\draw[Blue,thick] (8,-1.05) -- (8.2,-0.85);
\draw[Blue,thick] (8,-1.05) -- (7.8,-0.85);
\draw[Blue,thick] (8.25,-1.3) -- (8.45,-1.5);
\draw[Blue,thick] (8.25,-1.3) -- (8.45,-1.1);
\draw[Blue,thick] (8.,-1.3) -- (7.8,-1.5);

\end{tikzpicture}
    \caption{The original integration region in worldsheet coordinates (left) is ``blown up'' via local changes of variables into a geometry which exposes the complete set of singularities of the amplitude (right).}
    \label{fig:5pt-Blowup}
\end{figure}
This problem does not arise at $4$-points but is already present for $n=5$, in which case the amplitude is given by 
\begin{equation}
    \mathcal{A}_5[12345] = \int \frac{d^5 z_i}{\text{SL}(2,\mathbb{R})} \frac{1}{z_{1,2} z_{2,3} z_{3,4} z_{4,5} z_{5,1}} \prod_{i < j} z_{i,j}^{2 p_i \cdot p_j}.
    \label{eq:5pt-Zform}
\end{equation}

Since $\mathcal{A}_5$ is a color-ordered amplitude, we know that the amplitude has a pole when sums of consecutive momenta go on-shell. That is when the planar variables, $X_{i,j} = (p_i+p_{i+1} + \cdots p_{j-1})^2$, hit zero or negative integers. However, this basic fact is not obvious from the way the kinematics appear in the integral, which is through general dot products $p_i \cdot p_j$. This is intimately related to the need for blow-ups to manifest all singularities. 

Let's gauge-fix the SL$(2,\mathbb{R})$ by choosing $(z_1,z_2,z_3,z_4,z_5) = (0,z_2,z_3,1,\infty)$, which yields:
\begin{equation}
    \mathcal{A}_5[12345] = \int_{0<z_2<z_3<1} d z_2 dz_3\ z_2^{2 p_1 \cdot p_2-1} (z_3 -z_2)^{2 p_2\cdot p_3-1} (1-z_2)^{2 p_2 \cdot p_4} z_3^{2p_1 \cdot p_3} (1-z_3)^{2p_3\cdot p_4-1}.
\end{equation}
Plotting the branch cut locations of the integrand, we obtain five different lines. However, only three of these bound the integration domain (see figure \ref{fig:5pt-Blowup}, left). In particular, the lines $z_3=0$ and $z_2=1$, associated with the non-planar exponents $2p_1\cdot p_3$ and  $2p_2\cdot p_4$ respectively, only touch the integration region in a single point. 

If we make $X_{1,3}=-n_{1,3}$, for $n_{1,3} \in \mathbb{N}_0$, the integral develops a singularity along the line $z_2=0$, and similarly making $X_{3,5}=-n_{3,5}$, for $n_{3,5} \in \mathbb{N}_0$, produces a singularity in the boundary $z_{3}=1$. In this locus, the most divergent region is then the point $(0,1)$, where these two lines meet. This reflects the fact that the tree-level amplitude has a maximal divergence when both $X_{1,3}$ and $X_{3,5}$ go on-shell, which comes from the diagram containing these two propagators. Now, we know that at $5$-points there are $5$ different diagrams contributing to this color ordering. However, in addition to the one identified above, the region of integration only has two other codim-2 boundaries -- where now three lines meet rather than two. 

To see the remaining diagrams, one needs to blow-up these points. For instance, we can implement the blow-up at $(z_2,z_3) =(0,0)$ by doing a local change of variables:
\begin{equation}
z_2 = \epsilon z, \quad z_3 = \epsilon, 
\end{equation}
with $\epsilon \ll 1$ and $z\in [0,1]$, which correctly ensures that $0<z_2<z_3<1$ for all $z,\epsilon$. We can rewrite the contribution from the region near $(z_2,z_3)=(0,0)$, in terms of $(z,\epsilon)$ as 
\begin{equation}
    \mathcal{A}_5[1,2,3,4,5] = \int_{\epsilon \ll 1} d z d\epsilon \,  \epsilon^{X_{1,4}-1} z^{X_{1,3}-1} (1-z)^{X_{2,4}-1} (1-\epsilon)^{X_{3,5}-1}(1-\epsilon z)^{2p_2\cdot p_4},
\end{equation}
where we see that in the region $\epsilon \ll 1$, we can still access the singularities from $X_{1,4}$ at $\epsilon=0$, $X_{1,3}$ at $z=0$, and $X_{2,4}$ at $z=1$. So effectively, in these new coordinates the point $(0,0)$, where previously three lines intersected, is ``blown-up'' into a line corresponding to the $X_{1,4}$ singularity, which now intersects the two other lines pairwise. Equivalently, back in the original coordinates, as we approach the corner of the integration region $(z_2,z_3)=(0,0)$ the integrand actually depends on the ratio $z_2/z_3$. Because of this the point $(0,0)$ is effectively blown up into a line parametrized by $z_2/z_3 = z$. 

Following the same procedure near $(z_2,z_3)=(1,1)$, we get that the correct description of the boundary of the moduli space is given by a pentagon, depicted on the right of figure \ref{fig:5pt-Blowup}. In this pentagon we see five boundaries, each associated with the singularity coming from one of the five planar propagators, $X_{i,j}$. These boundaries meet pairwise in five vertices, each labeling one of the five planar diagrams, as depicted in figure \ref{fig:5pt-Blowup}. 

In general, the boundary structure of $\mathcal{M}_{0,n}$ is described by the $n$-point associahedron \cite{devadoss2000tessellationsmodulispacesmosaic}, which is an $(n-3)$-dimensional polytope whose vertices label all possible planar trees with $n$ external legs.

Thus, in the standard worldsheet parametrization, one needs to perform two different blow-ups -- one near $(z_2,z_3)=(0,0)$ and another for $(z_2,z_3)=(1,1)$\footnote{This is for the particular gauge-fixing we chose, but for any other gauge-fixing one would similarly need two blow-ups.} to fully expose the boundary structure of $\mathcal{M}_{0,5}$. However, as we increase $n$, the number of blow-ups required to access all the singularities grows exponentially. Crucially, there is \textit{no} gauge-fixing choice that gives access to all singularities \textit{simultaneously} (without the need to manually blow up). This is clearly a practical obstacle to studying string amplitudes.

Meanwhile, in the last couple of years a new formulation of field theory and string amplitudes in terms of \emph{surface variables} has been introduced \cite{Arkani-Hamed:2017mur,Arkani-Hamed:2019mrd,Arkani-Hamed:2023lbd,Arkani-Hamed:2023mvg,Arkani-Hamed:2024pzc}, revealing new features of string amplitudes \cite{Arkani-Hamed:2023jry,Arkani-Hamed:2023swr,Arkani-Hamed:2024fyd} and providing an elegant solution to this problem. In this formalism, string amplitudes are given as surface integrals defined in terms of the $u$-variables \cite{Koba:1969rw,Chan:1969ex,Bardakci:1968rse,Gross:1969db} and parametrized by positive coordinates $y$ which manifestly blow-up \textit{all} the singularities. As a result, this representation makes it especially simple to extract the low energy limit by ``tropicalizing'' the integrand. Taking this limit divides the integration domain into discrete regions -- the so-called cones in the Feynman fan -- associated with the different Feynman diagrams. In particular, by taking the tropical limit of the integrand in each region, one manifestly recovers the usual Schwinger parametrized version of the field theory integral corresponding to that diagram.

In section \ref{sec:surface review}, we start by reviewing the representation of string amplitudes in terms of $u$-variables and their parametrization in terms of the positive $y$ coordinates. We explain how the different singularities are encoded in $y$-space and provide a practical way to access them. Using this, we show how to extract residues of the amplitude on general massive poles via residues of the integrand in $y$-space, and use this to read off the spectrum and couplings of low lying string states \cite{Arkani-Hamed:2023jwn}.    

Having a way to automatically blow up all singularities suggests revisiting a simple and fundamental question about string amplitudes: what is the precise integration contour under which string amplitudes are defined? We find that working in $y$-space makes it especially simple to give a global contour prescription that renders a finite integral for all regions in kinematic space.

At tree-level, we realize this in two distinct ways. On the one hand, in section \ref{sec:TreeContour} we construct a contour that effectively implements the Feynman $i\varepsilon$ as proposed in \cite{Witten:2013pra}. Recall that the need for a contour deformation arises when we make a certain $X_{i,j}<0$. The worldsheet integrals then acquire a divergence from the region of the moduli space where parts of the worldsheet degenerate into a long tube -- corresponding to a boundary of $\mathcal{M}_{0,n}$. In these regions one should deform the contour, turning the Euclidean worldsheet into a Lorentzian one and thus defining the correct analytical continuation of the integral. In doing this, we can use the same periodicity arguments made in \cite{Eberhardt:2024twy} to reduce the integration domain to a finite region and repackage the singular behavior into explicit prefactors.

On the other hand, in section \ref{sec:Pochhammer} we propose a generalization of the classic Pochhammer contour \cite{Pochhammer1890} for the Beta function to arbitrary number of points. Although there were some attempts to define this in terms of worldsheet coordinates for amplitudes involving more than four particles (see \cite{Hanson} for an example at five points), constructing such a closed contour becomes increasingly complicated in practice. This is because one needs to have control over the branch cut structure of the integrand while performing the local variable changes to blow up the degenerate singularities. In contrast, our positive parametrization allows us to define the contour globally in one go. We provide a practical implementation of it, which remarkably reduces to a sum over direct products of circles in parameter space.

In appendix \ref{app:numerical checks}, we present some numerical tests on both contour prescriptions for $n=5$ and $n=6$. We compare them with known exact formulae\cite{Arkani-Hamed:2024nzc}, showing that both contours reliably reproduce the correct results even in reasonably extreme kinematic regimes, with the Pochhammer contour being the most accurate. 

We then move on to discussing one-loop stringy amplitudes, which are far less studied and understood than their tree-level counterparts. We will therefore not aim for an exhaustive treatment for all $n$. Instead, we will focus on studying the issues associated with understanding unitarity cuts and defining integration contours, which already arise in the simplest case of $n=2$.

At loop-level, the textbook worldsheet formulation only gives us access to loop-integrated amplitudes \cite{Green:2012pqa}, while surfaceology directly defines the loop-integrand \footnote{A loop integrand in string theory is also naturally provided by the chiral splitting formalism \cite{DHoker:1989cxq}}. This makes it simpler to study loop-level singularities, such as unitarity cuts. In addition, there is a great deal of freedom in defining ``stringy'' surface integrals, all of which reduce to the same field theory amplitudes, only differing in the details of their UV behavior. A specific choice in this space of possibilities exactly reproduces the usual tachyon amplitudes of the bosonic string, as we review in appendix \ref{app:GSW matching}. As for all the other regularizations, while they are unitary at all orders in the $\alpha^\prime$ expansion at low-energies, it has been an open question to test their consistency with unitarity on massive thresholds in the UV. 

In section \ref{sec:loop cuts}, we directly compute the discontinuity of the amplitude using the surface integrand (i.e. without the need to perform any integrations), which fixes the spectrum and $3$-point couplings. By matching with the spectrum and couplings found at tree-level (in section \ref{sec:surface review}), we derive a set of constraints that allow us to test the consistency of  different ``stringy'' surface integrals at one-loop.

As for the prescribing an integration contour, in section \ref{sec:loop contour} we show that in our positive parametrization, the integral in each cone is naturally separated into distinct contributions corresponding to the threshold expansion of the amplitude. Therefore, at any point in kinematic space, only finitely many terms are divergent on the original integration region, and thus need a contour deformation. Moreover, we find that each threshold contribution is identical to the Schwinger parametrization of a Feynman integral with massive internal propagators, and so we can make use of techniques developed for field theory amplitudes (see e.g. \cite{Hannesdottir:2022bmo}) to provide a contour prescription. Lastly, we conclude by mentioning some future directions in Section \ref{sec:outlook}.

\section{Positive parametrization and blowing up the singularities}\label{sec:surface review}

Let us start by reviewing the formulation of string amplitudes via surface integrals \cite{Arkani-Hamed:2019mrd,Arkani-Hamed:2023lbd,Arkani-Hamed:2023jry}. For simplicity, we will focus on the tree-level case, but the results summarized in this section generalize to loop-level in a simple way, and we will come back to it in sections \ref{sec:loop cuts}-\ref{sec:loop contour}. 

We start by considering the standard worldsheet integral,  
\begin{equation}
    \ah_n[12\cdots n] = \int_{z_1<\ldots<z_n} \frac{d^n z_i}{\text{SL}(2,\mathbb{R})}\ \frac{1}{z_{1,2}z_{2,3}\cdots z_{n,1}} z_{i,j}^{2 p_i\cdot p_j},
\end{equation}
with $p_i^2=0$, also called the $Z$-theory integral \cite{Huang:2016tag}. We can manifest the dependence on the kinematics associated with the singularities, $X_{i,j}$, by writing it in terms of the SL$(2,\mathbb{R})$-invariant cross ratios given by the so-called $u$-variables to obtain
\begin{equation}
    \ah_n[12\cdots n] =\int_{z_1<\ldots<z_n} \frac{d^n z_i}{\text{SL}(2,\mathbb{R})}\ \frac{1}{z_{1,2}z_{2,3}\cdots z_{n,1}}\prod_{(i,j) \in \mathcal{D}_n} u_{i,j}[\{z_{i,j}\}]^{X_{i,j}},
    \label{eq:stringZtheory}
\end{equation}
with $u_{i,j} = z_{i,j-1}z_{i-1,j}/(z_{i,j}z_{i-1,j-1}) \in [0,1]$. Given a pair of indices $(i,j)$, we associate a curve on the disk with $n$ marked points in the boundary, going from marked point $i$
to marked point $j$. By assigning an external momentum $p_i^\mu$ to each boundary component $(i,i+1)$ of the disk, we can read off the momentum of the curve $(i,j)$ by homology; $X_{i,j}$ is then simply this momentum squared. 

The $u$-variables satisfy a set of non-linear equations -- called the $u$-equations -- that read \cite{Arkani-Hamed:2017mur}:
\begin{equation}
    u_X + \prod_{X'} u_{X'}^{\#(X,X')}=1, \quad \text{ with }u_{X} \in [0,1],
    \label{eq:ueqns}
\end{equation}
where $X$ denotes some curve in disk and the product is over all other curves $X^\prime$, with the exponents $\#(X,X')$ being the number of times $X$ and $X^\prime$ intersect. At tree-level, these exponents are either $0$ or $1$, but at loop-level they can take any integer value (see section \ref{sec:loop cuts}).

While there are as many $u$-equations as $u$-variables, the dimensionality of the space of solutions of these equations precisely matches that of the Teichm\"uller space of the respective Riemann surface -- specifying the order in the topological expansion under consideration. In other words, the $u$-equations give a purely \textit{algebraic} definition of the Teichm\"uller space underlying the string amplitude. 

In addition, the $u$-equations imply the binary behavior \cite{Arkani-Hamed:2019plo} of the $u$-variables, which automatically ensures the correct factorization structure of the amplitude \eqref{eq:stringZtheory}. Concretely, from \eqref{eq:ueqns} we have that when we set $u_X=0$, the $u$-functions of all the curves that cross it must go to $u_{X'}=1$, since \textit{all} $u$'s are bounded as $u_{X'}\in [0,1]$. Now, when we set a given exponent $X \to 0$, the integral in \eqref{eq:stringZtheory} develops a singularity in the locus where $u_X \to 0$. Thus, by extracting the residue in this singularity, we precisely localize on the region where all $u_{X^\prime} \to 1$ for $X^\prime$ crossing $X$. As a result, the KN factor turns into the product of the KN factors of the subsurfaces we obtain by cutting the original one along chord $X$.

By writing the $u_X$ in terms of cross ratios of the $z$'s, we can check that the $u$-equations are satisfied as a result of the \textit{Pl\"ucker relations} satisfied by the $z_{i,j}$ -- which therefore defines one way of parametrizing the moduli space. We will now proceed to discussing a different parametrization of the space of solutions of the $u$-equations, in which the $u$'s are functions of the positive variables, $\{y_P\}$. It arises from a simple counting problem about the path of curves on the underlying surface \cite{Arkani-Hamed:2023lbd}. More extensive reviews on how to derive these parameterizations can also be found in \cite{Arkani-Hamed:2023lbd,Arkani-Hamed:2023jry,Arkani-Hamed:2024vna}. However, for the sake of this paper, we will simply describe the main features of this new representation of string amplitudes and focus on explaining how the singularities are blown-up in $y$-space as well as giving a practical prescription on how to access them. 

\subsection{Positive parametrization of string amplitudes}
\label{sec:FanReview}

A positive parametrization of an $n$-point string amplitude always relies on a choice of an underlying triangulation, $\mathcal{T}$, of the surface defining the amplitude -- at tree-level this is simply the disk with $n$-marked points in the boundary, $\mathcal{D}_n$. Such a triangulation specifies a fat graph by duality, such that each internal propagator on the fatgraph is dual to one of the chords in $\mathcal{T}$. This fat-graph plays a crucial role in defining all the ingredients entering the surface integral giving the string amplitude. To begin with, to each one of the chords in the triangulation $P \in \mathcal{T}$, we associate a positive variable $y_P$. Different choices of underlying triangulation will result in different representations of the string amplitude, which are related to each other by simple coordinate transformations \cite{Arkani-Hamed:2019mrd,Arkani-Hamed:2020tuz}. 

For simplicity, in this paper, when discussing tree-level amplitudes we pick the underlying triangulation to be \textit{ray-like}, i.e. we choose $\mathcal{T} = \{(1,3),(1,4),\cdots,(1,n-1)\}$. The dual diagram is then a ladder containing propagators, $\{X_{1,3},X_{1,4},\cdots,X_{1,n-1}\}$ (see top left of figure \ref{fig:GVectorFan5}, for $5$-point example). For this choice, we have that the tree-level string amplitude is written in terms of the positive variables $y_{1,k}$ ($k \in \{3,4,\cdots,n-1\}$) as follows:
\begin{equation}
    \mathcal{A}_n[12\cdots n]= \int_{\mathbb{R}_+^{n-3}} \prod_{k=3}^{n-1} \frac{d y_{1,k}}{y_{1,k}} \prod_{i<j} u_{i,j}[\{y_{1,k}\}]^{X_{i,j}},
    \label{eq:npt-u}
\end{equation}
so that the integration measure in $y$-space is a simple d$\log$ form \cite{Arkani-Hamed:2019mrd}. 

Next, we want to determine the explicit parametrization of $u_{i,j}[\{y_{1,k}\}]$. To do this, we start by drawing the curve $(i,j)$ in the dual fatgraph -- a curve going from $i$ to $j$ determines a path on the fatgraph which enters the graph on the edge between $i$ and $i+1$ and exits on that between $j$ and $j+1$ (see figure \ref{fig:GVectorFan5}, point 2). We then record the path of the curve via its word, $\mathcal{W}_{i,j}$, which lists the edges of the fatgraph the curve goes through, as well as the left and right $(L/R)$ turns at each intersection, as follows: 
\begin{equation}\label{eq:fatgraph word}
   \mathcal{W}_{i,j} : \quad (i,i+1) \xrightarrow{L/R} (P_{k_1}) \xrightarrow{L/R} (P_{k_2}) \xrightarrow{L/R}\cdots \xrightarrow{L/R} (j,j+1),
\end{equation}
where $(i,i+1)$ and $(j,j+1)$ denote the edges at which the curve enters/exits the graph and $(P_k)$ are the internal propagators the curves passes through, which by construction are propagators in the underlying triangulation $\mathcal{T}$ (see figure \ref{fig:GVectorFan5}, point 2 for some examples of curves and the respective words in the $n=5$ case).

Now, to extract the $u_{i,j}$ from $\mathcal{W}_{i,j}$, we define the following matrices associated to $L$ and $R$ turns from road $y_P$~\cite{Arkani-Hamed:2023lbd,Arkani-Hamed:2024vna}
\begin{equation} 
M_L(y_{P}) = \begin{pmatrix}
    y_{P}& y_{P}\\ 0&1
\end{pmatrix},\quad M_R(y_{P}) = \begin{pmatrix}
    y_{P}& 0\\ 1&1
\end{pmatrix}.
\end{equation}
To a given word, we associate the matrix corresponding to the product of $M_{L/R}(y_P)$ as dictated by the word, where for the case of the boundary edges we evaluate $M_{L/R}(1)$. As an example, for $\mathcal{W}_{i,j}$ in \eqref{eq:fatgraph word} we get the following product:
\begin{equation}
    M_{X_{i,j}} = M_{L/R}(1) M_{L/R}(y_{P_{k_1}})  M_{L/R}(y_{P_{k_2}}) \cdots = \begin{pmatrix}
        M^{(1,1)}_{i,j} & M^{(1,2)}_{i,j}\\
        M^{(2,1)}_{i,j} & M^{(2,2)}_{i,j}
    \end{pmatrix} ,
\end{equation}
where each entry of $M_{X_{i,j}}$ ends up being some polynomial in the $y_P$. Finally, the corresponding $u$-variable is given by the cross-ratio of the matrix elements:
\begin{equation} 
u_{i,j} = \frac{M_{i,j}^{(1,2)}M_{i,j}^{(2,1)}}{M_{i,j}^{(1,1)}M_{i,j}^{(2,2)}}.
\end{equation}

In particular, for the tree-level case under consideration, this leads to: 
\begin{equation}
    u_{1,k} = y_{1,k} \frac{\mathcal{F}_{1,k-1}(\{y_{1,k}\})}{\mathcal{F}_{1,k}(\{y_{1,k}\})}, \quad  \quad u_{i,j} = \frac{\mathcal{F}_{i-1,j}(\{y_{1,k}\})\mathcal{F}_{i,j-1}(\{y_{1,k}\})}{\mathcal{F}_{i,j}(\{y_{1,k}\})\mathcal{F}_{i-1,j-1}(\{y_{1,k}\})},
\end{equation}
where the $\mathcal{F}_{i,j}$ are the so-called $\mathcal{F}-$polynomials, which can be written simply as:
\begin{equation}
   \mathcal{F}_{i,n-1}(\{y_{1,k}\}) =1, \quad  \mathcal{F}_{i,j}(\{y_{1,k}\}) = 1 + \sum_{m=i+2}^j \prod_{k=m}^{j} y_{1,k}. 
    \label{eq:FPols}
\end{equation}

Plugging this back into the $n$-point tree-level amplitude \eqref{eq:npt-u}, we can write the tree amplitude directly in terms of the $y$'s \cite{Arkani-Hamed:2019mrd,Arkani-Hamed:2024nzc}:
\begin{equation}
    \mathcal{A}_n[12\cdots n]= \int_{\mathbb{R}_+^{n-3}} \prod_{k=3}^{n-1} \frac{d y_{1,k}}{y_{1,k}} y_{1,k}^{X_{1,k}} \times \prod_{i=1}^{n-3} \prod_{j=i}^{n-1} \mathcal{F}_{i,j}(\{y_{1,k}\})^{-c_{i,j}},
    \label{eq:StrAmpY}
\end{equation}
where the non-planar variables $c_{i,j} = -2p_i \cdot p_j$ are given by 
\begin{equation}\label{eq:non-planar variables}
    c_{i,j} = X_{i,j}+X_{i+1,j+1}-X_{i,j+1}-X_{i,j+1}.
\end{equation}

For the whole tree-level discussion of sections \ref{sec:TreeContour}-\ref{sec:Pochhammer}, we will always deal directly with the final form of the amplitude as given in \eqref{eq:StrAmpY}. However, at loop-level (see sections \ref{sec:loop cuts}-\ref{sec:loop contour}), we will use the procedure given above and give explicit formulae for the $u$'s of all curves on the punctured disk.

\subsection*{Location of singularities in $y$-space} 

Having understood how to obtain the $y$-space representation of the string integral, we now want to understand how the different degenerations of the worldsheet are mapped into the boundaries of the positive $y$-space.

Looking at \eqref{eq:StrAmpY} and knowing the form of $\mathcal{F}_{i,j}$, we can see that setting $X_{1,k}<0$ produces a divergence in the boundary of the integration domain where $y_{1,k} \to 0$. In particular, setting all $X_{1,k}=0$, we obtain a maximal divergence corresponding to the underlying ladder diagram, which forms a codim-$(n-3)$ part of the boundary at the origin of $\mathbb{R}_+^{n-3}$. If instead we are interested in the boundary region encoding the singularity associated to some other $X_{i,j}$ (not in the underlying triangulation), all we need to know is the \textit{$g$-vector} of curve $(i,j)$.
\begin{figure}[t]
    \centering    \includegraphics[width=\linewidth]{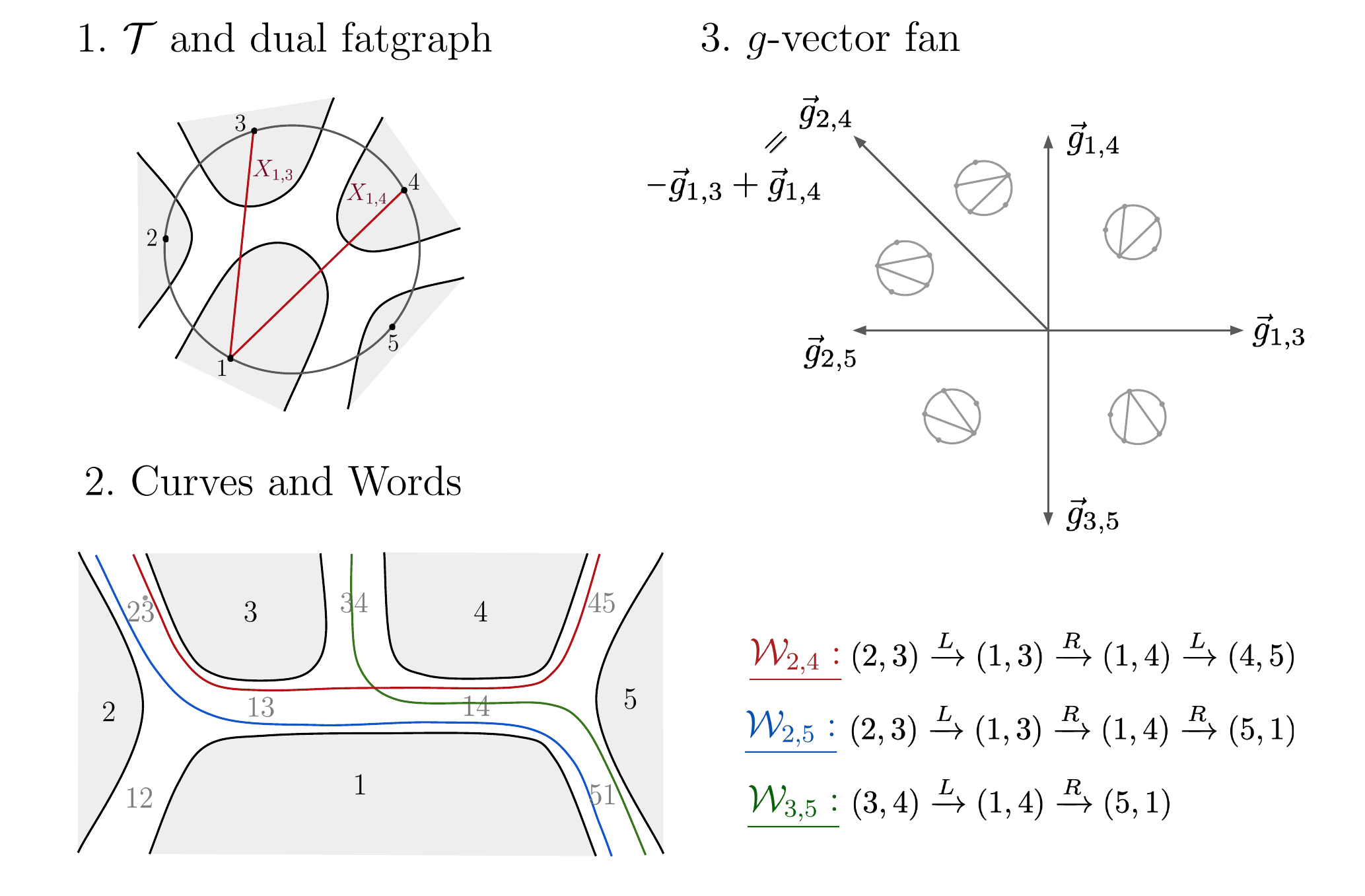}
    \caption{From curves on surfaces to the $g$-vector fan. In 1, we show a choice of the triangulation of the $5$-point disk containing chords $\{(1,3),(1,4)\}$ and the respective dual fatgraph. In 2, we show how the curves $(2,4)$ (red),  $(2,5)$ (blue), and $(3,5)$ (green) can be drawn as paths on the fatgraph, as well as the respective words we associate with each path. In 3, we present the $g$-vector fan which divides the plane into $5$-cones, each of which corresponding to one of the five cubic planar diagrams entering the $5$-point amplitude.}
    \label{fig:GVectorFan5}
\end{figure}

According to \cite{Arkani-Hamed:2023lbd}, to each chord $(i,j)$ on the disk $\mathcal{D}_n$ we can associate an $(n-3)$-dimensional vector (called the $g$-vector), which records the information about the boundary region associated with the dual propagator, $X_{i,j}$. We start by associating a unit vector to each internal propagator in the underlying fatgraph -- so for the case of the ladder fatgraph, we have $\{\vec{g}_{1,3},\vec{g}_{1,4}, \cdots, \vec{g}_{1,n-1}\}$ -- and take this to define a standard unit basis in $\mathbb{R}^{n-3}$.  Then, to obtain the $g$-vector $\vec{g}_{i,j}$ for the curve $(i,j)$, we collect all $P_i$ that appear in the word as \textit{peaks} ($\mathcal{P}_{i,j}$) and as \textit{valleys} ($\mathcal{V}_{i,j}$), which are respectively the subsets\footnote{A word can be turned into a mountainscape by drawing the left turns as up arrows and the right turns as down arrows. Then, the set of peaks $\mathcal{P}_{i,j}$ is simply the set of $P_i$ appearing as maxima in this mountainscape, and the set of valleys includes those appearing as the minima.}: 
\begin{equation}
\begin{aligned}
    \mathcal{P}_{i,j} : \, \,  \{ P_k : \, [\cdots \xrightarrow{L} (P_k)\xrightarrow{R} \cdots ] \subset \mathcal{W}_{i,j} \}, \\
    \mathcal{V}_{i,j} : \, \,  \{ P_k : \, [\cdots \xrightarrow{R} (P_k)\xrightarrow{L} \cdots ] \subset \mathcal{W}_{i,j} \}. \\
\end{aligned}
\end{equation}
With this, we can finally write the $g$-vector for the curve $(i,j)$ as:
\begin{equation}
    \vec{g}_{i,j} = \sum_{P_l \in \mathcal{V}_{i,j}} \vec{g}_{P_l}-\sum_{P_k \in \mathcal{P}_{i,j}} \vec{g}_{P_k},
\end{equation}
where for our ladder triangulation we have that $\vec{g}_{P_k} \in \{\vec{g}_{1,3},\vec{g}_{1,4}, \cdots, \vec{g}_{1,n-1}\}$. Collecting the $g$-vectors for all the curves in $\mathcal{D}_n$, we obtain the \emph{g-vector fan}, which divides $\mathbb{R}^{n-3}$ into cones. Each of them is bounded by $(n-3)$ $\vec{g}_{i,j}$'s, which specify a collection of $(n-3)$ curves defining a triangulation of $\mathcal{D}_n$. Therefore, a particular cone is associated to a diagram with the respective $(n-3)$ propagators $X_{i,j}$. In figure \ref{fig:GVectorFan5} (point 3), we show the $g$-vector fan at $5$-points, together with the labels of the different cones with the corresponding triangulations of $\mathcal{D}_5$.

Finally, we want to understand how we can use the $\vec{g}_{i,j}$ to determine the boundary of the integration domain in \eqref{eq:StrAmpY} associated with the singularity for $X_{i,j}$. To do this, we start by doing a change of coordinates in \eqref{eq:StrAmpY} to:
\begin{equation}
    y_{1,k} = e^{-t_{1,k}} \ \longrightarrow \ \mathcal{A}_n= \int_{\mathbb{R}^{n-3}} \prod_{k=3}^{n-1} d t_{1,k} e^{-t_{1,k} X_{1,k}} \times \prod_{i=1}^{n-3} \prod_{j=i}^{n-1} \mathcal{F}_{i,j}(\{e^{-t_{1,k}}\})^{-c_{i,j}}, 
\end{equation}
where now each $t_{1,k}$ ranges from $[-\infty,+\infty]$. These variables are the analogue of the Schwinger proper times for the corresponding ladder Feynman diagram in the field theory limit\footnote{This change of variables is not strictly needed for this part, but it will be useful when defining the integration contour in section \ref{sec:TreeContour}}. In these coordinates, the boundary of $\mathcal{M}_{0,n}$ associated with $X_{i,j}$ is defined as: 
\begin{equation}
   (\partial \mathcal{M}_{0,n})_{X_{i,j}} : (t_{1,3},t_{1,4},\cdots,t_{1,n-1}) \to z \vec{g}_{i,j} \quad \text{with } z\to +\infty. 
\label{eq:singMap}
\end{equation}

For example, for the curves in the underlying ray triangulation the entries of $\vec{g}_{i,j} = \vec{g}_{1,k}$ are all zero except for the one corresponding to $t_{1,k}$, which it is equal to $1$. Thus, we have that the boundary is located at $t_{1,k} \to +\infty$ or $y_{1,k} \to 0$, which is precisely what we saw earlier. 

Let's now look at the $n=5$ example to see how the remaining singularities can be accessed using \eqref{eq:singMap}. As shown in figure \ref{fig:GVectorFan5}, we pick the underlying fatgraph with propagators $\{X_{1,3},X_{1,4}\}$, in which case we can write the $5$-point amplitude in $t$-space as 
\begin{equation}
\mathcal{A}_5 = \int_{\mathbb{R}^2} dt_{1,3}dt_{1,4}\ e^{-t_{1,3}X_{1,3} - t_{1,4}X_{1,4}}(1 + e^{-t_{1,3}})^{-c_{1,3}}  (1 + e^{-t_{1,4}})^{-c_{2,4}}  (1 + e^{-t_{1,4}} + e^{-t_{1,3} -t_{1,4}})^{-c_{1,4}}.
 \label{eq:A5_tspace}
\end{equation}

Now, to see how the different singularities/diagrams are mapped into $t$-space according to \eqref{eq:singMap}, let's expand \eqref{eq:A5_tspace} in the different cones given by the $g$-vector fan in figure \ref{fig:GVectorFan5}. In each cone, we have that different terms in the $\mathcal{F}$-polynomials dominate, and so to make the respective singularities manifest, we factor out the dominating term from each $\mathcal{F}$-polynomial. For instance in the $\{(1,3),(1,4)\}$ cone, where $t_{1,3},t_{1,4} \in \mathbb{R}^+$, the leading term in each $\mathcal{F}$-polynomial is the $1$, and therefore the representation in \eqref{eq:A5_tspace} already makes the singularities in $X_{1,3}$ and $X_{1,4}$ manifest. If instead we look at cone $\{(1,3),(3,5)\}$, now since $t_{1,4}<0$ the term $e^{-t_{1,4}}$ dominates. Factoring it out, we can write the integrand $\mathcal{I}_5$ as 
\begin{equation}
\begin{aligned}
    &\underline{\text{Cone }\{(1,3),(3,5)\}:} \quad (t_{1,3}>0,t_{1,4}<0) \\[1em]
    & \mathcal{I}^{C_{13,35}}_5(t_{1,3},t_{1,4})= e^{-t_{1,3}X_{1,3} + t_{1,4}X_{3,5}} (1 + e^{-t_{1,3}})^{-c_{1,3}}  (1 + e^{t_{1,4}})^{-c_{2,4}}  (1 + e^{t_{1,4}} + e^{-t_{1,3}})^{-c_{1,4}},
\end{aligned}
\end{equation}
where each new $\mathcal{F}$-polynomial goes to $1$ in the limit where $t_{1,3}\to +\infty$ and $t_{1,4} \to -\infty$. In addition, it is now obvious that the integral develops a singularity in these limits when $X_{1,3}<0$ and $X_{3,5}<0$, respectively. 

As we will see in section \ref{sec:TreeContour}, when implementing the contour deformation in some boundary of the integration domain, it will be crucial to use the appropriate representation of the integrand, as it allow us to deform contour without crossing any branch cuts coming from the different $\mathcal{F}$-polynomials. 

Proceeding in the same way in the remaining cones, we get:
\begin{align}
    &\underline{\text{Cone }\{(2,5),(3,5)\}:} \quad (t_{1,3}<0,t_{1,4}<0) \\[1em]
&\mathcal{I}^{C_{25,35}}_5(t_{1,3},t_{1,4})= e^{t_{1,3}X_{2,5} + t_{1,4}X_{3,5}} (1 + e^{t_{1,3}})^{-c_{1,3}}  (1 + e^{t_{1,4}})^{-c_{2,4}}  (1 + e^{t_{1,3}} + e^{t_{1,3}+t_{1,4}})^{-c_{1,4}}, \nonumber
\end{align}
\begin{align}
&\underline{\text{Cone }\{(2,5),(3,5)\}:} \quad (t_{1,3}<0,t_{1,4}<0) \\[1em]
&\mathcal{I}^{C_{25,35}}_5(t_{1,3},t_{1,4})= e^{t_{1,3}X_{2,5} + t_{1,4}X_{3,5}} (1 + e^{t_{1,3}})^{-c_{1,3}}  (1 + e^{t_{1,4}})^{-c_{2,4}}  (1 + e^{t_{1,3}} + e^{t_{1,3}+t_{1,4}})^{-c_{1,4}}, \nonumber\\ 
\nonumber\\
&\underline{\text{Cone }\{(2,4),(2,5)\}:} \quad (t_{1,3}<0,t_{1,4}>0,t_{1,4}<-t_{1,3}) \\[1em]
    & \mathcal{I}^{C_{24,25}}_5(t_{1,3},t_{1,4})= e^{(t_{1,3}+t_{1,4})X_{2,5} - t_{1,4}X_{2,4}} (1 + e^{t_{1,3}})^{-c_{1,3}}  (1 + e^{-t_{1,4}})^{-c_{2,4}}  (1 + e^{t_{1,3}+t_{1,4}} + e^{t_{1,3}})^{-c_{1,4}}, \nonumber \\
    \nonumber\\
&\underline{\text{Cone }\{(1,4),(2,4)\}:} \quad (t_{1,3}<0,t_{1,4}>0,t_{1,4}>-t_{1,3}) \\[1em]
    & \mathcal{I}^{C_{14,24}}_5(t_{1,3},t_{1,4})= e^{-(t_{1,3}+t_{1,4})X_{1,4} +t_{1,3}X_{2,4}} (1 + e^{t_{1,3}})^{-c_{1,3}}  (1 + e^{-t_{1,4}})^{-c_{2,4}}  (1 + e^{-t_{1,4}} + e^{-t_{1,3}-t_{1,4}})^{-c_{1,4}}. \nonumber 
\end{align}

From the expressions above, together with the definitions of each cone, it is simple to see that one can find linear transformations that map each cone into the positive orthant. For general $n$, given a cone $C$ bounded by $(n-3)$ compatible chords $X \in C$, we can map it into the positive orthant through the following change of variables: 
\begin{equation}
    t_{1,k} = \sum_{X\in C} (\vec{g}_X)_{1,k}  \, \hat{t}_X,
    \label{eq:cone_map}
\end{equation}
where $(\vec{g}_X)_{1,k}$ is the component of $\vec{g}_X$ corresponding to $\vec{g}_{1,k}$, and $\hat{t}_X$ are the coordinates in the new cone. Note that, due to the cones always being of unit volume \cite{Arkani-Hamed:2023lbd}, the Jacobian of this transformation is one. For example, in the $n=5$ case studied above, this change of variables applied to cone $\{(2,4),(2,5)\}$ would give us:
\begin{equation}\label{eq:cone X24X25}
    t_{1,3} = -\hat{t}_{2,4} -\hat{t}_{2,5}, \quad  t_{1,4} = \hat{t}_{2,4},
\end{equation}
which precisely turns the prefactor into $e^{-\hat{t}_{2,5} X_{2,5} -\hat{t}_{2,4} X_{2,4}}$.

Similarly, we can define the change of variables \eqref{eq:cone_map} directly in terms of the $y_{1,k}$ to map them to $\hat{y}_X$ with $X\in C$. In this case, we have
\begin{equation}
    y_{1,k} = \prod_{X \in C} (\hat{y}_X)^{-(\vec{g}_X)_{1,k}},
    \label{eq:coneMapy}
\end{equation}
which allow us to write the original integral in terms of $\hat{y}_X$, and therefore the singularities associated to the $X\in C$ are now located at $\hat{y}_X=0$.

\subsection{Extracting cuts and fixing string couplings}\label{sec:tree cuts}

Let us now proceed with explaining how the $y$-space representation of string amplitudes lets us trivially extract its residues at arbitrary mass levels. As mentioned earlier, when we go on the locus $X_{i,j} = -n$, the string integral develops a singularity in the integration region where the respective $u_{i,j} \to 0$. Now remarkably, we can directly extract the residue of the amplitude at $X_{i,j} = -n$ by computing a residue of the integrand in $y$-space. Concretely, say we are interested in computing the residue when $X_{1,l} =-n$, with $X_{1,l}$ being one of the chords in the base triangulation. Then, we simply need to take a residue of the integrand at $y_{1,l}=0$, that is~\cite{Arkani-Hamed:2023jry,Arkani-Hamed:2023jwn,Arkani-Hamed:2024nzc}
\begin{equation}
  \mathop{\mathrm{Res}}_{X_{1,k}=-n} \mathcal{A}_n[12\cdots n] =  \int_{\mathbb{R}_+^{n-2}} \prod_{k\neq l} \frac{d y_{1,k}}{y_{1,k}} y_{1,k}^{X_{1,k}} \times  \mathop{\mathrm{Res}}_{y_{1,l}=0} \left[ y_{1,l}^{-n-1} \prod_{i=1}^{n-3} \prod_{j=i}^{n-1} \mathcal{F}_{i,j}(\{y_{1,k}\})^{-c_{i,j}} \right].
  \label{eq:res_tree}
\end{equation}
In particular, since the $\mathcal{F}-$polynomials have a nice recursive structure, hinted by \eqref{eq:FPols}, one can use the above to derive explicit formulae for these residues in terms of shifted lower-point amplitudes \cite{Arkani-Hamed:2024nzc}.

If, instead, we are interested in extracting a residue in a chord $X_{i,j}$ which is not in the underlying triangulation, we can simply use the map \eqref{eq:coneMapy} to turn the integral in $\{y_{1,k}\}$ to one where the integration variables are $\{\hat{y}_{P}\}$, 
with $P$ in some triangulation $\mathcal{T}^\prime$ containing chord $X_{i,j}$. By performing this map to the string integral, we can then write it in $\hat{y}_{P}$-space and extract the residue in $X_{i,j}=-n$ exactly as in \eqref{eq:res_tree} -- by taking a residue of the integrand at $\hat{y}_{i,j}=0$.

Following this procedure, we can extract the maximal cuts -- also known as the leading singularities -- by taking $(n-3)$ residues that kill off all integration variables. If we want the leading singularity of the ladder diagram, that is simply given by the maximal residue of the integrand when all $y_{1,k}=0$ for $k=3,\ldots,n-1$. For the leading singularity associated to a different diagram, we start by performing the change of variables \eqref{eq:coneMapy} into $\hat{y}_P$, where the $P$'s correspond to the propagators entering the diagram under consideration, and similarly extract the maximal residue when all $y_{P}=0$.

In addition, by looking directly at the string residues and matching them with a general tree-level gluing \emph{ans\"atze}, one can derive the string spectrum and respective couplings from the bottom-up. This was done explicitly in \cite{Arkani-Hamed:2023jwn}, and here we summarize some of the relevant results for our analysis. In section \ref{sec:loop cuts}, we will use the string couplings presented here to check the consistency of loop-cuts.

For this part of the analysis, in order to allow for more general theories, we consider the version of the string amplitude in \eqref{eq:StrAmpY} where each $X_{i,j}$ is shifted with $-\alpha_0$ -- the Regge intercept -- so that the level $0$ states have $m^2 = -\alpha_0$.

Already just looking at the $4$-point amplitude, we can extract the $3$-point amplitudes where the external states have masses $m_1^2=-\alpha_0-n,\ m_2^2=m_3^2=-\alpha_0$. In other words, leg $1$ is in level $n$, while legs $2$ and $3$ are in level zero. For instance, for $n=1$, our \emph{ansatz} for the 3-point interaction looks like:
\begin{equation}
    A_{1,0,0} = \lambda_{1,0,0}^{0,0,0} + \lambda_{1,0,0}^{1,0,0} (\varepsilon_1 \cdot p_2),
    \label{eq:3ptA100}
\end{equation}

where $\lambda_{n_1,n_2,n_3}^{l_1,l_2,l_3}$ is the coupling for the 3-point interaction involving particles at mass levels $n_i$ and spins $l_i$ ($i=1,2,3$). Using the vector propagator $\Pi_{\mu\nu} = \eta_{\mu\nu}+p_\mu p_\nu/M^2$ to sum over intermediate states of the internal particle\footnote{If $\alpha_0=-1$, level $n=1$ is massless and we have $\Pi_{\mu\nu}\to\eta_{\mu\nu}$ for tree residues.}, the residue of the 4-point amplitude at mass-level $n=1$ is then given by:
\begin{equation}
    R_1^{(4)} = \left(\lambda_{1,0,0}^{0,0,0} \right)^2 + \frac{1}{4}\left(\lambda_{1,0,0}^{1,0,0} \right)^2\left( 1 - 3\alpha_0 - 2X_{2,4} \right).
\end{equation}
Meanwhile, taking the residue at the level of the 4-point string integrand results in:
\begin{equation}
    \mathop{\mathrm{Res}}_{y=0}  \frac{dy}{y^{2}} (1+y)^{1-X_{2,4}-\alpha_0} = 1-X_{2,4}-\alpha_0.
\end{equation}
Matching both expressions, the couplings are fixed up to a sign:
\begin{equation}
    \left(\lambda_{1,0,0}^{0,0,0} \right)^2 = \frac{1}{2}(1+\alpha_0),\quad \left(\lambda_{1,0,0}^{1,0,0} \right)^2 = 2.
    \label{eq:3ptsCouplingsn1}
\end{equation}

If instead we wanted to read off couplings where two states where at higher levels, $\lambda_{0,n_2,n_3}^{0,l_2,l_3}$, we would need to look at the $5$-point leading singularity, as this coupling first appears in the middle vertex of the ladder diagrams. Meanwhile, to get the general $\lambda_{n_1,n_2,n_3}^{l_1,l_2,l_3}$-type couplings, we need the $6$-point leading singularity where we have a purely internal vertex (which comes from the diagram with propagators $X_{1,3},X_{3,5}$ and $X_{1,5}$). 

In general, at sufficiently high levels (starting at level $l_1=3$), the string spectrum famously has degenerate states, and therefore the matching procedure gets more intricate. Nonetheless, as we show in section \ref{sec:loop cuts}, for our purposes we are able to put stringent constraints on the allowed loop integrands using solely information about low level cuts (without degeneracies).

\section{Tree-level contour in $y$ space}
\label{sec:TreeContour}

In this section, we discuss the implementation of the $i\varepsilon$ prescription \cite{Witten:2013pra} for tree-level amplitudes in $y$-space. Using the techniques described in the previous section on how to access the different singular boundary regions, we provide a new and simple contour prescription for general $n$.

In practical terms, the need for a contour deformation near the boundaries of the moduli space comes from the presence of power divergences when the real part of any of the $X_{i,j}$'s is negative. It is well known by now that this arises as a consequence of working with Euclidean worldsheets, while our target space should be Lorentzian. The solution is to analytically continue the contour to the complex plane in the appropriate manner in the regions where parts of the worldsheet degenerate into long tubes.

As we explain below, this behavior is also the origin of a numerical limitation for any deformation that only differs from the naive one near the boundaries of the moduli space. The contour we obtain via the $i\varepsilon$ prescription is naturally divided into two regions: one where the worldsheet remains Euclidean, and one where the worldsheet is deformed into a Lorentzian one. As we make the $X_{i,j}$'s more negative, the size of the divergence on the Euclidean part increases, and the final result rendered by this contour comes from a fine cancellation between these two large (but finite) contributions. Therefore, for large and negative $X_{i,j}$ it becomes practically intractable to numerically evaluate this contour in any computer with finite precision. 

We illustrate this phenomenon in detail at $4$- and $5$-points, and discuss to which extent we can avoid these cancellations within our framework. Nonetheless, this practical limitation is manifestly unavoidable for extreme kinematics. In appendix \ref{app:numerical checks}, we present some numerical checks on the $i\varepsilon$ contour by comparing it to available exact results (as well as to the Pochhammer contour described in section \ref{sec:Pochhammer}). We show that even for negative kinematics, these contours work better than naively expected, with the loss of accuracy only happening when a surprisingly large number of resonances are relevant.

\subsection{4-point warm up}
\label{sec:4ptcontour}

Let's start by studying what the $i\varepsilon$-deformation looks like at 4-points. Of course, for this case, implementing a contour deformation to define the analytic continuation of the worldsheet integral is unnecessary since in this case we actually have a closed form formula for the integral -- the Beta function \cite{Veneziano:1968yb} -- which correctly gives us the analytic continuation. Nonetheless, this simplest example is still a useful starting point to explain how the standard prescription given in \cite{Witten:2013pra} translates into this representation, as well as to illustrate the practical problems that arise when we try to numerically evaluate these integrals at large kinematics. 

Choosing the underlying triangulation of $\mathcal{D}_4$ to contain curve $(1,3)$, we obtain the fat-graph corresponding to the standard $s$-channel diagram, and we can write the $4$-point amplitude in $y$-space as
\begin{equation}    \mathcal{A}_4[1234]=\int_{\mathbb{R}^+} \frac{dy_{1,3}}{y_{1,3}} y_{1,3}^{X_{1,3}} (1+y_{1,3})^{-c_{1,3}} = \frac{\Gamma(X_{1,3}) \Gamma(X_{2,4})}{\Gamma(X_{1,3}+X_{2,4})},
\label{eq:4ptY13}
\end{equation}
where $c_{1,3}=X_{1,3}+X_{2,4}$, and in terms of the standard $4$-point Mandelstams we have $X_{1,3}=s$, $X_{2,4}=t$ and $c_{1,3}=-u$.
As mentioned earlier, at $4$-points no blow-ups are needed because the singularities corresponding to the $s$ and $t$ channels are located at $y_{1,3} \to 0$ and $y_{1,3} \to +\infty$, respectively. Going to $t_{1,3}$-space we obtain the following integral over the real line
\begin{equation}   \mathcal{A}_4[1234]=\int_{\mathbb{R}} dt_{1,3} e^{-t_{1,3} X_{1,3}} (1+e^{-t_{1,3}})^{-c_{1,3}},
\label{eq:4pt-tspace}
\end{equation}
where the singularities are now instead located at $t_{1,3} \to \pm \infty$. In this case, the fan which divides $t_{1,3}$-space into the two diagrams is one-dimensional, with the $s$-channel cone being the region $t_{1,3}>0$ and the $t$-channel cone, $t_{1,3}<0$. 

According to \cite{Witten:2013pra}, when $X_{1,3}$ or $X_{2,4}$ are negative one should analytically continue the integral by changing to Lorentzian signature and making the Schwinger parameters imaginary near the degeneration regions ($t_{1,3} \to +\infty$ and $t_{1,3} \to -\infty$, respectively). Let's start by considering $X_{1,3}<0$. In this case, in the region $t_{1,3}>0$, we want to integrate in the standard contour until a certain point $t_{1,3}=R_\star$, and after this point deform the contour to make $t_{1,3}$ imaginary. Concretely, we integrate $t_{1,3}$ along
\begin{equation}
    \Gamma_{(1,3)}(\tau)=R_\star + i\tau, \text{ for }\tau \in [0,+\infty],
\end{equation} 
which yields the following integral
\begin{equation} 
\begin{aligned}
\mathcal{A}_4^{\Gamma_{(1,3)}}&= i e^{-R_\star X_{1,3}}\int_{0}^{\infty} d\tau e^{- i\tau X_{1,3}} (1+e^{-R_\star - i\tau})^{-c_{1,3}},\\
&= i e^{-R_\star X_{1,3}} \times \left(\sum_{k=0}^\infty e^{-2\pi i\, k X_{1,3}} \right) \times \int_{0}^{2\pi} d\tau e^{- i\tau X_{1,3}} (1+e^{-R_\star - i\tau})^{-c_{1,3}}.
\end{aligned}
\end{equation}
In the second line, we have used the almost-periodic nature of the integrand under $\tau\to\tau+2\pi$ to resum the contributions over the positive $\tau$ axis into an integral over a single period. We do this by factoring out a geometric series in the phase $\exp\left(2\pi i\, X_{1,3}\right)$, which formally only converges if we give $X_{1,3}$ a small imaginary part, $X_{1,3} \to X_{1,3} - i\epsilon$, with $\epsilon>0$. In addition, note that the integral over this new contour is well defined, since by taking $R_\star>0$ we can freely vary $\tau$ without ever crossing any branch cuts of the $\fh$-polynomial $(1+e^{-R_\star - i\tau})$.

In the case where the sum converges, we can finally write:
\begin{equation}    
\mathcal{A}_4^{\Gamma_{(1,3)}} = \frac{i e^{-R_\star X_{1,3}}}{1-e^{-i2\pi X_{1,3}}} \times \int_{0}^{2\pi} d\tau e^{- i\tau X_{1,3}} (1+e^{-R_\star - i\tau})^{-c_{1,3}}.
\label{eq:Gamma13}
\end{equation}
We can now simply define the analytic continuation of the geometric series as usual to extend this to all values of $X_{1,3}$, which is done before any integration.

Likewise, when $X_{2,4}<0$, we need to do a similar deformation in the region $t_{1,3}<0$. We start by following the original contour until we reach some $t_{1,3}=-R_\star$, and then deform the contour in the imaginary direction, 
\begin{equation}
    \Gamma_{(2,4)}(\tau)=-R_\star - i\tau, \text{ for }\tau \in [0,+\infty],
\end{equation} 
where this time we go along the negative imaginary axis. Now, since $t_{1,3}<0$, if we just flat-out deform the contour using the representation of the integrand as given in \eqref{eq:4pt-tspace}, then we will cross multiple branch cuts of the $\mathcal{F}$-polynomial. As explained in section \ref{sec:surface review}, we can avoid the branch cuts by using the representation of the integrand suited to this cone, $i.e.$ where we factor the leading term in the $\fh$-polynomial:
\begin{equation}
    \mathcal{I}_4^{C_{24}}= e^{t_{1,3} X_{2,4}} (1+e^{t_{1,3}})^{-c_{1,3}},  
\end{equation}
We can now freely integrate along $\Gamma_{(2,4)}$, from which we get the following contribution:
\begin{equation}    
\mathcal{A}_4^{\Gamma_{(2,4)}} = \frac{i e^{-R_\star X_{2,4}}}{1-e^{-i2\pi X_{2,4}}} \times \int_{0}^{2\pi} d\tau e^{- i\tau X_{2,4}} (1+e^{-R_\star - i\tau})^{-c_{1,3}}.
\label{eq:Gamma24}
\end{equation}

\begin{figure}[t]
    \centering    \includegraphics[width=\linewidth]{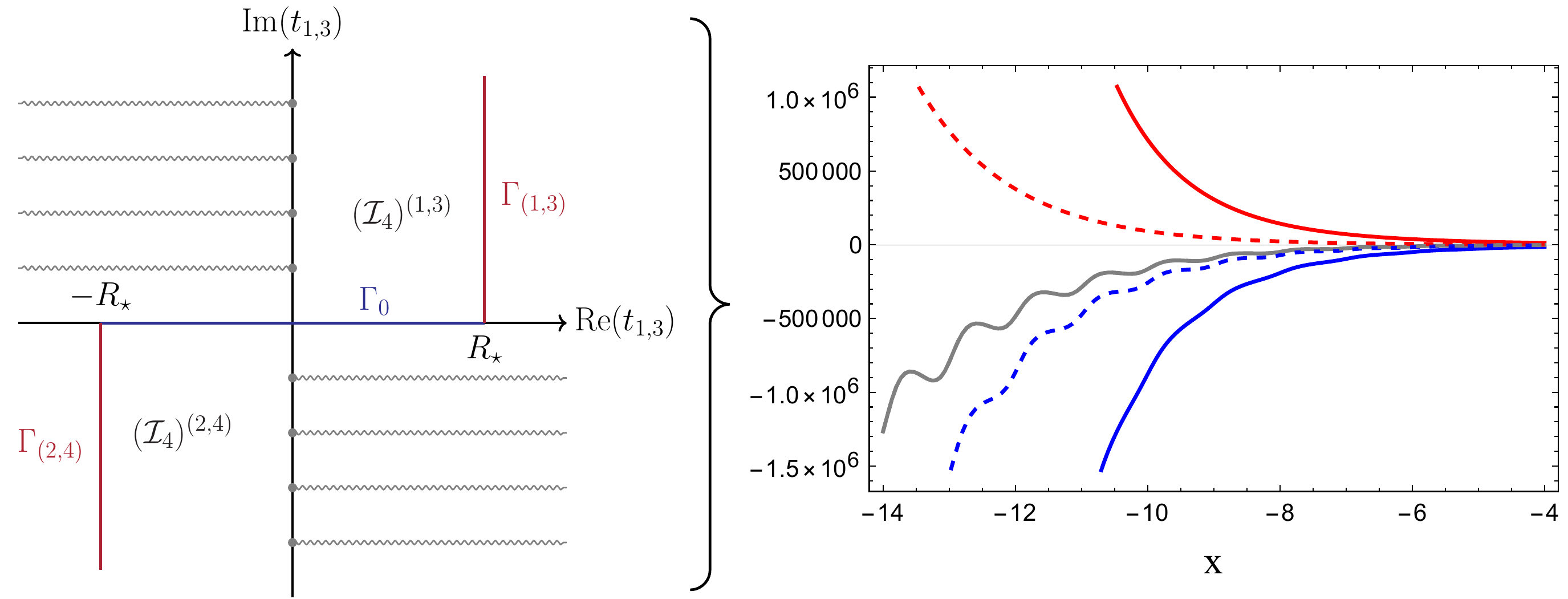}
    \caption{(Left) Integration contour at $4$-points.$\ \Gamma_0$ (in blue) is the region where we integrate along the original contour, and $\Gamma_{(1,3)} \cup \Gamma_{(2,4)}$ (in red) are the deformations where $t_{1,3}$ becomes imaginary. (Right) Illustration of the cancellation between the real part of the contributions from $\Gamma_0$ and $\Gamma_{(1,3)}\cup \Gamma_{(2,4)}$, for $X_{1,3} = x+0.4i$ and $X_{2,4} = -8 + 0.4i$. In solid blue and red we show the result from $\Gamma_0$ and $\Gamma_{(1,3)} \cup \Gamma_{(2,4)}$, respectively, for $R_\star =0.8$. In dashed we represent the analogous plots but for $R_\star =0.2$. In gray we show the real part of the $4$-point amplitude, given by the sum of the red and blue contributions.}
    \label{fig:4pt-Contour}
\end{figure}

The final contour of integration is depicted in figure \ref{fig:4pt-Contour} (left), so that the final $4$-point amplitude is given by the sum over the three pieces
\begin{equation}
    \mathcal{A}_4(X_{1,3},X_{2,4}) = \mathcal{A}_4^{\Gamma_{(1,3)}} +\mathcal{A}_4^{\Gamma_{(2,4)}} + \int_{\Gamma_0} e^{-t_{1,3} X_{1,3}} (1+e^{-t_{1,3}})^{-c_{1,3}}.
    \label{eq:A4FinCont}
\end{equation}
Here, $\Gamma_0$ is just the interval $[-R_\star,R_\star]$, where the worldsheet remains Euclidean. Of course, this is the contour for the case in which both $X_{1,3}<0$ and $X_{2,4}<0$. If instead one of these is positive, then we can integrate along the standard contour in the respective cone, as the integral is totally finite. 

Now, simply by looking at the form of $\mathcal{A}_4^{\Gamma_{(1,3)}}$ \eqref{eq:Gamma13} and $\mathcal{A}_4^{\Gamma_{(2,4)}}$ \eqref{eq:Gamma24}, we see that in both cases, if we keep $R_\star$ fixed and study kinematics such that $|X_{1,3}|,|X_{2,4}|\gg1$, then these contributions grow exponentially. A similarly large contribution also comes from the $\Gamma_0$ term in the boundaries of integration for $t_{1,3}=\pm R_\star$. Quite remarkably, these two contributions \textit{cancel} each other out in such a way that the total amplitude $\mathcal{A}_4(X_{1,3},X_{2,4})$ ends up being much smaller than each piece in the RHS of \eqref{eq:A4FinCont}! We show an example of this phenomenon on the right of figure \ref{fig:4pt-Contour} for the case where we fix $X_{2,4}$ and vary $X_{1,3}$.

At $4$-points, since the only constraint in $R_\star$ is for it to be positive, we can always consider adjusting $R_\star$ such that $|R_\star X_{1,3}|,|R_\star X_{2,4}|\ll1$. As we can see from figure \ref{fig:4pt-Contour}, by reducing the $R_\star$, the size of the cancellations between the two parts of the contour reduces significantly (dashed curves for smaller $R_\star$, and solid curves for larger $R_\star$).  However, eventually for extremely large kinematics, this requires such a small value of $R_\star$, that the finite integral from $[0,2\pi]$ becomes highly oscillatory, as we have that $|1+e^{\pm t_{1,3}}|$ gets closer and closer to the branch point at $\tau=\pi$. This means that, even though this is a finite integral over a compact integration domain, achieving high precision in numerical computations is complicated.

As we will see in the next subsection, at higher points we can no longer make $R_\star$ arbitrarily small due to the presence of different $\mathcal{F}$-polynomials, and therefore the cancellation aforementioned is \textit{unavoidable}.  

\subsection{Contour at 5-points}\label{sec:5-pt periodic}

Let's now proceed to understanding how the contour prescription described at $4$-points generalizes to $5$- and higher points. We first study the $5$-point amplitude, for which the parametrization in $(t_{1,3},t_{1,4})$-space is given in \eqref{eq:A5_tspace}.

Just like in the $4$-point case, we start by integrating over the original contour in some finite region centered at $(0,0)$ in $(t_{1,3},t_{1,4})$-space. Outside this region, in each cone (using the correct representation of the integrand), we perform the contour deformation where both $t_{1,3}$ and $t_{1,4}$ become purely imaginary. 

\begin{figure}[t]
    \centering
   \begin{tikzpicture}

\coordinate (T1) at (0,3);
\coordinate (T2) at (3,3);
\coordinate (T3) at (3,0);
\coordinate (C) at (0,0);

\fill[Blue!15] (C) -- (1.5,0) -- (1.5,1.5) -- (0,1.5)  --cycle;
\node[below] at (1.5,0) {$R_\star$};
\draw[->,thick] (-1,0) -- (4.2,0) node[right] {$t_{1,3}$}; 
\draw[->,thick] (0,-1) -- (0,4.2) node[above] {$t_{1,4}$}; 
\draw[thick] (0,0) -- (-1,1); 
\node[] at (0.75,0.75) {$\mathcal{R}^{S}$};
\node[] at (0.75,2.75) {$\mathcal{R}^{2}$};
\node[] at (2.75,0.75) {$\mathcal{R}^{1}$};
\node[] at (2.75,2.75) {$\mathcal{R}^{1,2}$};
\coordinate (P1) at (11,2.6);
\coordinate (P2) at (10,2.6);
\coordinate (P3) at (9.3,3.3);
\coordinate (P4) at (9.3,2.3);
\coordinate (P5) at (8.3,2.3);
\coordinate (P6) at (9,1.6);
\coordinate (P7) at (9,0.6);
\coordinate (P8) at (11,0.6);

\fill[Blue!15] (P1) -- (P2) -- (P3) -- (P4)  --(P5)  --(P6)  --(P7)  --(P8) --cycle;

\node[below] at (11.5,1.6) {$R_\star$};

\draw[Gray,dashed,thick] (1.5,0) -- (1.5,4); 
\draw[Gray,dashed,thick] (0,1.5) -- (4,1.5); 

\draw[<->,thick] (7,1.6) node[left] {$X_{2,5}$} -- (13,1.6) node[right] {$X_{1,3}$}; 
\draw[<->,thick] (10,-1) node[below] {$X_{3,5}$}-- (10,4.6) node[above] {$X_{1,4}$}; 
\draw[->,thick] (10,1.6) -- (7,4.6) node[above] {$X_{2,4}$};
\draw[Gray,dashed,thick] (11,-1) -- (11,4.6); 
\draw[Gray,dashed,thick] (10,2.6) -- (13,2.6); 
\draw[Gray,dashed,thick] (10,2.6) -- (8,4.6);
\draw[Gray,dashed,thick] (9,1.6) -- (7,3.6);
\draw[Gray,dashed,thick] (9.3,2.3) -- (9.3,4.6);
\draw[Gray,dashed,thick] (9.3,2.3) -- (7,2.3);
\draw[Gray,dashed,thick] (9,1.6) -- (9,-1);
\draw[Gray,dashed,thick] (13,0.6) --  (7,0.6);
\end{tikzpicture}
    \caption{Left: division of the $\{(1,3),(1,4)\}$ cone into region with deformed/undeformed variables as in \eqref{eq:regions 13,14}. Right: Division of the whole Feynman fan. The shaded octagon corresponds to the region where all variables are undeformed.}
    \label{fig:5pt-ConeDivision}
\end{figure}
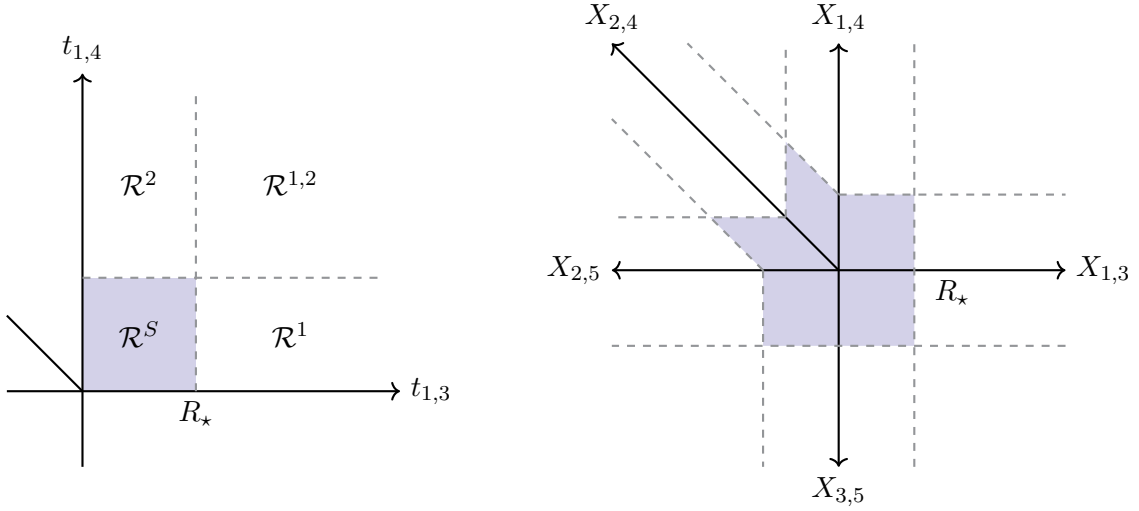

To begin with, we look at the cone $\{t_{1,3}>0,t_{1,4}>0\}$. We integrate along the original contour in the square $(t_{1,3},t_{1,4}) \in [0,R_\star]^2$, and outside this square we perform the contour deformation. By defining this square, we naturally divide the integration domain into the following four regions:
\begin{equation}\label{eq:regions 13,14}
\begin{aligned}
&\mathcal{R}^{S} =\{(t_{1,3},t_{1,4}) : (t_{1,3},t_{1,4})\in [0,R_\star]^2\}, \\
&\mathcal{R}^{1} =\{(t_{1,3},t_{1,4}):t_{1,3}>R_\star, t_{1,4} \in [0,R_\star]  \} , \\
&\mathcal{R}^{2} =\{(t_{1,3},t_{1,4}): t_{1,3} \in [0,R_\star], t_{1,4}>R_\star \}, \\
&\mathcal{R}^{1,2} =\{(t_{1,3},t_{1,4}): (t_{1,3},t_{1,4}) > R_\star\} ,
\end{aligned}
\end{equation}
represented in the left of figure \ref{fig:5pt-ConeDivision}. In the regions $\mathcal{R}^{1}/\mathcal{R}^{2}$ we deform the contour in $t_{1,3}/t_{1,4}$, respectively, and in $\mathcal{R}^{1,2}$ we deform both integration variables. 

Now, as pointed out earlier, we need to make $R_\star$ sufficiently large to ensure that none of the $\mathcal{F}$-polynomials vanishes for any of the deformations above. For the $\mathcal{F}$-polynomials $\mathcal{F}_{1,3}=(1+e^{-t_{1,3}})$ and $\mathcal{F}_{2,4}=(1+e^{-t_{1,4}})$ (which are the same as the one appearing at $4$-points), this is guaranteed as long as $R_\star>0$. However, we have an additional $\mathcal{F}$-polynomial that mixes the dependence in $t_{1,3}$ and $t_{1,4}$, $\mathcal{F}_{1,4} = (1+e^{-t_{1,4}} +e^{-t_{1,4} - t_{1,3}})$, for which we must have 
\begin{equation}
    |e^{-t_{1,4}} +e^{-t_{1,4} - t_{1,3}}|<1 \Leftrightarrow |e^{-t_{1,4}}| < \frac{1}{|1 +e^{ - t_{1,3}}|}, \, \, \, \text{ for any }(t_{1,3},t_{1,4})\in \mathbb{R}^2_+.
    \label{eq:condRmax}
\end{equation}
For this to hold in the three regions where we apply the deformations, we must have 
\begin{equation}
    e^{-R_\star} < 1/2 \quad \Leftrightarrow \quad R_\star > \log{2}.
\end{equation}

In summary, due to the presence of $\mathcal{F}_{1,4}$, we find that there is a \textit{non-zero} lower-bound on how small we can make $R_\star$. At higher-points, we see the appearance of more interesting $\mathcal{F}$-polynomials which similarly imply different lower bounds on $R_\star$, but we postpone this discussion for the next section. 

Having understood what values $R_\star$ is allowed to take, let's proceed to giving an explicit expression for the contour in the different regions outside the square. Starting with $\mathcal{R}^{1}$, in this region we integrate over $t_{1,4}$ along the original contour, but deform $t_{1,3}$ into the imaginary direction to get:
\begin{equation}
    \Gamma_{1}(\tau_{1,3},\tau_{1,4}) = (R_\star + i\tau_{1,3},\tau_{1,4}), \quad \text{for } \tau_{1,3}\in[0,+\infty), \tau_{1,4} \in [0,R_\star]. 
\end{equation}

Just like at $4$-points, we have that the integrand is periodic (up to a phase) in $\tau_{1,3} \to \tau_{1,3} + 2\pi$. Using this fact, we can rewrite the integration over $\tau_{1,3}$ as a compact integration from $[0,2\pi]$ times an overall factor, which just like at $4$-points gives
\begin{equation}    \mathcal{A}_5^{\Gamma_{1}} = \frac{i e^{-R_\star X_{1,3}}}{1-e^{-i2\pi X_{1,3}}} \times \int_{0}^{2\pi} d\tau_{1,3}  \int_{0}^{R_\star} d\tau_{1,4} \, \, \mathcal{I}_5^{C_{13,14}}[ \Gamma_{1}(\tau_{1,3},\tau_{1,4})].
\end{equation}
Here, we have again analytically continued the geometric series to all values of $X_{1,3}$. $\mathcal{I}_5^{C_{13,14}}$ is simply the representation of the integrand suited to cone $\{(1,3),(1,4)\}$, given in \eqref{eq:A5_tspace}. Prescribing a similar contour for region $\mathcal{R}^{2}$, we obtain:
\begin{equation}
\begin{aligned}
    &\Gamma_{2}(\tau_{1,3},\tau_{1,4}) = (\tau_{1,3},R_\star + i\tau_{1,4}), \quad \text{for } \tau_{1,3}\in[0,R_\star], \tau_{1,4} \in [0,+\infty), \\[1em]
  &\Rightarrow \, \, \mathcal{A}_5^{\Gamma_{2}} = \frac{i e^{-R_\star X_{1,4}}}{1-e^{-i2\pi X_{1,4}}} \times \int_{0}^{R_\star} d\tau_{1,3} \int_{0}^{2\pi} d\tau_{1,4}   \, \, \mathcal{I}_5^{C_{13,14}}[ \Gamma_{2}(\tau_{1,3},\tau_{1,4})].
\end{aligned}
\end{equation}

Finally, in $\mathcal{R}^{1,2}$ we deform both $t_{1,3}$ and $t_{1,4}$ in the imaginary direction, and so factoring out the phase from both integrations we get:
\begin{equation}
\begin{aligned}
    &\Gamma_{1,2}(\tau_{1,3},\tau_{1,4}) = (R_\star + i\tau_{1,3},R_\star + i\tau_{1,4}), \quad \text{for } \tau_{1,3},\tau_{1,4}\in[0,+\infty),  \\[1em]
  &\Rightarrow \, \, \mathcal{A}_5^{\Gamma_{1,2}} =\frac{(i)^2 e^{-R_\star (X_{1,3}+X_{1,4})}}{(1-e^{2\pi i\, X_{1,3}})(1-e^{2\pi i\, X_{1,4}})} \times \int_{[0,2\pi]^2} d\tau_{1,3} d\tau_{1,4}   \, \, \mathcal{I}_5^{C_{13,14}}[ \Gamma_{1,2}(\tau_{1,3},\tau_{1,4})].
  \end{aligned}
\end{equation}
Thus, we can write the contribution from cone $\{(1,3),(1,4)\}$ as the sum of the contributions from the four different regions:
\begin{equation}
    \mathcal{A}_5^{C_{13,14}} = \mathcal{A}_5^{\Gamma_{S}}+\mathcal{A}_5^{\Gamma_{1}} + \mathcal{A}_5^{\Gamma_{2}} + \mathcal{A}_5^{\Gamma_{1,2}}, 
\end{equation}
where $\Gamma_S$ is the original contour, but only covering the square $[0,R_\star]^2$ (see figure \ref{fig:5pt-ConeDivision}, left).

Having defined the contour in the cone $\{(1,3),(1,4)\}$, we can trivially extend it to the remaining cones in the fan by using the maps defined in section \ref{sec:FanReview}. That is, starting in some cone, we can apply the linear transformation given by \eqref{eq:cone_map} to map it to the positive quadrant. Provided we use the appropriate cone-representation of the integrand (where we have pulled out the leading monomials), we can apply the contour deformations above to the respective regions. To give a concrete example, for cone $\{(1,4),(2,4)\}$ we define the integrand to be
\begin{equation}
    \mathcal{I}_5^{C_{14,24}} (\hat{t}_{1,4},\hat{t}_{2,4}) = \mathcal{I}^{C_{14,24}}_5\left(   t_{1,k} = \medmath{\sum_{X\in\{(1,4),(2,4)\}} }(\vec{g}_X)_{1,k}  \, \hat{t}_X \right),
\end{equation}
where $\hat{t}_{1,4},\hat{t}_{2,4} \in \mathbb{R}_+$, and so we can similarly divide the domain into the square region $\mathcal{R}^S = (\hat{t}_{1,4},\hat{t}_{2,4}) \in [0,R_\star]^2$  and the remaining regions $\mathcal{R}^1$,$\mathcal{R}^2$ and $\mathcal{R}^{1,2}$, where we perform the contour deformation accordingly. Proceeding in the same way for the remaining cones, we can write the final answer as follows:
\begin{equation}
    \mathcal{A}_5 = \sum_{\mathcal{T} \text{ triang. } \mathcal{D}_5} \mathcal{A}_5^{C_{\mathcal{T}}}, \quad \text{with } \mathcal{A}_5^{C_{\mathcal{T}}} = \int_{\Gamma_S} \mathcal{I}^{C_{\mathcal{T}}}_5(\hat{t}_{P_\mathcal{T}}) + \sum_{i=1,2}\int_{\Gamma_i} \mathcal{I}^{C_{\mathcal{T}}}_5(\hat{\tau}_{P_\mathcal{T}}) + \int_{\Gamma_{1,2}} \mathcal{I}^{C_{\mathcal{T}}}_5(\hat{\tau}_{P_\mathcal{T}}),
\end{equation}
where $\mathcal{T}$ is a triangulation of the disk, labeling a particular diagram with propagators $X_P$, and $\mathcal{I}^{C_{\mathcal{T}}}$ is the suitable representation of the integrand in the respective cone $C_{\mathcal{T}}$, parametrized with coordinates $\hat{t}_P/\hat{\tau}_P$. Hence, we have that each cone is broken up into four regions, which are depicted on the right of figure \ref{fig:5pt-ConeDivision}. The shaded blue region is the part in which there is \textit{no} contour deformation (corresponding to the union of all the $\mathcal{R}^S$ of the different cones), and so the worldsheet remains Euclidean. Note that in our construction this finite regions ends up not being an associahedron (in contrast with \cite{Eberhardt:2024twy}). 

All in all, to implement the contour we have to sum over the different cones of the fan -- each of which corresponding to a diagram -- and each cone is further divided into $2^{n-3}$ regions. As a result we get that the region of the moduli space we integrate over \textit{without} any deformation ends up being the octagon shaded in \ref{fig:5pt-ConeDivision}.

Of course, there is nothing unique about this shape. For instance, one could instead define a contour by taking a circle centered at the origin with radius $R_{\star}$, integrating along the real plane inside the circle and deforming outside of it as follows: 
\begin{equation}
    \begin{gathered}
	\begin{tikzpicture}[line width=0.6,scale=0.3,line cap=round,every node/.style={font=\normalsize}]
		\coordinate (T1) at (0,3);
\coordinate (T2) at (3,3);
\coordinate (T3) at (3,0);
\coordinate (C) at (0,0);

\node[below,style={font=\tiny}] at (2.3,0) {$R_\star$};
\fill[Blue!15] (0,0) circle[radius=1.7];
\draw[Gray,dashed,thick] (0,0) circle[radius=1.7];
\draw[<->,thick] (-4.2,0) node[left] {$X_{2,5}$}-- (4.2,0) node[right] {$X_{1,3}$}; 
\draw[<->,thick] (0,-4.2) node[below] {$X_{3,5}$}-- (0,4.2) node[above] {$X_{1,4}$}; 
\draw[->,thick] (0,0) -- (-4,4) node[left] {$X_{2,4}$}; 
\draw[->,Gray,thick] (2,2) to[out=45,in=120] (10,1);
	\end{tikzpicture}
\end{gathered} \, \, \Gamma(t_{1,3},t_{1,4}) = ((R_\star + (1+i m)\delta)\cos{\theta},(R_\star + (1+i m)\delta )\sin{\theta}),
\end{equation}
where $\delta \in [0,+\infty)$, $\theta \in [0,2\pi]$, and $m>0$ quantifies some imaginary slope which ensures that the integral in $\delta$ converges as long as for all $X_{i,j} \to X_{i,j} -i\epsilon$ we have 
\begin{equation}
    \text{Re}[(X_{i,j}-i\epsilon)(1+i m)]>0 \Leftrightarrow X_{i,j}+\epsilon\, m>0.
    \label{eq:convDelta}
\end{equation}
In order to avoid crossing any branch cuts outside the circle, we take $R_\star > \log{2}$ and pick the appropriate representations of the integrand in each cone, which now map to different ranges in $\theta$. 

In the limiting case where $m\gg1$, we recover the situation where outside the circle the contour is purely imaginary. However, since in this case we parametrize the integrand in polar coordinates $(\delta,\theta)$, we have that it is \textit{no longer} periodic as $\delta \to \delta + 2 \pi$. As a result, we can't factor out the sum over phases and use it to define the analytic continuation of the contour to \textit{all} kinematics. Instead in this case, since we have to integrate over the full domain $\delta = [0,\infty)$, if a some $X_{i,j}<0$ we must give it a small imaginary part, $X_{i,j} - i\epsilon$ to make the integral converge, and pick $m$ such that \eqref{eq:convDelta} holds.

Similarly, if we consider any other division of the undeformed and deformed regions that is naturally implemented via a non-linear transformation of the coordinates $(t_{1,3},t_{1,4})$, then we loose the periodicity of the integrand, and therefore the ability to find a contour prescription which applies for \textit{all} kinematics. 

From a practical perspective of actually evaluating this integral, we find that as we take kinematics $X_{i,j}$ in the physical region and let $|X_{i,j}| \gg 1$, we once more run into the cancellation pointed out in the $4$-point case in section \ref{sec:4ptcontour}. While at $4$-points we can change $R_\star$ to reduce the cancellation, at $5$-points there is less freedom in $R_\star$ due to the presence of a new $\mathcal{F}$-polynomial. However, there is nothing preventing us from having a different pair of $R_\star^{(i,j)}$ for each cone in figure \ref{fig:5pt-ConeDivision}, and we can use this fact to optimize the computation.  
\subsubsection{Changing $R^{ij}_{\star}$}

\begin{figure}[t]
    \centering
    \begin{tikzpicture}[line width=0.8,scale=1.6,line cap=round,every node/.style={font=\normalsize}]
		\coordinate (T1) at (0,3);
\coordinate (T2) at (3,3);
\coordinate (T3) at (3,0);
\coordinate (C) at (0,0);
\coordinate (P1) at (0,1.7);
\coordinate (P2) at (0.7,1.7);
\coordinate (P3) at (0.7,-1);
\coordinate (P4) at (-1.8,-1);
\coordinate (P5) at (-1.8,0);
\coordinate (P6) at (-2.3,0.5);
\coordinate (P7) at (-0.5,0.5);
\coordinate (P8) at (-0.5,2.3);

\fill[Blue!15] (P1) -- (P2) -- (P3) -- (P4)  --(P5)  --(P6)  --(P7)  --(P8) --cycle;

\draw[Blue] (P1) -- (P2) -- (P3) -- (P4)  --(P5)  --(P6)  --(P7)  --(P8) --cycle;
 
\node[style={font=\small}] at (2.5,2.5) { $\displaystyle
      \begin{aligned}
       &(1+e^{-t_{1,3}}), (1+e^{-t_{1,4}}) \\
    &(1+e^{-t_{1,4}}+e^{-t_{1,3}-t_{1,4}})
      \end{aligned}$};
\node[style={font=\small}] at (2.5,-2.3) { $\displaystyle
      \begin{aligned}
       &(1+e^{-\hat{t}_{1,3}}), (1+e^{-\hat{t}_{3,5}}) \\
    &(1+e^{-\hat{t}_{1,3}}+e^{-\hat{t}_{3,5}})
      \end{aligned}$};
\draw[->,Maroon,thick] (1.3,-2.5) to[out=180,in=-120] (1.1,-1.4) ;
\node[style={font=\small}] at (-2.5,-2.3) { $\displaystyle
      \begin{aligned}
       &(1+e^{-\hat{t}_{2,5}}), (1+e^{-\hat{t}_{3,5}}) \\
    &(1+e^{-\hat{t}_{2,5}}+e^{-\hat{t}_{2,5}-\hat{t}_{3,5}})
      \end{aligned}$};
\node[style={font=\small}] at (-3.5,1.5) { $\displaystyle
      \begin{aligned}
       &(1+e^{-\hat{t}_{2,4}})\\
       &(1+e^{-\hat{t}_{2,5}-\hat{t}_{2,4}}) \\
    &(1+e^{-\hat{t}_{2,5}}+e^{-\hat{t}_{2,5}-\hat{t}_{2,4}})
      \end{aligned}$};
\node[style={font=\small}] at (-1.7,3.2) { $\displaystyle
      \begin{aligned}
    (1+e^{-\hat{t}_{1,4}}+e^{-\hat{t}_{1,4}-\hat{t}_{2,4}})&\\
       (1+e^{-\hat{t}_{1,4}-\hat{t}_{2,4}})&\\
     (1+e^{-\hat{t}_{2,4}})&
      \end{aligned}$};

\draw[<->,thick] (-2.7,0) node[left] {$X_{2,5}$}-- (2.7,0) node[right] {$X_{1,3}$}; 
\draw[<->,thick] (0,-2.7) node[below] {$X_{3,5}$}-- (0,2.7) node[above] {$X_{1,4}$}; 
\node[Blue,style={font=\footnotesize},rotate=-90]
  at (0.9,0.6)             
  {$R_\star^{13}>0$};
\draw[->,thick] (0,0) -- (-2.5,2.5) node[left] {$X_{2,4}$};
\node[Blue,style={font=\footnotesize},rotate=90]
  at (-0.7,1.5)             
  {$R_\star^{24}>0$};
\draw (0,1.5) -- ++(0.1,0) node[right,style={font=\small}] {$\ln{2}$};
\draw (-1.5,0) -- ++(0,-0.1) node[below,style={font=\small}] {$\ln{2}$};
\node[style={font=\footnotesize},Blue,rotate=90]
  at (-2,-0.6)   
  {$R_\star^{25}>\ln{2}$};
\node[style={font=\footnotesize},Blue,left]
  at (0,-1.2)   
  {$R_\star^{35}>0$};
\node[style={font=\footnotesize},Blue,right]
  at (0,1.9)   
  {$R_\star^{14}>\ln{2}$};

\node[style={font=\small},Maroon,right]
  at (1.1,-1.2)   
  {$e^{-R_\star^{25}}+e^{-R_\star^{13}}<1$};

	\end{tikzpicture}
    \caption{Conditions on the minimum values for the region boundaries $R^{ij}_\star$ in each cone, which are determined by the $\fh$-polynomials appearing in the corresponding representation of the integrand.}
    \label{fig:Fan_FPoly}
\end{figure}
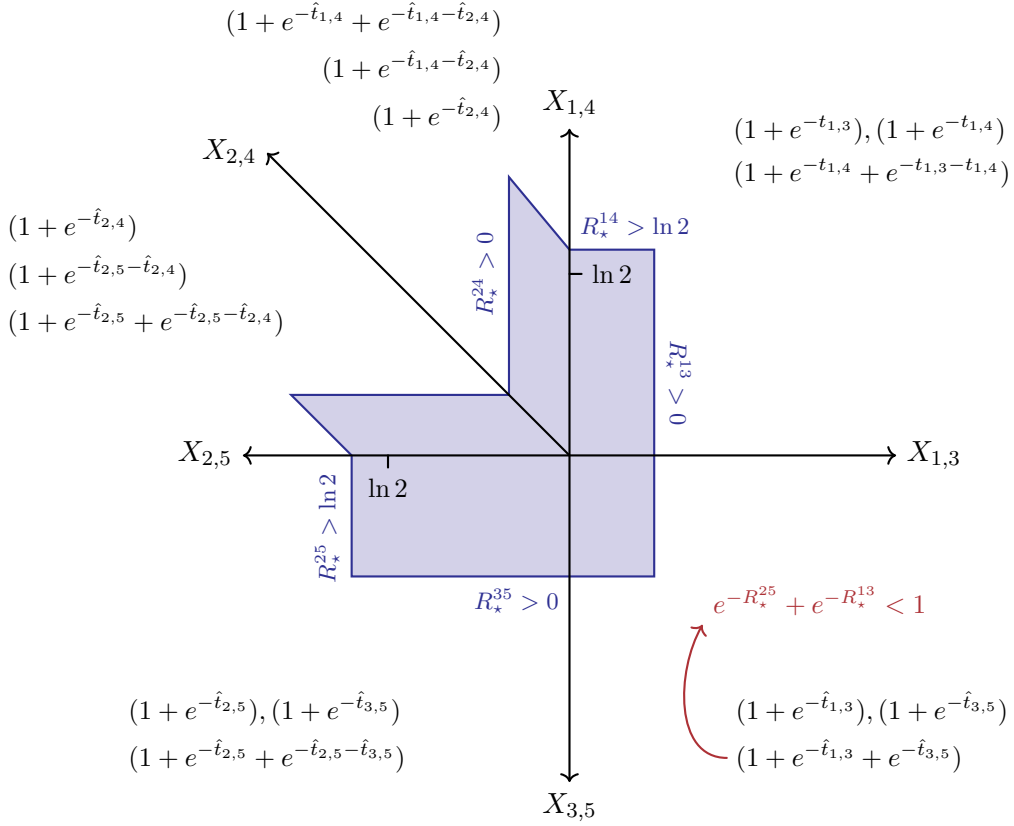

As explained earlier, we derive the constraint in $R_\star$ by looking at the $\mathcal{F}$-polynomial $(1 +e^{-t_{1,4}} + e^{-t_{1,3}-t_{1,4}})$, and ask for it to never be zero when we make the contour deformation in $t_{1,3}/t_{1,4}$, which leads to the condition \eqref{eq:condRmax}. This implies that in the $t_{1,4}$ direction, in particular in region $\mathcal{R}_2$ of cone $\{(1,3),(1,4)\}$, we can only deform the contour to make $t_{1,4}$ imaginary for $t_{1,4}> \ln{2}$. For simplicity, we imposed that all regions $\mathcal{R}_i$ started when the respective $t>\ln{2}$. 

However, this is not a requirement. Already for region $\rh_1$ in the cone $\{(1,3),(1,4)\}$, we can start deforming $t_{1,3}$ at any point, since we won't cross any branch cuts. In addition, when we do the change of coordinates \eqref{eq:cone_map} to map the different cones into the positive quadrant, we find new  $\mathcal{F}$-polynomials which yield different bounds on the respective $R_\star^{ij}$. 

In figure \ref{fig:Fan_FPoly}, we show the $5$-point fan as well as the $\mathcal{F}$-polynomials entering in each cone, in terms of the coordinates $\hat{t}_i$ suited to the cone. By doing the same analysis we did in cone $\{(1,3),(1,4)\}$, we find that we need $R^{25}_\star > \ln{2}$, while $R^{24}_\star,R^{35}_\star > 0$. Additionally, due to the presence of the $\fh$-polynomial $(1+e^{-\hat{t}_{1,3}}+e^{-\hat{t}_{3,5}})$, we have that 
\begin{equation}
    e^{-R_\star^{13}} +   e^{-R_\star^{35}} <1.
    \label{eq:Const1335}
\end{equation}

Therefore, the contour as prescribed in the previous section holds for any octagon, as long as each $R_\star^{ij}$ satisfies the conditions above (also shown in figure \ref{fig:Fan_FPoly}). In particular, just like at $4$-points, we can use this extra freedom to mitigate the big cancellations happening when we take $|X_{i,j}|\gg1$. Of course, this freedom is not enough solve the problem in all kinematic regions, e.g. when $X_{1,4}\gg 1$ or $X_{2,5}\gg 1$ we can't make the respective $R_\star$ arbitrarily small. Nonetheless, when $X_{2,4}$ is large, we can make $R_\star^{24}$ as small as we want, and the same is true when either $X_{1,3}$ or $X_{3,5}$ get large. If instead $X_{1,3}$ and $X_{3,5}$ get large simultaneously we can't avoid the cancellation, since \eqref{eq:Const1335} prevents us from taking $R_\star^{13}$ and $R_\star^{35}$ arbitrarily small at the same time. 

As we increase $n$, the number of cones in the fan grows exponentially, and it becomes harder to do a similar analysis for all the $\mathcal{F}$-polynomials appearing in the different cones. For this reason,  what we will care about for general $n$ is finding the minimum common $R_\star$, and applying the contour deformation as defined previously. Nonetheless, the analysis presented here is useful at higher $n$ in case one wants to study a specific kinematic limit where only a subset of the variables are taken large. In such cases, one can simply ask about the constraints on the relevant subset of $R_\star^{ij}$, and use them to find the optimal (with least cancellations) division of regions to define the contour.

\subsection{General $n$}
\label{sec:Generaln}

Given the definition of the contour at $5$-points, the generalization to higher multiplicities is now straightforward. At $n$-points, we parametrize the full string amplitude in terms of the $(n-3)$ $t_{1,j}$'s for $j\in\{3,\cdots,n-1\}$, and by collecting the $g$-vectors of all curves in $\mathcal{D}_n$, we divide $t$-space into the respective cones (each of which labeling a diagram). Using \eqref{eq:cone_map}, we can map each cone $C_{\mathcal{T}}$ to the positive orthant of $\mathbb{R}^{n-3}$ parametrized in terms of the $\hat{t}_P$'s, where $P$ corresponds to the labels of the curves which enter in $C_{\mathcal{T}}$, $P\in \mathcal{T}$. Each positive orthant is further divided into $2^{n-3}$ regions, coming from the division of the variables $\hat{t}_P$ into $\hat{t}_P<R_\star$ and  $\hat{t}_P>R_\star$, for $P\in \mathcal{T}$. Therefore, a particular region is characterized by the subsets of deformed and undeformed variables, which we call $C_\mathcal{T}^{>}$ and $C_\mathcal{T}^{<}$, respectively. More precisely, they are defined as:  
\begin{equation}
    C_\mathcal{T}^{>} = \{ P \in \mathcal{T} : \hat{t}_P>R_\star\}, \quad C_\mathcal{T}^{<} = \{ Q \in  \mathcal{T} : \hat{t}_Q<R_\star\}.
\end{equation}
Denoting by $\mathcal{R}^{C_\mathcal{T}^{<}}$ the corresponding region, we can write its contribution for the $n$-point amplitude as:
\begin{equation}
    \mathcal{A}_n^{\mathcal{R}^{C_\mathcal{T}^{<}}} = \left(\prod_{P \in C^{>}_\mathcal{T}} \frac{i e^{-R_\star X_P}}{1-e^{-i2\pi X_P}}\right) \times \int_{0}^{2\pi} \prod_{P \in C^{>}_\mathcal{T}} d \hat{\tau}_P  \int_{0}^{R_\star} \prod_{Q \in C^{<}_\mathcal{T}} d \hat{t}_Q\  \mathcal{I}_n^{C_\mathcal{T}}\left[\{\hat{t}_P = R_\star + i \hat{\tau}_P\},\{\hat{t}_Q \}\right],
\end{equation}
where once again $\mathcal{I}_n^{C_\mathcal{T}}$ is the representation of the integrand having factored out the leading terms out of all $\mathcal{F}$-polynomials. So, just like at $5$-points, we have that for all the coordinates $\hat{t}_P>R_\star$ we implement the contour deformation that makes the respective Schwinger parameters imaginary, and due to the quasi-periodicity of the integrand we can factor our a phase and integrate over a single copy of the in $[0,2\pi]$. On the other hand, for the variables $\hat{t}_Q<R_\star$ we just integrate over the original contour. 

The final answer is then simply obtained by summing over all possible cones, as well as all such regions inside each cone. The one thing left to discuss is how to determine $R_\star$ uniformly, ensuring that we do not cross any branch cuts in the different regions in which we make the Schwinger parameters imaginary. To do this, let's start by considering the original positive orthant, given by $\{t_{1,k}>0\}$, and look at the different $\mathcal{F}$-polynomials entering in this cone -- which correspond to those in the original representation of the integrand as given in \eqref{eq:FPols}. Concretely, let's focus on the one containing the biggest number of monomials in $y_{1,k}$, which in $t$-space becomes 
\begin{equation}
    \mathcal{F}_{1,n-1}[\{t_{1,k}\}] = 1 + e^{-t_{1,n-1}} + e^{-t_{1,n-1} -t_{1,n-2}} + \cdots + e^{-\sum_{j=3}^{n-1}t_{1,j}}. 
\end{equation}

Now, the map which allows us to rotate the different cones into the positive orthant is a simple linear transformation between the $t_{1,k}$ and the corresponding $\hat{t}_P$. Therefore, the image of $\mathcal{F}_{1,n-1}$ in the different cones can be written as follows:
\begin{equation}
   \mathcal{F}_{1,n-1} = 1 + R_1 e^{i\theta_1} +R_2 e^{i\theta_2} + \cdots+ R_{n-3}e^{i\theta_{n-3}}.
   \label{eq:F1nm1}
\end{equation}
We have allowed our Schwinger parameters to have some imaginary parts, which are encoded by the $\theta_i$, and the $R_i$ are exponentials of weighted sums over the real part of $\hat{t}$, $i.e.$
\begin{equation}
    R_i = \exp\left(-\sum_{P \in \mathcal{T}} w_P \text{Re}[\hat{t}_P]\right), \quad \text{with }w_P =\{0,1\}.
\end{equation}

In order to make $\mathcal{F}_{1,n-1}$ in \eqref{eq:F1nm1} always positive, it is enough to have $R_i>1/(n-3)$ for all $i$. Taking into account the form of each $R_i$ in the $\hat{t}_P$, we have that this is always satisfied if
\begin{equation}
    \text{Re}[\hat{t}_P] > \ln{(n-3)} \quad \text{ for all } P\in \mathcal{T}.
\end{equation}

Of course, just like we saw at $5$-points, this condition is overly restrictive for some $\hat{t}_P $, while still being the best bound for others. Now, since all the remaining $\mathcal{F}-$polynomials have fewer terms, we have that the bound above will also ensure these won't vanish, and therefore if we choose
\begin{equation}
    R_\star > \ln{(n-3)},
    \label{eq:boundRmax}
\end{equation}
we have a global cutoff $R_\star$ for all the cones contributing to the $n$-point amplitude.

\section{Pochhammer contour in $y$ space}\label{sec:Pochhammer}

So far, we have exploited the parametrization of the string integral in terms of positive coordinates to effectively implement the Feynman $i\varepsilon$ prescription on the string amplitudes, as described in \cite{Witten:2013pra}. This is well motivated from a physical viewpoint, since it transforms the worldsheet signature from Euclidean to Lorentzian in the regions where it degenerates. However, there are other approaches to define a convergent integral over the moduli space. In the case of the 4-point tree amplitude, which can be written in terms of the Euler Beta function, an analytic continuation has long been known by use of the Pochhammer contour \cite{Pochhammer1890}. This is a closed contour defined in the complex plane with the branch points of the Koba-Nielsen form removed. Since it is compact, the resulting integral is clearly finite for any value of the kinematics, and thus can be defined as an analytic continuation of our string integral as long as the values of both coincide in the regions where the latter converges.

Here, we introduce a generalization of the classic Pochhammer contour for the Beta function to arbitrary number of points. A version of the contour in terms of worldsheet coordinates was constructed at five points in \cite{Hanson}. However, as we have seen, in these variables it's necessary to expose the degenerations of the integrand by manually blowing up the singularities. This makes it increasingly complicated to explicitly realize a closed contour at higher points. This problem once again disappears when we use the positive parametrization of string amplitudes, in $y$-space. 

\begin{figure}[t]
    \centering
    \includegraphics[width=.5\textwidth,valign=c]{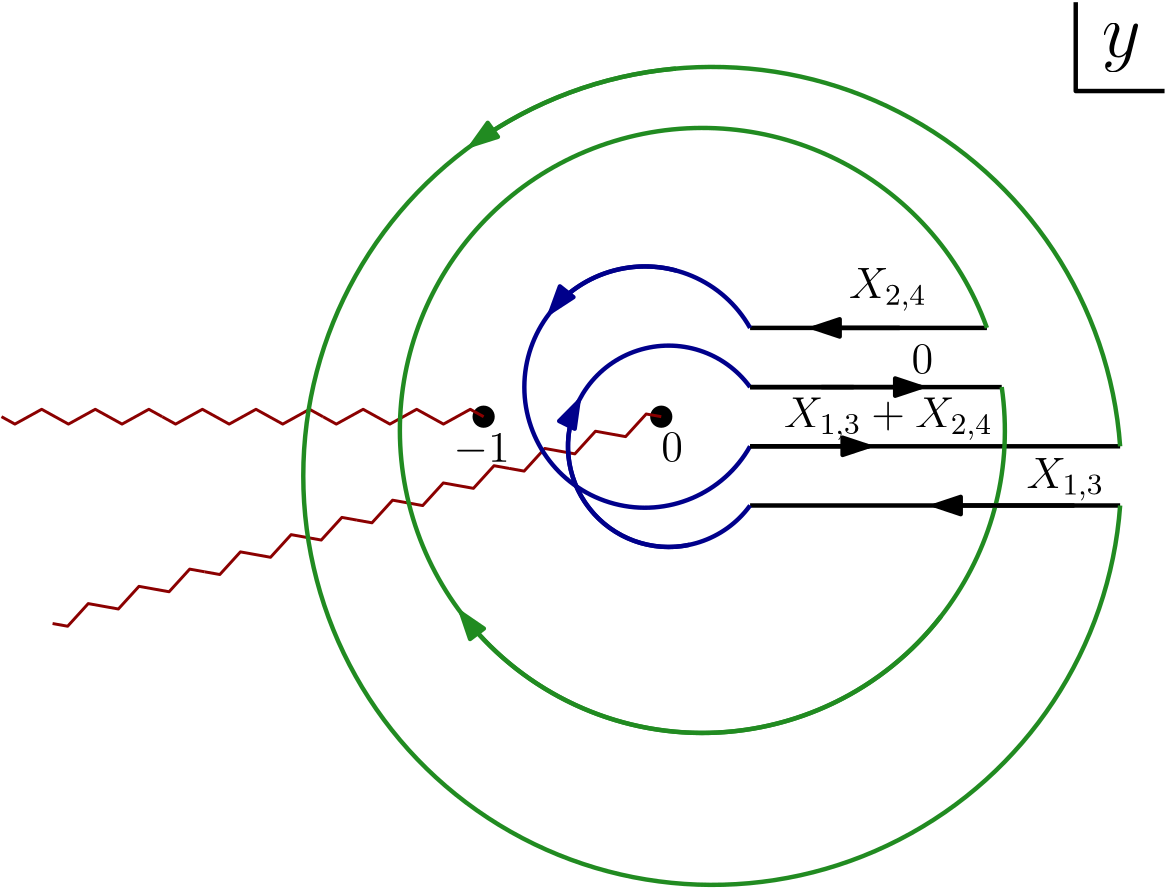} 
    \caption{Pochhammer contour for the 4-point amplitude in $y$-space. At $y=-1$ and $y=0$, we show the two branch points and respective branch cuts (in red) of $I_4$. The closed contour is made out of $4$ intervals along the real line, each in a different sheet, and $4$ circles, two crossing the branch cut from $y=0$ (in blue), and two crossing both brach cuts (in green). Each of the real intervals has a phase factor given by $\exp{-2\pi i \phi}$, with $\phi \in \{0,X_{1,3},X_{2,4},X_{1,3}+X_{2,4}\}$.}
    \label{fig:Pochhammer 4-pts} 
\end{figure}

Let's start by considering the simplest example of the 4-point amplitude. In $y$-variables, the integrand is given in \eqref{eq:4ptY13}, but repeated here for convenience:
\begin{equation}\label{eq:4-pt integrand Poch}
    \mathcal{A}_4 = \int_0^\infty dy \ I_4(y), \quad \text{with} \quad I_4(y)=  y^{X_{1,3}-1}(1+y)^{-X_{1,3}-X_{2,4}}.
\end{equation}
So the integrand, $I_4$, defined in $y$ space has branch cuts anchored at $y=0$, $y=-1$ and $y=\infty$ (represented in red in figure \ref{fig:Pochhammer 4-pts}), so that when going around each of these points we pick up the phases, $e^{\pm 2\pi i X_{1,3}}$, $e^{\pm 2\pi i (X_{1,3}+X_{2,4})}$ and  $e^{\pm 2\pi i X_{2,4}}$, respectively. 

We now want to define a closed contour $\gamma$ on which the $4$-point integral is finite. To do that, we consider four copies of a finite horizontal path along the positive real line -- black lines in figure \ref{fig:Pochhammer 4-pts}. Each line will be defined on a different branch of the complex plane, and we assign one of the phases $\{0,\, X_{1,3},\, X_{2,4},\,X_{1,3}+X_{2,4}\}$ to each of them. Their precise length, shape or positioning isn't important, since we can always deform them freely within the same branch. We then connect the endpoints of these lines with circles around $y=0$ or $y=-1$ (given in blue and green in figure \ref{fig:Pochhammer 4-pts}), in such a way that they precisely pick up the appropriate phase difference when crossing the branch cuts. The resulting contour is clearly closed, and the integral is guaranteed to be finite.

In analogy to the usual argument, when $X_{1,3},\,X_{2,4}>0$ the integrand \eqref{eq:4-pt integrand Poch} is regular at $y=\{0,\,\infty\}$, so we can deform the small circles (blue in figure \ref{fig:Pochhammer 4-pts}) to a point $y=0$ and the big circles (green) towards infinity. Then, the integral is given only by the contributions along the straight lines, such that:
\begin{equation}
\begin{aligned}
    \int_\gamma dy \ I_4(y) &= \left(1-e^{-2\pi i X_{1,3}}\right)\left(1-e^{-2\pi i X_{2,4}}\right)\int_0^\infty  dy \ I_4(y) \\ &= \left(1-e^{-2\pi i X_{1,3}}\right)\left(1-e^{-2\pi i X_{2,4}}\right)\ah_4.
\end{aligned}
\end{equation}
Thus, since the integral over the closed contour $\gamma$ agrees with the 4-point amplitude (up to some phase factors) for this safe kinematic region, we can define it as an analytic continuation of $\ah_4$ for all values of $X_{1,3},\, X_{2,4}$:
\begin{equation}
    \ah_4 := \frac{1}{\left(1-e^{-2\pi i X_{1,3}}\right)\left(1-e^{-2\pi i X_{2,4}}\right)}\int_\gamma \frac{dy}{y}\ y^{X_{1,3}}(1+y)^{-X_{1,3}-X_{2,4}}.
\end{equation}
Notice that the pole structure for (negative) integer values of the kinematics is encoded solely in the prefactor. For positive integer values, the integral over $\gamma$ gives zero since one of the branch points disappears and therefore all the circles cancel, and so the final result is finite, as expected. 

We can now apply exactly the same procedure to define a closed contour for higher-point amplitudes. At five points, the integral is two-dimensional, and for the base triangulation $\mathcal{T}=\{(1,3),\,(1,4)\}$ it can be written as:
\begin{equation}
\begin{aligned}
    \ah_5 = \int_0^\infty \frac{dy_{1,3}}{y_{1,3}}\frac{dy_{1,4}}{y_{1,4}}\ y_{1,3}^{X_{1,3}}y_{1,4}^{X_{1,4}} (1+y_{1,3})^{-X_{1,3}+X_{1,4}-X_{2,4}}(1+y_{1,4})^{-X_{2,4}+X_{2,5}-X_{3,5}}\\\times (1+y_{1,4}+y_{1,3}y_{1,4})^{-X_{1,4}+X_{2,4}-X_{2,5}},
\end{aligned}
\end{equation}
where we have explicitly written the non-planar variables $c_{i,j}$ in terms of their planar counterparts.
\begin{figure}[t]
    \centering
    \includegraphics[width=.5\textwidth,valign=c]{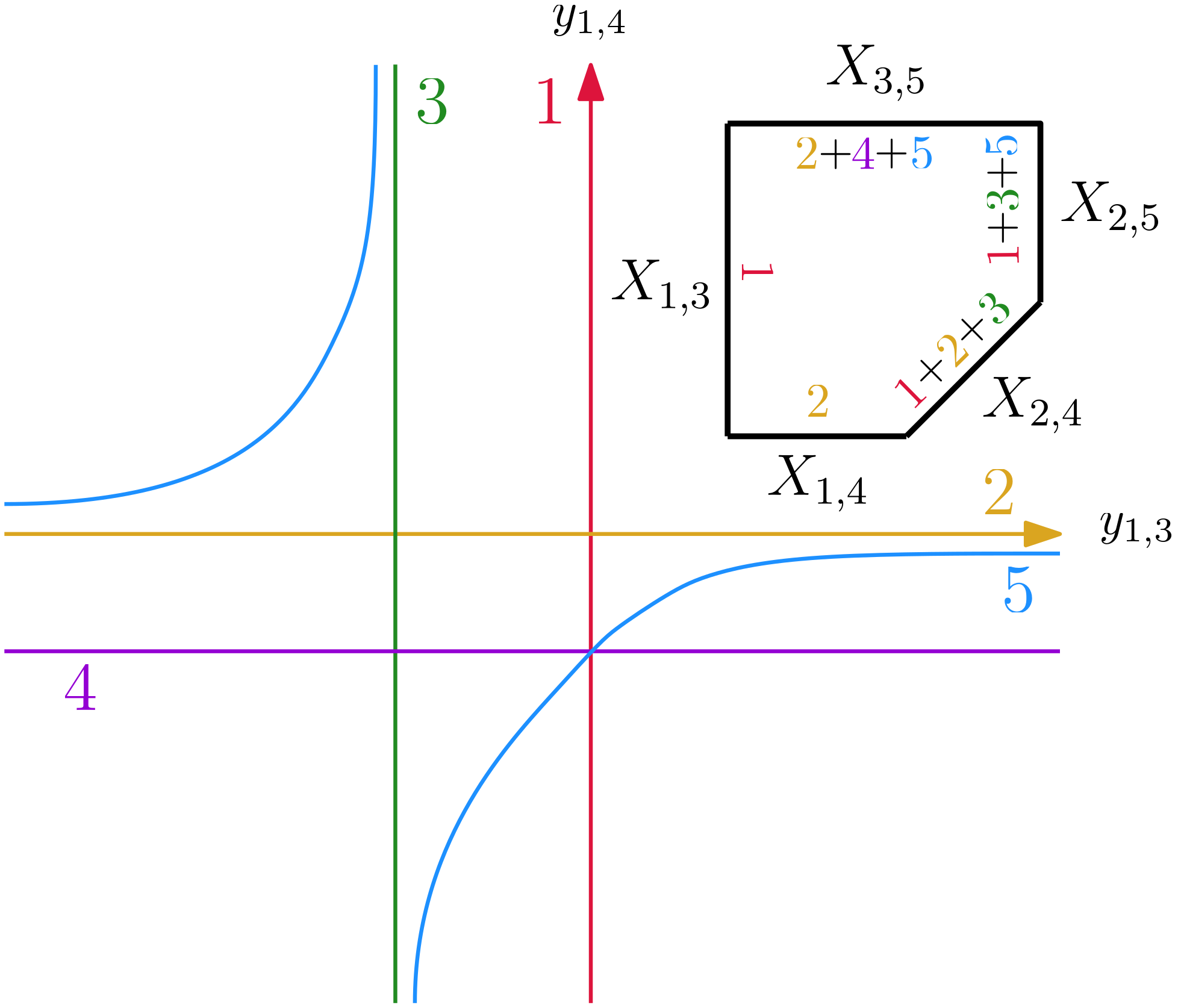} 
    \caption{Pentagonal sheet in the real two-dimensional $y$-plane. We represent the different brach lines of the integrand in different colors and label them from $1$ to $5$. In addition, on each boundary of the pentagon associated to a given chord $X_{i,j}$, we specify the set of branch cuts one has to cross to pick up the corresponding phase, $\exp(-2\pi i X_{i,j})$.}
    \label{fig:5-pt branches} 
\end{figure}
To start with, we consider $2^5=32$ copies of a finite, two-dimensional sheet in the positive real plane in $\mathbb{C}^2\backslash\, \{\text{branches}\}$, where we remove the points corresponding to branch lines of the integrand -- represented in different colors and labeled $(1,2,3,4,5)$,  in figure \ref{fig:5-pt branches}; by going around each line we pick up a phase
\begin{equation}
 \left(e^{\pm 2\pi i X_{1,3}},e^{\pm 2\pi i X_{1,4}},e^{\pm 2\pi i (X_{1,3}-X_{1,4}+X_{2,4})},e^{\pm 2\pi i (X_{2,4}-X_{2,5}+X_{3,5})},e^{\pm 2\pi i (X_{1,4}-X_{2,4}+X_{2,5})}\right).
\end{equation}

The precise shape of the sheet is not important, but it should be represented with five distinct boundaries, i.e. being isomorphic to a two-dimensional $5$-point associahedron. To each such sheet, we associate a phase $\varphi_i\ (i=1,\ldots,32)$ given by:
\begin{equation}
    \varphi_i = \vec{\varphi}_i\cdot(X_{1,3},X_{1,4},X_{2,4},X_{2,5},X_{3,5}),
\end{equation}
where $\vec{\varphi}_i$ is a five-dimensional vector with entries being 0 or 1. It is then obvious that there are precisely 32 distinct phases. 

We now need to connect these two-dimensional sheets along their boundaries by tracing tubes that cross the appropriate branch cuts (here a tube can be defined as a direct product of a circle going around a branch cut and a straight line). For instance, two sheets which differ by a phase of $X_{2,5}$ should be connected via the associated boundary (labeled by $X_{2,5}$ in figure \ref{fig:5-pt branches}) by a tube that crosses the branch cuts $y_{1,3}=0$, $(1+y_{1,3})=0$ and $(1+y_{1,4}+y_{1,3}y_{1,4})=0$ (lines $1$, $3$ and $5$, respectively, in figure \ref{fig:5-pt branches}). Such a contour will precisely pick up a phase $e^{2\pi i X_{2,5}}$, as expected. Once we have connected all the codimension-1 boundaries of our sheets in this way (there will be =80 such tubes), we need to also glue together the vertices (codimension-2 boundaries) in order to completely close the contour. Since each vertex in the sheet is associated to two compatible planar variables $X_{i,j}$ (as expected from the associahedron structure), we can simply connect them with direct products of two circles that cross the appropriate set of branch cuts, which are homeomorphic to two dimensional-tori. For instance, there will be a torus made of two circles crossing the $y_{1,3}=0$ and the $y_{1,4}=0$ branch cuts, which will connect the vertex $(X_{1,3},X_{1,4})$ of the sheets labeled by the phases $\{0,\,X_{1,3},\,X_{1,4},\, X_{1,3}+X_{1,4}\}$.

All in all, the set of 32 sheets, $5\times (32/2) =80$ tubes and $5\times (32/4) =40$ tori form a closed two-dimensional contour in $\mathbb{C}^2\backslash\,\{\text{branches}\}$. Now, for the case where all $X_{i,j}>0$ the integrand will be well behaved at $y_{i,j}=\{0,\,\infty\}$, so we can deform the contour in such a way that the circles around $y_{1,3},\,y_{1,4}=0$ shrink to points, while the circles crossing any other branch cut get extended towards infinity. Thus, the only contribution to the integral will be from the flat sheets, which will have been stretched to span the whole positive real plane. However, this is nothing but our original definition for the 5-point amplitude! Taking into account the pertinent phase factors, we find:
\begin{equation}
\begin{aligned}
    \int_\gamma& dy_{1,3}dy_{1,4} \ I_5 = \left(\prod_X 1-e^{-2\pi i X_{i,j}}\right) \int_0^\infty dy_{1,3}dy_{1,4} \ I_5 \\ &= \left(1-e^{-2\pi i X_{1,3}}\right)\left(1-e^{-2\pi i X_{1,4}}\right)\left(1-e^{-2\pi i X_{2,4}}\right)\left(1-e^{-2\pi i X_{2,5}}\right)\left(1-e^{-2\pi i X_{3,5}}\right)\ah_5,
\end{aligned}
\end{equation}
where the product in the first line is over all propagators that appear in the 5-point amplitude. As in the 4-point case, we can extend this definition for $\ah_5$ to all regions in kinematic space as an analytic continuation of the integral that is finite everywhere. We stress again that defining such a contour in a global manner is only possible because our positive parametrization doesn't include any degenerate singularities, which allows us to join the different boundaries of the integration sheets without the need for blowups. Note also that this contour is not exactly analogous to the one presented in \cite{Eberhardt:2024twy} as a \emph{generalized Pochhammer}, which more closely resembles the periodic resummation in section \ref{sec:5-pt periodic}.

By now, it should clear how to generalize this construction to an $n$-point amplitude. We first consider $2^{n(n-3)/2}$ copies of an $(n-3)$-dimensional, $n$-point associahedron in the positive real orthant in $\mathbb{C}^{n-3}\backslash\,\{\text{branches}\}$. To each of them, we associate a phase (which places them on different branches of the complex space) given by:
\begin{equation} 
\varphi_i = \vec\varphi_i\cdot \vec{X},\quad i=1,\ldots,2^{n(n-3)/2},
\end{equation}
where $\vec{X}=(X_{1,3},\,X_{1,4},\,\ldots)$ is a vector containing all possible chords in the surface, and $\vec{\varphi}_i$ only has entries 0 or 1. In other words, the phases are in a one-to-one correspondence with subsets of propagators.

The codimension-$d$ boundaries of the associahedra are then connected together by direct products of $d$ circles and $(n-3-d)$ straight lines, where the circles cross the appropriate set of branch cuts to pick up the phase difference between the copies (e.g. at six points, one would have to join the faces, edges and vertices of the three-dimensional associahedra).

To see that this is the right construction, we only have to shrink the radii of the circles around the branches $y_{1,k}=0$ to points (we assume that we work in the ray-like triangulation given in \eqref{eq:StrAmpY}), and stretch all the others towards infinity. For all $X_{i,j}>0$, this then recovers $2^{n(n-3)/2}$ copies of our original integration region, weighed by a number of phases. More specifically, we have:
\begin{equation}\label{eq:Pochhammer contour}\int_\gamma\ \prod_{i=3}^{n-1} dy_{1,i} I_{n}\ = \Bigg(\prod_C \left(1-e^{2\pi i X_C}\right)\Bigg) \int_0^\infty \ \prod_{i=3}^{n-1} dy_{1,i}  I_{n}\  = \Bigg(\prod_C \left(1-e^{2\pi i X_C}\right)\Bigg) \ah_n,
\end{equation}
where the product is over all chords $C=(i,j)$ we can draw on the surface. Extending this definition to all values of the $X_{i,j}$'s, we obtain a suitable analytic continuation of the $n$-point string amplitude that converges everywhere.

So far, we have only given a topological description of this generalized Pochhammer contour. It would be possible to straightforwardly realize it on $y$-space, but there actually is an easier way to implement this contour by working in $t$-space instead. 

\subsection{Explicit construction of the Pochhammer contour for general $n$}
\begin{figure}[t]
    \centering
    \includegraphics[width=\linewidth,valign=c]{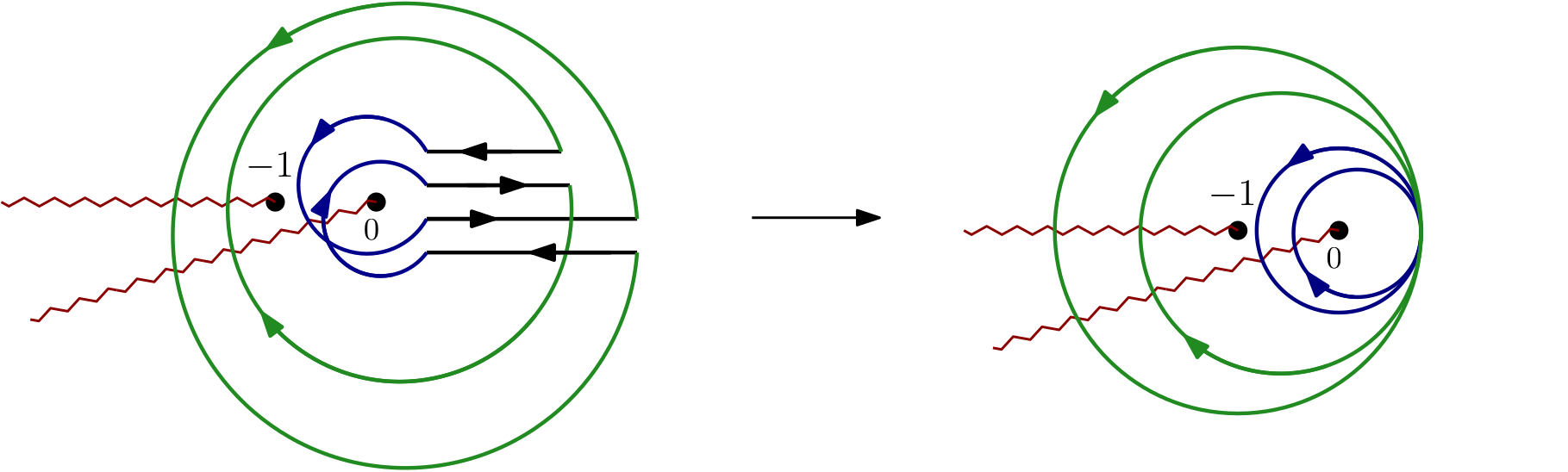} 
    \caption{Deformation of the 4-point Pochhammer by shrinking the original integration region to a point, leaving only the circles.}
    \label{fig:Pochhammer 4-pts transform} 
\end{figure}

In this section, we provide a concrete (non-unique) way to realize this generalized Pochhammer contour in order to evaluate an $n$-point tree-level string integral. We start by noticing that, although the argument for matching to the original contour involved extending the $2^{n(n-3)/2}$ associahedra to the region $0\leq y<\infty$, nothing stops us from doing the opposite. In the most extreme case, we can shrink them to a single point in $y$-space, which we can place anywhere in the positive orthant. By doing this,
the only contributions of the contour that won't be of measure zero correspond to the maximal codimension-$(n-3)$ boundaries, i.e. the vertices. For instance, in the 5-point Pochhammer contour, shrinking the pentagon sheets in figure \ref{fig:5-pt branches} to a point will reduce the tubes connecting the facets to one-dimensional circles, which don't contribute to the two-dimensional integral. All in all, the Pochhammer contour gets reshapen into a collection of tori that loop around the different branches. The example for the 4-point case is illustrated in figure \ref{fig:Pochhammer 4-pts transform}.

\begin{figure}[t]
    \centering
    \includegraphics[width=.9\linewidth,valign=c]{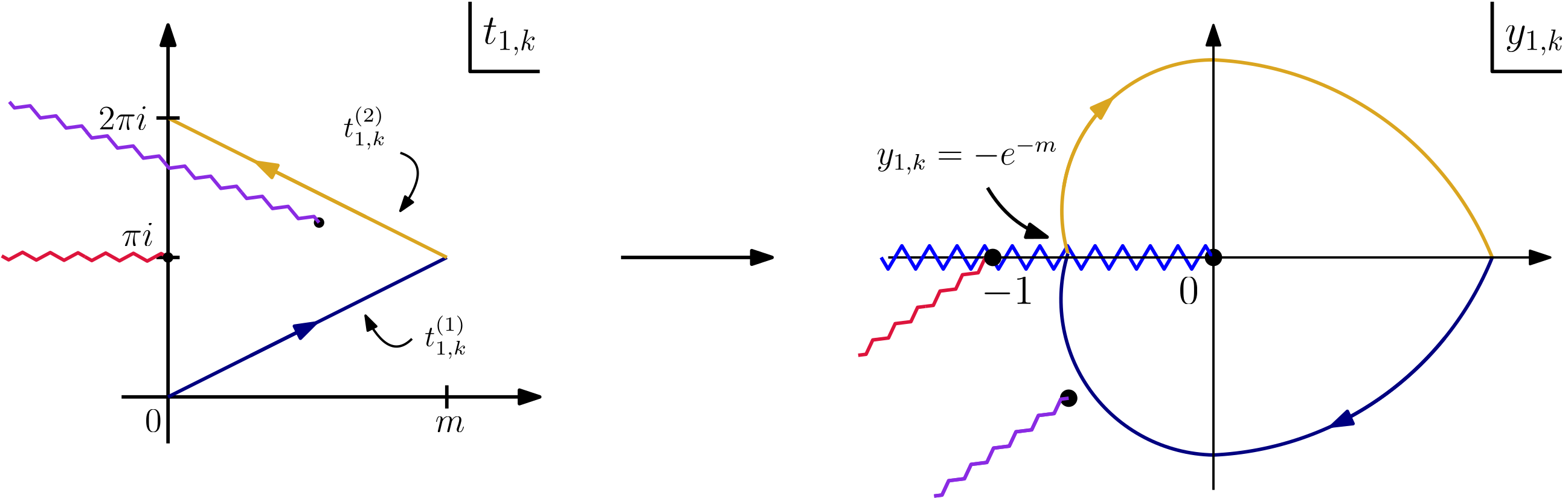} 
    \caption{The curve starting at $y=1$ and crossing the $y=0$ branch cut (right) can be realized as two straight lines in $t$-space (left). The parameter $m$ has to be large enough to avoid potential branch cuts coming from other $\fh$-polynomials (purple).}
    \label{fig:curve y branch cut} 
\end{figure}

To make this even simpler, we make a change of variables and express the integrand in terms of the Euclidean proper times, $t_{1,k}$, with $y_{1,k}=e^{-t_{1,k}}$. Let's focus on the torus associated to the vertex where the facets of the chords $X_{1,k}$ on the base triangulation, $\mathcal{T}$, meet. By construction, it will be a direct product of curves that loop around the branches associated to $y_{1,k}=0$. Let's imagine we have placed our shrunken copies of the associahedra at the point $y_{1,k}=1$ for all $k=3,\ldots,n-2$. Then, in $t$-space, each component of the torus can be realized by e.g. two straight lines:
\begin{equation} \label{eq:Pochhammer straight lines}
t_{1,k}^{(1)}(\sigma_k) = (m_k+i\pi)\sigma_k,\quad t_{1,k}^{(2)}(\sigma_k) = (m_k+i\pi) + (-m_k+i\pi)\sigma_k,\quad 0\leq\sigma_k\leq1,
\end{equation}
where $m_k>0$ are arbitrary parameters that quantify how close to the $y_{1,k}=0$ axis we are getting when we cross the cut. A complete torus is then simply a direct product of $(n-3)$ of these pairs of straight lines.
A graphical representation each curve in one of the variables is presented in figure \ref{fig:curve y branch cut}. In addition to the branch-cut starting at $y_{1,k} = -1$ (represented in red in figure \ref{fig:curve y branch cut}), coming from the presence of the simplest type of $\mathcal{F}$-polynomial, $(1+y_{1,k})$, we also have other branch cuts, coming from the remaining $\mathcal{F}$-polynomial. As explained in detail in section \ref{sec:TreeContour}, since these involve different $y_{1,j}$'s, their location in $y_{1,k}$ space depends on the precise value of the remaining integration variables (a schematic example of such a branch cut is given in purple in figure \ref{fig:curve y branch cut}). Therefore, when defining the circles, or equivalently the two lines in $t$-space, the only restriction on $m_k$ is to be large enough so as to guarantee that we don't cross any of these branch cuts. However, from the analysis in section \ref{sec:TreeContour}, we know that this can be ensured minimally by requiring $m_k > \ln(n-3)$, as derived in \eqref{eq:boundRmax}!

Now, since one torus connects $2^{(n-3)}$ sheets and we have a total of $2^{n(n-3)/2}$ sheets, the total contribution associated with the vertex is given by a sum over $2^{(n-2)(n-3)/2}$ such tori, each weighed by the phase of the initial sheet. To determine the phase of each tori let us look at simplest case of $4$-points, and in particular the two blue circles crossing branch cut at $y=0$ (see fig. \ref{fig:Pochhammer 4-pts transform}, right). By looking at the phases of each black line (given in fig. \ref{fig:Pochhammer 4-pts}) that were shrunken into a point, we can read off the phases of each half-circle to be:
\begin{equation}
\begin{aligned}
   &\underline{\text{Inner circle}}: \quad   e^{-2\pi i X_{1,3}} \int_{(1)} dy \, I_4(y) + \int_{(2)} dy \, I_4(y), \\  
   &\underline{\text{Outer circle}}:  \quad  -e^{-2\pi i (X_{1,3}+X_{2,4})}\int_{(1)} dy \, I_4(y) -e^{-2\pi i X_{2,4}}   \int_{(2)} dy \, I_4(y) , \\ 
\end{aligned}
\end{equation}
where here we assume that $\int_{(1)}$ and $\int_{(2)}$ are being computed in the same branch. Putting together these four terms we obtain
\begin{equation}
    (1 - e^{-2\pi i X_{2,4}})\times  \left[ 
  e^{-2\pi i X_{1,3}}\int_{(1)} dy \, I_4(y) +  \int_{(2)} dy \, I_4(y) \right] 
  \label{eq:phases4pt}
\end{equation}

Let's now understand how the parametrization in $t$-space given in \eqref{eq:Pochhammer straight lines}, precisely gives us the piece between brackets above. Let's start with the part along $(2)$, depicted in yellow in figure \ref{fig:curve y branch cut}. This piece does not have any overall phase and therefore defines the branch in which we want to compute both $\int_{(1)}$ and $\int_{(2)}$. According to the parametrization in $t$-space, the yellow part corresponds to $\text{Im}(t) \in [\pi,2\pi]$, and therefore, since $y=e^{-t}$, we have that $\text{Im}(y)$ goes from $-\pi$ to $-2\pi$. On the other hand, the part along $(1)$ is defined in $t$-space as the blue line in figure \ref{fig:curve y branch cut}, where $\text{Im}(t) \in [0,\pi]$, and therefore $\text{Im}(y)$ goes from $0$ to $-\pi$. Thus, the $t$-space parametrization is computing the integral along $(1)$ in a different sheet to that of $(2)$, and so to write it as an integral in the same sheet we must multiply by a factor of $e^{-2\pi i X_{1,3}}$ -- which is precisely what we find in the brackets in \eqref{eq:phases4pt}! Therefore, we can write the total contribution from both circles simply as: 
\begin{equation}
    (1 - e^{-2\pi i X_{2,4}})\times \int_{t^{(1)} \cup t^{(2)}} dt \ I_4(t).
\end{equation}

In general, at $n$ points, we then have that the integral over the $(n-3)$ circles multiplies a phase factor corresponding to all possible subsets, $S$, of the remaining $X^{\overline{\mathcal{T}}}_{i,j}$ that are not in our base triangulation (where  $\overline{\mathcal{T}}$ denotes the complement of our triangulation $\mathcal{T}$). In other words, the contribution from this specific vertex can be written as:
\begin{equation}\Bigg(\sum_{S\subset \overline{\mathcal{T}}}(-1)^{|S|}\exp\left[2\pi i \sum_{X_{i,j}\in S}X_{i,j}\right]\Bigg)\times \prod_{k=1}^{n-3} \int_{t_{1,k}^{(1)} \cup t_{1,k}^{(2)}} dt_{1,k} \, I_{n}(\{t_{1,k}\}),
\end{equation}
where $|S|$ is the number of chords in $S$ and the sum also includes the case where $S$ is the empty subset. 

Now, naively one could expect that the parametrization of the tori in the remaining vertices of the associahedron would be much more complicated than this one. Not only do the curves have different radii for the same torus, but the branch cuts that we must cross have an increasingly complex structure that we need to take into account. However, once again, we can use the rotation of the Feynman fan so solve this problem. Using  \eqref{eq:cone_map}, we can map the cone of the corresponding vertex to the positive orthant. Then, we have a situation that is identical to the one discussed above, and we can parametrize the curves by the same two straight lines. The only thing that changes is the overall phase and the expression for the integrand, as the $\fh$-polynomials change under this map. Nonetheless, by the reasoning in section \ref{sec:Generaln},we can assure that no branch cuts are crossed by asking for $m_P>\ln(n-3)$, for all $X_P$ entering in the vertex. 

For example, at 5-points, if we wanted to construct the torus associated to the vertex $\{X_{2,4},X_{2,5}\}$, we would rotate the fan according to \eqref{eq:cone X24X25}, which results in the integrand:
\begin{equation}
\begin{aligned}
    \mathcal{I}^{C_{24,25}}_5(t_{1,3},t_{1,4})= \int d\hat{t}_{2,4}d\hat{t}_{2,5}\ &e^{-\hat{t}_{2,5}X_{2,5} - \hat{t}_{2,4}X_{2,4}} (1 + e^{-\hat{t}_{2,4}-\hat{t}_{2,5}})^{-c_{1,3}}\\ \times  &(1 + e^{-\hat{t}_{2,4}})^{-c_{2,4}}  (1 + e^{-\hat{t}_{2,5}} + e^{-\hat{t}_{2,4}-\hat{t}_{2,5}})^{-c_{1,4}}.
\end{aligned}
\end{equation}
Now it is trivial to trace the part of the contour that picks up the phases $X_{2,4}$ and $X_{2,5}$: we can simply parametrize $\hat{t}_{2,4}$ and $\hat{t}_{2,5}$ as in \eqref{eq:Pochhammer straight lines}, because the associated branch cuts have moved to $\hat{y}_{2,4}=0$ and $\hat{y}_{2,5}=0$, respectively (where $\hat{y}=e^{-\hat{t}}$). 

In summary, using the positive parametrization and the map in \eqref{eq:cone_map}, we have reduced a very complicated Pochhammer contour in arbitrary dimensions to a direct product of pairs of straight lines in $t$-space. In appendix \ref{app:numerical checks}, we present some numerical test on this construction of the Pochhammer contour up to $n=6$, and show how remarkably accurate this contour is, even in extreme kinematical regimes.

\subsection{Avoiding huge cancellations with Pochhammer}
\begin{figure}[t]
    \centering
    \includegraphics[width=.6\textwidth,valign=c]{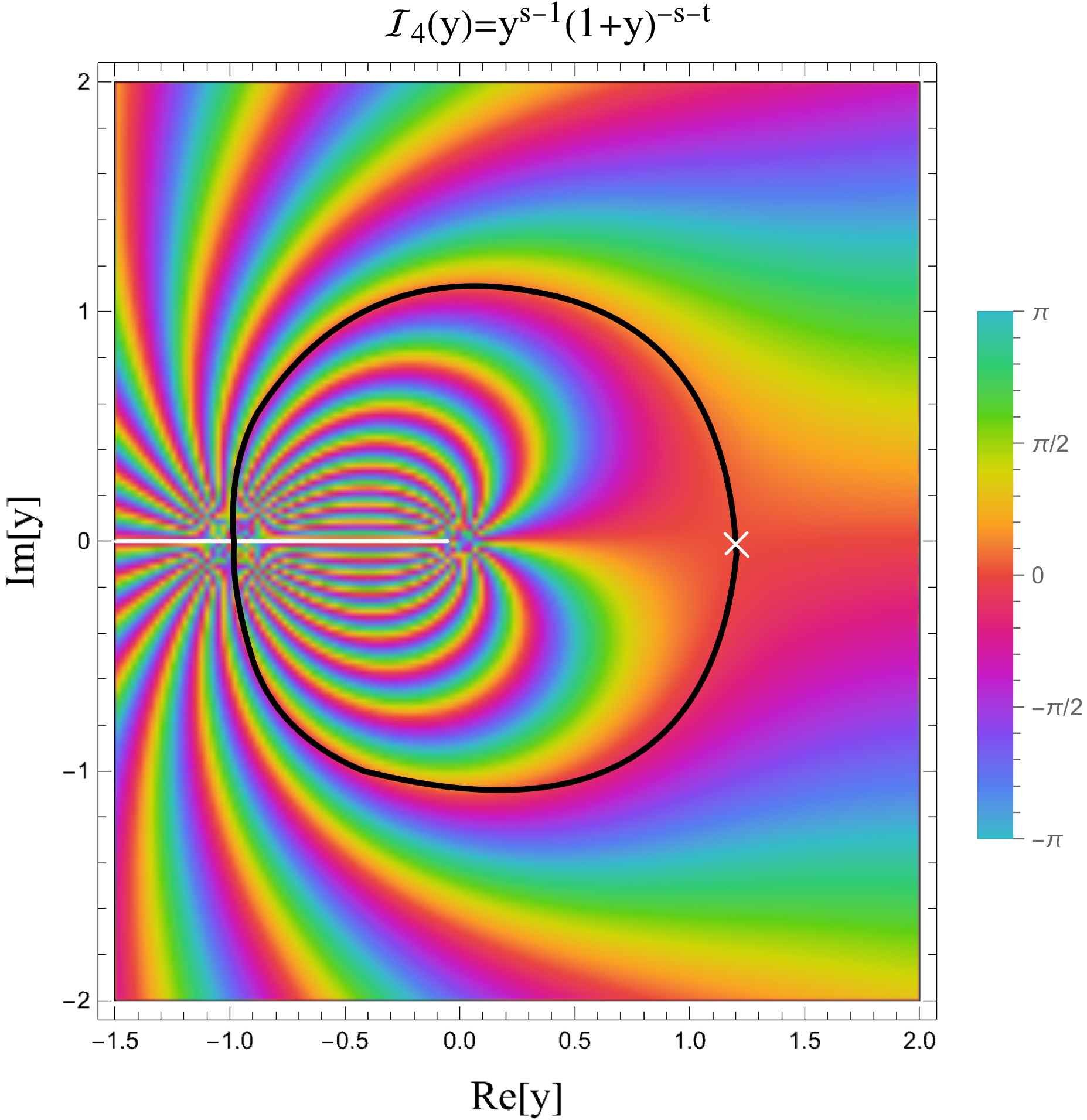} 
    \caption{Complex plot of the 4-point integrand for hard scattering kinematics ($s=t=-15.3$). The contour of steepest descent (black) is drawn emanating from the saddle point (white cross).}
    \label{fig:hard scattering 4pt} 
\end{figure}

The construction of the generalized Pochhammer contour we have provided above is just one possible realization of it. In theory, one could customize the point in the positive orthant where we choose to place the shrunken associahedra, as well as the particular shape that each curve takes when looping around a branch cut. By crafting a smart enough contour, one could even avoid the problem of huge cancellations that seems to be ubiquitous when evaluating string integrals at large values of the kinematics, as discussed in section \ref{sec:TreeContour}. However, as we will see, this would imply having arbitrarily high control on the behavior of the integrand, which in practice would be more cumbersome than simply performing a saddle point approximation. We will illustrate this idea with an example at four points, where it is easy to visualize the contours, but it is plausible that this can be done for any number of external particles.

Let's consider the situation of hard scattering, i.e. both $X_{1,3}$ and $X_{2,4}$ are negative and large. The integrand:
\begin{equation}
    \int\frac{dy}{y}\ y^{X_{1,3}}(1+y)^{-X_{1,3}-X_{2,4}},
\end{equation}
will be very large close to $y=0$ and $y=\infty$, while decaying rapidly close to $y=-1$. Moreover, the saddle point, which is given by:
\begin{equation} 
y_s = \frac{X_{1,3}-1}{X_{2,4}+1},
\end{equation}
is located somewhere in the positive real half plane, $0< \text{Re}(y_s)<\infty$. In the plot in figure \ref{fig:hard scattering 4pt}, we have schematically drawn the contour of steepest descent, which starts at the saddle point and ends at $y=-1$.

To trace our Pochhammer contour, we would need two types of curves, one crossing the two branch cuts starting at $y=0$ and $y=-1$, and the other crossing only the former. Doing this while also avoiding large oscillatory behavior is achieved by keeping both curves along the contour of steepest descent until they are close to $y=-1$. There, we can move freely because the integrand is very small, so we would simply take one of the curves to pass to the left of the branch point, and the other to the right. In this way, although the contour passes through regions where the integrand is large, the fact that we follow a path of stationary phase means that there won't be any cancellations due to oscillatory behavior.

However, in order to practically trace this contour we would need to numerically compute the curve of stationary phase starting from the saddle point, which even in the one-dimensional case can be challenging if the kinematics are very large. For higher-point amplitudes, the presence of multiple saddle points and the complicated branch structure makes an implementation of this contour practically unfeasible.

\section{Integrand cuts at 1-loop}\label{sec:loop cuts}

After analyzing the structure of string amplitudes at tree-level, we now switch to studying features of one-loop amplitudes that are made manifest via their surface integral formulation. Both in this section as well as in section \ref{sec:loop contour}, we won't concern ourselves with providing a general all $n$ approach to studying loop-amplitudes. Instead we take a first step at understanding them by studying the   most interesting features arising at loop level which are already present in the simplest $n=2$ example.

A key distinction between the surface formulation and the textbook approach to loop‐level string integrals is that the former directly yields the loop \textit{integrand}, whereas the latter produces the loop‐integrated amplitude.
As a result, using surface integrals, we can make exact statements about the discontinuities of the amplitude without even performing any loop or worldsheet integration, simply by looking at the integrand! 

As reviewed in section \ref{sec:tree cuts}, at tree-level we can extract the residue of the amplitude on a certain pole $X_{i,j}=-n$ by substituting these kinematics in the integrand and taking the residue when the corresponding $y_{i,j}=0$ (see eqn. \eqref{eq:res_tree}). 
Following the exact same steps with
the curves associated to loop propagators then lands us on the leading singularities of
the amplitude. Note that this residue
corresponds to the product of tree-level amplitudes, which are glued together by summing
over all possible intermediate states, at fixed values of the cut loop momentum.  The discontinuity/imaginary
part results from performing a further integration of this cut  over the remaining Lorentz invariant phase space
of the loop momenta, as we show explicitly in the next section. Following the exact same steps with the curves associated to loop propagators would then land us on the leading singularities of the amplitude. 

Another feature of the surface picture is that there are many different surface
integrals that suitably UV regulate the low-energy theory, connected with a variety of choices for the kinematic exponents associated with the curves on the surface. One particular choice matches the surface integral to the tachyon amplitude in bosonic string theory, as we review in appendix~\ref{app:GSW matching}. However, it is interesting to ask
whether this particular choice defines the \textit{only} surface integral consistent with unitarity at all massive levels. 

In section \ref{sec:reviewsurfaceInt}, we start by reviewing the $2$-point integrand defined via the surface integral for the punctured disk. Using this integrand, in section \ref{sec:consistencyCuts}, we extract the low-level leading singularities for the bubble diagram. Asking for consistency with the tree-level spectrum, we derive the conditions on the curve exponents necessary to ensure unitarity at the respective levels. In some cases, these are enough to completely fix all exponents,
while in others we can prove that certain simple surface integrals are inconsistent with unitarity on massive cuts. In section \ref{sec:MasslessLS}, we show a couple of examples of the latter: using the lowest level residues we show the unitarity inconsistency of the $\hat D_n$ stringy integral, and \textit{baby string integrals}~\cite{NimasTalk}.

\subsection{Surface integral at 2-points one-loop}
\label{sec:reviewsurfaceInt}
\begin{figure}[t]
    \centering
    \includegraphics[width=\textwidth,valign=c]{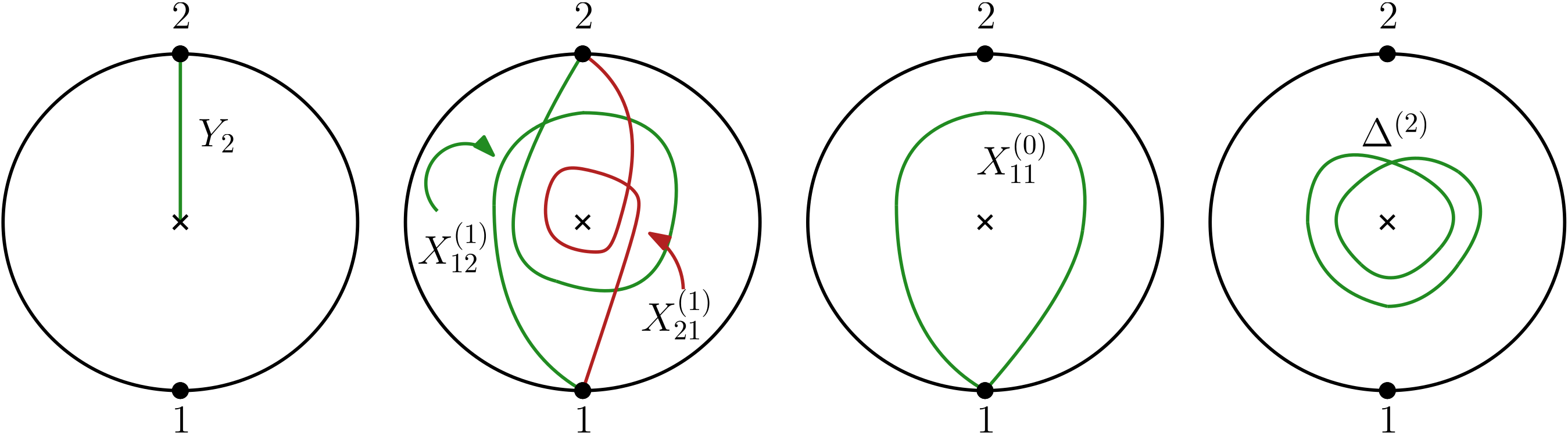} 
    \caption{Types of curves on the 1-loop 2-point surface. The superscript on the ones on the left indicate the number windings.}
    \label{fig:1-loop 2-pt curves} 
\end{figure}
Let's start by reviewing the new ingredients entering loop-level surface integrals by looking at the simplest example of the 2-point integrand:
\begin{equation}
    \mathcal{I}_2 = \int \frac{d y_{1,0}}{y_{1,0}} \frac{d y_{2,0}}{y_{2,0}} \prod_{\mathcal{C} \in \mathcal{S}} u_{C}[y_{1,0},y_{2,0}]^{X_\mathcal{C}}
    \label{eq:surfaceIntegrand1loop}
\end{equation}
where the surface $\mathcal{S}$ is a punctured disk with two marked points on the boundary, and $\mathcal{C} \in \mathcal{S}$ are all the non-homotopic curves on the surface. For each such curve we have the respective $u$-variable, $u_\mathcal{C}$, as well as the kinematic variable $X_\mathcal{C}$ \cite{Arkani-Hamed:2023lbd}. We label the boundary points with indices $1$ and $2$, and the puncture with $0$. The curves $\mathcal{C}$ are then divided into four categories: the ones going between marked points 1 and 2 ($X_{1,2}^{(q)}$ and $X_{2,1}^{(q)}$), the ones starting at a boundary point and ending in the internal puncture ($X_{i,0}\equiv Y_i$), the ones starting and ending at the same a boundary point ($X_{i,i}^{(q)}$), and the closed curves $\Delta^{(q)}$ winding around the puncture. An example of each is provided in figure \ref{fig:1-loop 2-pt curves}. Here, the superscript $q$ indicates the number of times the curve winds around the puncture.

\begin{figure}[t]
    \centering
    \includegraphics[width=0.95\linewidth]{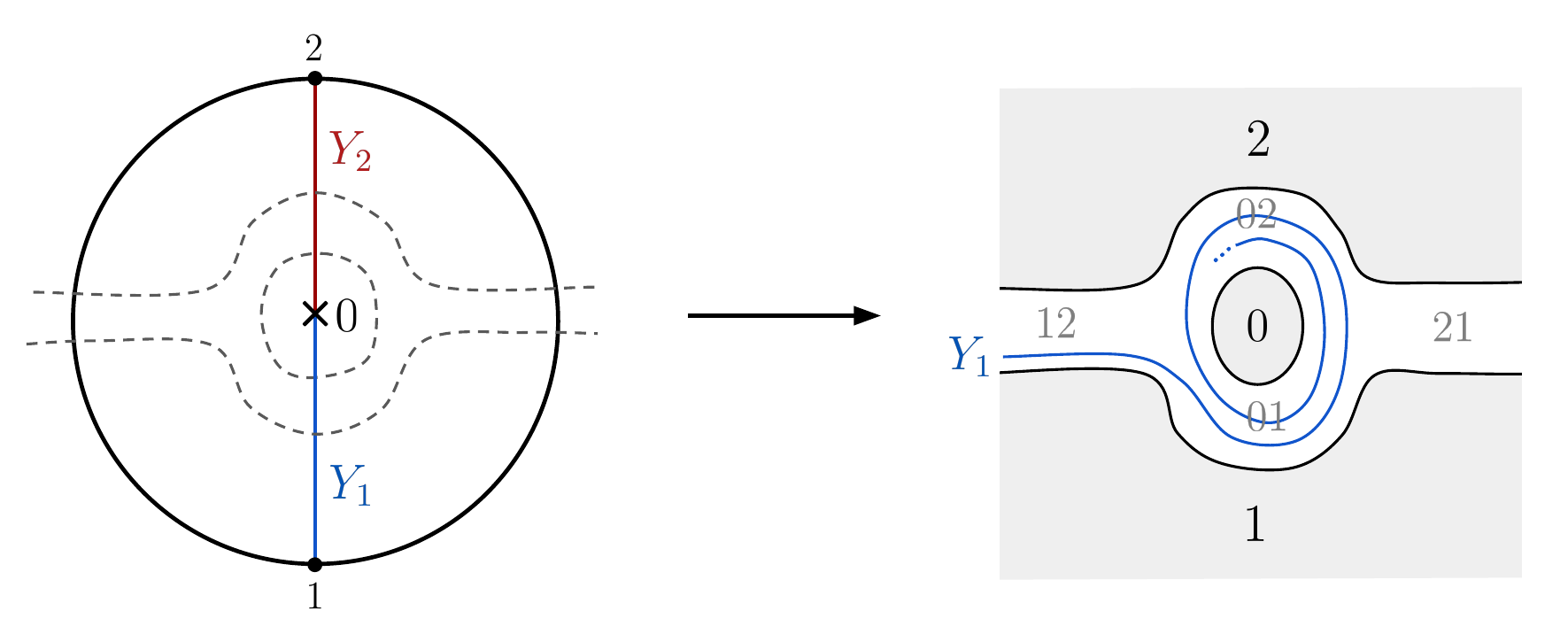}
    \caption{(Left) Triangulation of the $2$-point punctured disk containing curves $Y_1$ and $Y_2$, dual to the bubble diagram. (Right) Bubble fatgraph with curve $Y_1$ in blue -- this curve spirals around the puncture counterclockwise forever.}
    \label{fig:1loopFatgraph}
\end{figure}

As our base triangulation of the surface, we choose the chords $\{Y_1,Y_2\}$, which correspond to the bubble diagram (see figure \ref{fig:1loopFatgraph}). On the fatgraph, these curves correspond to paths that start at an external leg and end up spiraling counter-clockwise around the loop. Since the curves are not homologous to any boundary, we need to assign a momentum to one of them, say $P_{1,0}^\mu = \ell^\mu$ (this is precisely the loop momentum). By momentum conservation (or equivalently, by homology), this implies $P_{2,0}^\mu = (\ell+p)^\mu$, where $p^\mu$ is the external momentum. Therefore, the dependence on the loop momentum in \eqref{eq:surfaceIntegrand1loop} enters via the exponents $X_{1,0} = Y_1 = \ell^2$ and $X_{2,0} = Y_2 = (\ell+p)^2$.

As explained in \cite{Arkani-Hamed:2023jry}, neither the self-intersecting curves $X^{(q)}_{i,j}$ ($q\geq1$) nor the closed curves $\Delta^{(q)}$ contribute to the leading order amplitude in the low-energy limit. However, in order to match full string theory, one needs to include them as well as fine-tune their exponents in the Koba-Nielsen factor.

By performing the loop integrals in \eqref{eq:surfaceIntegrand1loop}, we obtain the 2-point amplitude:
\begin{equation}\label{eq:bubble integrand}
\begin{aligned}
    \ah_2^{(1)} &= \int\frac{d^d\ell}{\pi^{d/2}}\int_D\frac{dy_{1,0}dy_{2,0}}{y_{1,0}y_{2,0}}\ u_{1,0}^{Y_1+\alpha_0}u_{2,0}^{Y_2+\alpha_0} \\ &\times \Bigg(\prod_{q=0}^\infty \left(u_{1,2}^{(q)}\right)^{X_{1,2}+\alpha_0}\left(u_{1,1}^{(q)}\right)^{X_{1,1}+\alpha_0}\left(u_{2,2}^{(q)}\right)^{X_{2,2}+\alpha_0} \left(u_{\Delta}^{(q)}\right)^{\Delta(q)} \Bigg),
\end{aligned}
\end{equation}
where we have left an arbitrary Regge intercept $\alpha_0$ determining the external states' mass. For tachyon and massless surface amplitudes, we would set $\alpha_0=-1,0$, respectively. The domain of integration $D$ in $y$-space is $D = \{y_{i,0}\geq 0,\, y_{1,0}y_{2,0}\leq 1 \}$.

In order to extract the residues, we need to determine the explicit expressions for the $u$-variables of the different curves. Following the procedure outlined in section \ref{sec:FanReview}, we first obtain the words for each curve, using the fatgraph (shown on the right of figure \ref{fig:1loopFatgraph}). Starting with the closed curved $\Delta^{(0)}$ we have:
\begin{equation}
\mathcal{W}_\Delta^{(0)} = y_{1,0}\xrightarrow[]{L}y_{20}\xrightarrow[]{L}.
\end{equation}
The words for the remaining curves can then be compactly written in terms of this one:
\begin{equation}
\begin{aligned} 
    \mathcal{W}_{1,2}^{(q)} &= (1,2)\xrightarrow[]{R}\left(\mathcal{W}_\Delta^{(0)}\right)^q\xrightarrow[]{} y_{1,0}\xrightarrow[]{R}(2,1),\\ \mathcal{W}_{1,1}^{(q)} &= (1,2)\xrightarrow[]{R}\left(\mathcal{W}_\Delta^{(0)}\right)^q\xrightarrow[]{} y_{1,0}\xrightarrow[]{L}y_{2,0}\xrightarrow[]{R}(1,2),\\ \mathcal{W}_{2,2}^{(q)} &= (2,1)\xrightarrow[]{R}y_{2,0}\xrightarrow[]{L}\left(\mathcal{W}_\Delta^{(0)}\right)^q\xrightarrow[]{} y_{1,0}\xrightarrow[]{R}(2,1),\\ \mathcal{W}_{1,0} &= (1,2)\xrightarrow[]{R}\left(\mathcal{W}_\Delta^{(0)}\right)^\infty,\\ \mathcal{W}_{2,0} &= (2,1)\xrightarrow[]{R}y_{2,0}\xrightarrow[]{L}\left(\mathcal{W}_\Delta^{(0)}\right)^\infty.
\end{aligned}
\end{equation}
For the infinite products, $\left(\mathcal{W}_\Delta^{(0)}\right)^\infty$, to be convergent we must have $y_{1,0}y_{2,0}<1$, which is ensured by the modified integration domain $D$.  This reflects the fact that the $u$-variables for infinitely spiraling curves have a restricted domain of convergence, which is also why we can only consider one of the two possible orientations for the spirals (only turning left around the punctured, rather than only turning right).
From these, we can extract the $u$-variables for all the open curves entering the Koba-Nielsen factor in the integrand \eqref{eq:surfaceIntegrand1loop}:
\begin{equation} \label{eq:self-int u's}
\begin{aligned} 
    u_{1,0} &= \frac{y_{1,0}(1+y_{2,0})}{1+y_{1,0}},\quad \quad  u_{2,0} = \frac{y_{2,0}(1+y_{1,0})}{1+y_{2,0}},\\ u_{1,1}^{(q)} &= \frac{\left[ 1+y_{1,0}\left( 1-(1+y_{2,0})(y_{1,0}y_{2,0})^{q+1} \right) \right]\left[ 1+y_{2,0}\left( 1-(1+y_{1,0})(y_{1,0}y_{2,0})^{q} \right) \right]}{(1+y_{1,0})(1+y_{2,0})\left( 1-(y_{1,0}y_{2,0})^{q+1} \right)^2},\\ u_{2,2}^{(q)} &= \frac{\left[ 1+y_{2,0}\left( 1-(1+y_{1,0})(y_{1,0}y_{2,0})^{q+1} \right) \right]\left[ 1+y_{1,0}\left( 1-(1+y_{2,0})(y_{1,0}y_{2,0})^{q} \right) \right]}{(1+y_{1,0})(1+y_{2,0})\left( 1-(y_{1,0}y_{2,0})^{q+1} \right)^2}, \\ u_{1,2}^{(q)} &= u_{2,1}^{(q)} =\frac{(1+y_{1,0})(1+y_{2,0})\left( 1-(y_{1,0}y_{2,0})^q\right)\left( 1-(y_{1,0}y_{2,0})^{q+1}\right)}{\left[ 1+y_{1,0}\left( 1-(1+y_{2,0})(y_{1,0}y_{2,0})^q \right) \right]\left[ 1+y_{2,0}\left( 1-(1+y_{1,0})(y_{1,0}y_{2,0})^q \right) \right]}.
\end{aligned}
\end{equation}
Meanwhile, the $u$-variables for the closed curves are computed in a slightly different way, which is explained in \cite{CurvyU}. The upshot is that we only need the specific case where the exponent is taken as $\Delta(q) = \Delta_1\, q+\Delta_2\, q^2$, with:
\begin{equation}
\begin{aligned}
    \prod_{q=1}^\infty u_{\Delta(q)}^q &= 1-y_{1,0}\,y_{2,0},\\  \prod_{q=1}^\infty u_{\Delta(q)}^{q^2} &= (1-y_{1,0}\,y_{2,0})\prod_{n-2}^\infty (1-(y_{1,0}\,y_{2,0})^n)^2 \\&= \frac{1}{(y_{1,0}\,y_{2,0})^{1/24}(1-y_{1,0}\,y_{2,0})}\eta\left[ \frac{1}{2\pi i}\log(y_{1,0}\,y_{2,0}) \right],
\end{aligned}
\end{equation}
where $\eta[\tau]$ is the Dedekind eta function.

\subsection{Consistency conditions from integrand cuts}
\label{sec:consistencyCuts}
As it turns out, the concrete expressions for the different $u$-variables are not particularly important in order to analyze the general behavior of the residues. What really matters is the fact that a self-intersecting curve of degree $q$, $X^{(q)}$, only has a non-trivial contribution at order $\oh(y^q)$ or higher. Specifically, one can check that:
\begin{equation} 
\begin{aligned} 
    u_{1,2}^{(q)} = 1+\oh(y_{1,0}^q y_{2,0}^q),\\ u_{1,1}^{(q)} = 1+\oh(y_{1,0}^q y_{2,0}^{q+1}),\\ u_{2,2}^{(q)} = 1+\oh(y_{1,0}^{q+1} y_{2,0}^{q}).
\end{aligned}
\label{eq:SelfIntOrd}
\end{equation}
Recall from section \ref{sec:tree cuts} that we can extract the leading residues from the surface integral via a residue of the integrand in $y$-space. Therefore, it is straightforward to extract the bubble leading singularity for mass levels $m_1^2 = \alpha_0-n_1$ and $m_2^2=\alpha_0-n_2$ from \eqref{eq:surfaceIntegrand1loop}, by taking the residue at $y_{1,0}=y_{2,0}=0$:
\begin{equation}\label{eq:Res LS}
    \begin{aligned}
       \text{LS}_{n_1,n_2} &= \mathop{\mathrm{Res}}_{y_{1,0}=y_{2,0}=0}\frac{1}{y_{1,0}^{1+n_1}}\frac{1}{y_{2,0}^{1+n_2}}\ \left(\frac{1+y_{1,0}}{1+y_{2,0}}\right)^{n_1-n_2} \\ &\times \Bigg(\prod_{q=0}^{\min(n_1,n_2)} \left(u_{1,2}^{(q)}\right)^{2(X_{1,2}+\alpha_0)}\left(u_{1,1}^{(q)}\right)^{X_{1,1}+\alpha_0}\left(u_{2,2}^{(q)}\right)^{X_{2,2}+\alpha_0} \left(u_{\Delta}^{(q)}\right)^{\Delta(q)} \Bigg).
    \end{aligned}
\end{equation}

Therefore, from \eqref{eq:SelfIntOrd} it is clear that the leading singularity at levels $(n_1,n_2)$ only receives contributions from curves that wind around the puncture at most $q_\star\equiv\min(n_1,n_2)$ times -- and so in \eqref{eq:Res LS}, we can truncate the infinite product over $q$. In other words, our positive parametrization allows us to compute discontinuities of the amplitude without even considering the complete surface integrand, since it manifestly separates the contributions to each threshold. Another straightforward property we can see from \eqref{eq:self-int u's} is that none of the $u$-functions for $q\geq 1$ can ever go to zero for $y_{1,0},y_{2,0} \in D$, which reflects the fact that they don't appear as poles in the amplitude and thus don't contribute in the field theory limit.

The leading singularity residues, as given in \eqref{eq:Res LS}, will of course depend on the exponents we assign to the remaining curves. We can use this fact to constrain the expression for our surface integrand by imposing factorization conditions. This turns out to fix our theory to a remarkable degree, already at the level of this simple 2-point amplitude.

\subsection*{Loop cut and tree-level matching}

\begin{figure}[t]
    \centering
    \includegraphics[width=.7\linewidth,valign=c]{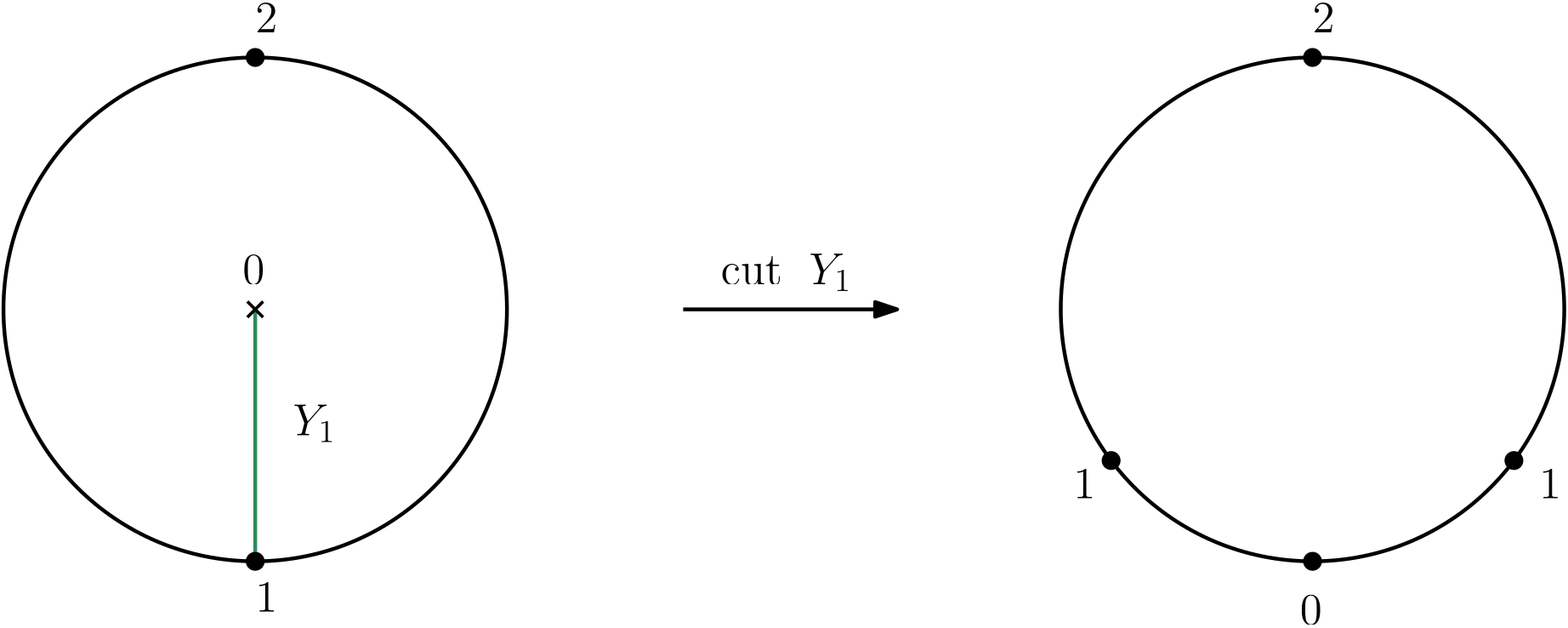} 
    \caption{On the loop cut $Y_1=0$, the 2-point function reduces to forward limit of the 4-point tree-level amplitude.}
    \label{fig:loop-cut} 
\end{figure}

To see this, we consider a loop-cut $Y_{i}=-\alpha_0-n$. In this limit, the integrand should reduce to the forward limit of a 4-point tree amplitude, as depicted in figure \ref{fig:loop-cut}. This matching should hold regardless of the specific stringy UV regularization under consideration, and thus provides strict constraints on the integrand at loop-level.

To implement these conditions, we work following a bottom-up approach. Starting from the tree-level integral \eqref{eq:StrAmpY} and extracting its residues we are able to derive the $3$-point couplings $\lambda^{l_1,l_2,l_3}_{n_1,n_2,n_2}$ -- as explained in section \ref{sec:tree cuts}, and done explicitly in \cite{Arkani-Hamed:2023jwn}. 
Once the tree-level couplings have been fixed, we can build an \emph{ansatz} for the leading singularity of the bubble cut on the 2-point amplitude and substitute the couplings with the values found at tree-level. The result must then match the residue we obtain from the loop integrand \eqref{eq:Res LS}.

Let's illustrate this procedure with some examples. In order to study the leading singularity on the levels $n_1=n,\ n_2=0$, we only need to keep the $u$-functions for the curves $Y_1,\, Y_2,\, X_{2,2}^{(0)}$, since the pole in $y_{2,0}$ is of zeroth order and all the other curves contribute at least as $\oh(y_{2,0})$. Meanwhile, the \emph{ansatz} residue is constructed by gluing together two 3-point amplitudes $A_3^{n,0,0}$, where the external states have masses $m_1^2=-\alpha_0-n$ and $m_2^2=m_3^2=-\alpha_0$. 

For the case $n_1=1$, the 3-point amplitude $A_3^{1,0,0}$ is given in \eqref{eq:3ptA100}, and the relevant couplings are shown in \eqref{eq:3ptsCouplingsn1}. Using these, the \emph{ansatz} for the leading singularity of the bubble cut at level $n_1=1,\, n_2=0$ is
\begin{equation}
    R^{bub}_{1,0} = \left[ \begin{gathered}
    \begin{tikzpicture}[line width=0.6,scale=0.7,baseline={([yshift=0.0ex]current bounding box.center)}]
        \coordinate (1L) at (-2.5,0);
        \coordinate (1M) at (-1.5,0);
        \coordinate (2R) at (2.5,0);
        \coordinate (2M) at (1.5,0);
        \draw[] (1M) to[out=90,in=-180] (-0.5,1);
        \draw[] (2M) to[out=-270,in=0] (0.5,1);
        \draw[Black,very thick] (1M) to[out=-90,in=-180] (-0.5,-1);
        \draw[Black,very thick] (2M) to[out=-90,in=0] (0.5,-1);
        \draw[] (1L) -- (1M);
        \draw[] (2R) -- (2M);
        
       \node[scale=0.7] at (0,1) {$\times$};
       \node[scale=0.7,above] at (-0.6,1) {$0$};
       \node[scale=0.7,above] at (0.6,1) {$0$};
       \node[scale=0.7,above] at (1L) {$0$};
       \node[scale=0.7,above] at (2R) {$0$};
       \node[scale=0.7] at (0.1,-0.8) {$\Pi^{\mu \nu}$};

        \node[scale=0.7,below] at (-0.7,-1) {$1 \, \mu$};
        \node[scale=0.7,below] at (0.7,-1) {$\nu \, 1$};
    \end{tikzpicture}
\end{gathered}  \right] =   1-\alpha_0,
\label{eq:ResBub}
\end{equation}
with $\Pi^{\mu \nu} = \eta^{\mu \nu} + p^\mu p^\nu/M^2$ for the case in which $M\neq 0$, and $\Pi^{\mu \nu} = \eta^{\mu \nu} - (p^\mu  q^\nu + q^\mu  p^\nu)/(p\cdot q)$ if $M=0$, for $q^\mu$ some reference momenta which drops out of the final answer. 

The result from \eqref{eq:ResBub} then has to be matched by the residue of the surface integrand. Keeping only the relevant curves, we have:
\begin{equation}
\begin{aligned}
    \text{LS}_{1,0} = \mathop{\mathrm{Res}}_{y_{1,0}=y_{2,0}=0} \frac{1}{y_{1,0}^2y_{2,0}} & (1+y_{1,0})^{1-X_{2,2}^{(0)}-\alpha_0}(1+y_{2,0})^{-1-X_{2,2}^{(0)}-\alpha_0}\\ &\times(1+y_{2,0}+y_{1,0}y_{2,0})^{X_{2,2}^{(0)}+\alpha_0} = 1- X_{2,2}^{(0)} - \alpha_0.
\end{aligned}
\end{equation}
Thus, for any value of the Regge intercept, the matching constrains the exponent of the curve $X_{2,2}^{(0)}$ to be zero. This is also consistent with its definition on the surface, since (by homology) this curve should carry zero momentum. In a similar fashion, we also infer that $X_{1,1}^{(0)} = 0$ by looking at the mass level $n_1=0,\, n_2=1$.

Following these steps in a recursive manner and correctly accounting for the degeneracies at higher levels, one can in principle obtain an infinite tower of constraints on the different parameters that characterize our surface loop integrals. In particular, for the tachyon amplitude, we observe that these unitarity constraints seem to be enough to sequentially fix the exponents of all the different curves that enter the integrand -- we have checked this up to level $2$, where there are still no degeneracies. However, in this case we can directly read off all the exponents by matching \eqref{eq:bubble integrand} to the exact expression for the moduli space integral, since this is well known in the literature (see e.g. \cite{Green:2012pqa}). A complete derivation is presented in Appendix \ref{app:GSW matching}. To summarize, the different exponents are fixed to be:
\begin{equation}
    X_{1,2}^{(q)} = -\alpha_0 = 1,\quad  X_{1,1}^{(q)} = X_{2,2}^{(q)} = 0,\quad \Delta(q) = \left(1-\frac{d}{2}\right)q^2 - \frac{d}{2}q,
\end{equation}
as expected by homology of the curves in the surface.

We now proceed to explain how these constraints let us study the validity of other UV regularization of the field theory amplitude via surface integrals of the type \eqref{eq:surfaceIntegrand1loop}.

\subsection{Massless surface amplitudes and $\hat D_n$}
\label{sec:MasslessLS}

In order to concretely show the power of the integrand residues as unitarity constraints, we now analyze the 1-loop 2-point amplitude in $\hat{D}_n$. This is a different stringy UV regularization of $\Tr(\Phi^3)$ theory defined on the surface, which arises from uplifting the canonical form of the 1-loop polytopes defined in \cite{Salvatori:2019ehb} (also studied in \cite{Backus:2025hpn}) to finite $\alpha'$. The main difference with respect to our usual picture is that for curves connecting external points with the puncture (identified with spirals around the loop on the fatgraph), we now distinguish between clockwise and anti-clockwise orientation. In other words, we consider two copies $Y^+,\, Y^-$ of these curves. This is a slight modification to the process of tagging curves in the context of type $\mathcal{D}$ cluster algebras, which allows to obtain a polytope that presents the correct factorization behavior at 1-loop \cite{Arkani-Hamed:2019vag,Salvatori:2018aha}. In addition, in this picture we don't include any closed or self-intersecting curves, which makes $\hat{D}_n$ a particularly simple extension of the field theory limit to finite values of $\alpha'$.

To illustrate these objects, we briefly consider the even simpler example of the one-point function $\hat D_1$. Since there is only one propagator in the fatgraphs, it's convenient to just work in $u$-parametrization, with $u_{1,0}^+=u$ and $ u_{1,0}^- = 1-u$, such that the integral is given by:
\begin{equation}
   \mathcal{A}_1^{\hat D_1} = \int \frac{d^d\ell}{\pi^{d/2}}\, \int_0^1 du\ u^{Y^+ -1}(1-u)^{Y^- -1} =  \int \frac{d^d\ell}{\pi^{d/2}}\, \int_0^1 du\ \left(u(1-u)\right)^{\ell^2 + m^2 -1},
\end{equation}
where we have provided a mass to the loop propagator $Y^\pm = \ell^2 + m^2$ to add some scale to the integral. We can now perform the loop integration, leaving us with:
\begin{equation}
   \mathcal{A}_1^{\hat D_1} =  \int_0^1 du\ \frac{\left(u(1-u)\right)^{m^2-1}}{\left[ -\log(u(1-u)) \right]^{d/2}}.
\end{equation}
Not only is this integral perfectly UV finite, but is also exponentially suppressed as $m^2\to\infty$. However, despite displaying safe high-energy behavior, this amplitude still fails to satisfy unitarity constraints, as we now proceed to showing.
To do this, let's go back to our 2-point example. Defining 
\begin{equation}
[i,j] = 1+ y_j+y_j\, y_{j-1}+\ldots+ y_j\, y_{j-1}\cdots y_i,
\end{equation}
the $u$-variables for the different curves in $\hat{D}_n$ are given by:
\begin{equation}
    u_{i,j} = \frac{[i+2,j-1][i+1,j]}{[i+2,j][i+1,j-1]},\quad u_{i,0}^+ = \frac{y_i [i+1,i-1]}{[i+1,i]},\quad u_{i,0}^- = \frac{[i+2,i]}{[i+1,i]}.
\end{equation}
Then, in our case the integrand takes the following form:
\begin{equation}
\begin{aligned}
   \mathcal{I}_2^{\hat{D}_2} = \int_{\mathbb{R}^+} \prod_{i=1}^2 \frac{dy_{i,0}}{y_{i,0}}\ y_{i,0}^{Y_i^+} (1+y_{1,0})^{Y_1^- + Y_2^+ - X_{1,1} - X_{2,2}}(1+y_{2,0})^{Y_2^- + Y_1^+ - X_{1,1} - X_{2,2}}\\ \times(1+y_{1,0}+y_{1,0}y_{2,0})^{X_{1,1}-Y_1^+ - Y_1^-}(1+y_{2,0}+y_{1,0}y_{2,0})^{X_{2,2}-Y_2^+ - Y_2^-}.
\end{aligned}
\end{equation}
In the same manner as before, we are looking to extract the leading singularity at some mass level $(n_1,n_2)$, which is simply:
\begin{equation}
\begin{aligned}
    \hat{\text{LS}}_{n_1,n_2} = \mathop{\mathrm{Res}}_{y_{1,0}=y_{2,0}=0} \frac{1}{y_{1,0}^{1+n_1}y_{2,0}^{1+n_2}}\ (1+y_{1,0})^{-n_1-n_2-X_{1,1}-X_{2,2}}(1+y_{2,0})^{-n_1-n_2-X_{1,1}-X_{2,2}}
\\ \times(1+y_{1,0}+y_{1,0}y_{2,0})^{X_{1,1}+2n_1}(1+y_{2,0}+y_{1,0}y_{2,0})^{X_{2,2}+2n_2}.
\end{aligned}
\end{equation}
While we can easily give a closed formula for this residue for general $(n_1,n_2)$, we only need a couple of mass levels to show that it fails the unitarity constraints. The explicit expressions for both the surface integral and the set of \emph{ansatz} residues are shown in table \ref{tab:LS_table}, where in the latter we have already substituted the values of the couplings obtained by matching to the tree-level string amplitude:
\begin{table}[t]
\centering
\begin{tabular}{c|c|c}
$(n_1,n_2)$ & $\hat{\text{LS}}_{n_1,n_2}$                                          & $\text{LS}_{n_1,n_2}$ \\[3pt] \hline 
(0,0)       & 1                                                                    & 1                     \\[5pt]
(1,0)       & $\scriptstyle1-X_{2,2}$                                                          & 1                     \\[5pt]
(2,0)       & $\scriptstyle\frac{1}{2}(1-X_{2,2})(2-X_{2,2})$                                  & 1                     \\[5pt]
(1,1)       & $\scriptstyle4+X_{1,1}+X_{2,2}+X_{1,1}X_{2,2}$                                   & $d$                   \\[5pt]
(2,1)       & $\scriptstyle\frac{1}{2}(4-9X_{2,2}-3X_{2,2}^2-X_{1,1}X_{2,2}-X_{1,1}X_{2,2}^2)$ & $\frac{29}{32}d-\frac{13}{16}$  
\end{tabular}
\caption{\label{tab:LS_table} Comparison of the leading singularities at levels $(n_1,n_2)$ from the $\hat{D}_2$ integral (middle), and the glued \emph{ansatz} 3-point amplitudes with the couplings matched to the $\alpha_0=0$ string amplitude (right). The mirror levels $n_1 \leftrightarrow n_2$ are simply obtained by setting $1\leftrightarrow2$ in the curve exponents.}
\end{table}

Just from the massless-massive residues, we can fix $X_{1,1}=X_{2,2}=0$, which then requires that the space-time dimension is $d=4$ to match the $(1,1)$ massive level. However, when we fix these values, there is no possible way to match the $\text{LS}_{2,1}$ leading singularity, which ultimately means that our $\hat{D}_2$ integral is inconsistent with unitarity conditions already at the second mass level.

Another candidate UV completion of Tr$(\phi^3)$ amplitudes is given by the standard surface integrals with the Regge intercept, $\alpha_0 =0$. As it has been pointed out in \cite{Arkani-Hamed:2023lbd,Arkani-Hamed:2023jry,NimasTalk,combinatString}, these integrals actually define an infinite class of apparently healthy UV regularizations: this is because, strictly speaking, to UV regularize the low energy theory we don't need the infinite product over all possible self-intersections $q$. Instead, we can consider the truncated integrals where we keep open self-intersecting curves up to a certain $q_{max}$. 

From our analysis, since the unitarity constraints impose an infinite number of constraints from the tower of massive residues, we expect that \textit{any} truncated integral will fail to satisfy these for high enough levels. For instance, this can be seen for the simplest case of the ``baby'' surface integrals, which only include curves that don't self-intersect (and potentially closed curves with exponents $\Delta(q)=\Delta_1\,q+\Delta_2q^2$). One can then easily check that:
\begin{equation}
    \text{LS}^{\text{baby}}_{1,1} = \text{LS}^{\text{baby}}_{1,2} = \text{LS}^{\text{baby}}_{2,1} = -\Delta_1-\Delta_2,
\end{equation}
which clearly contradicts the results in the right column in table \ref{tab:LS_table}.

If we instead keep the infinite number of self-intersecting curves, naively this could add enough degrees of freedom to satisfy all the unitarity constraints. We have checked that this is the case for the first few massive levels, which are free of degeneracies ($n\leq 2$). Nonetheless, the matching procedure at low levels already shows that the exponents of the self-intersecting curves look significantly more fine-tuned than in the tachyon case, where all the curve momenta were consistent with their homology assignments (see Appendix \ref{app:GSW matching}). In the case where $\Delta(q) = (a-d/2)q+(b-d/2) q^2$, the exponents of the curves up to level 2 are fixed to be:
\begin{equation}
\begin{aligned}
    X_{1,1}^{(0)} &= X_{2,2}^{(0)} = 0,\\
    X_{1,2}^{(1)} &= -a-b,\quad X_{1,1}^{(1)} = X_{2,2}^{(1)} = -\frac{3d+26}{32}-a-b,\\
    X_{1,2}^{(2)} &= \frac{13d-10-48a-80b}{16}.
\end{aligned}
\end{equation}
In general, these values don't appear as natural as in the tachyon case (except maybe for $d=2$). This of course is no definitive reason to discard the theory, but it might hint at the fact that the massless surface integrals should break down when performing consistency checks involving higher point/loop amplitudes.

\section{Loop-level contour}\label{sec:loop contour}

We now switch to discussing the integration contour for surface amplitudes at 1-loop. Just like in the previous section, we do not aim at providing a complete implementation of the $i\varepsilon$ prescription at arbitrary $n$. Instead, we stick to the $n=2$ case, and explain how to deal with the new general features entering at loop-level.

In \ref{sec:IntegrationLoopMom}, we show that after performing the loop-integrations on the surface integral, we are left with a global Schwinger representation of the $2$-point 1-loop integral in $y$-space where, just like at tree-level, the different diagrams are encoded in different regions of $y$-space -- the cones of the 1-loop fan \cite{Arkani-Hamed:2023lbd}. It is therefore natural to expect that the $i \varepsilon$ definition of the contour extends to loop-level in the obvious way: in each cone of the fan, we integrate over the original domain up to a certain point and then deform the Schwinger parameters along the imaginary direction, which allows us to regulate the divergences coming from the region where the worldsheet degenerates.

To test this contour prescription, in section \ref{sec:Discontinuities} we start by computing the discontinuities of the amplitude. We observe that each term in the threshold expansion is automatically given by the usual Schwinger parametrization for Feynman integrals with different integer $m^2$ for the internal propagators. At a given level, only a finite subset of the curves on the surface contribute to the discontinuity, making it simpler to extract the result. We explicitly compare the discontinuities given by the integration contour with the leading singularities obtained in section \ref{sec:MasslessLS}, and find that they precisely agree after accounting for the appropriate Lorentz invariant phase space integral.

Finally, we discuss the implementation of the contour on the full fan. In the bubble region, it is useful to use the threshold expansion to see that for fixed external momenta, only finitely many terms are divergent and need a contour deformation. Using this, we can treat these divergent parts (which take the form of simple Feynman integrals) separately, and regulate them as we do in field theory. In the remaining cones of the fan, where there are tree-like propagators -- namely the tadpoles -- the contour prescription ends up being identical to the one at tree-level. 

The results in this section suggest that the surface formalism could be used to straightforwardly implement the numerous tools recently developed for Landau analysis of field theory amplitudes \cite{Hannesdottir:2022bmo,Caron-Huot:2023ikn,Caron-Huot:2023vxl,Hannesdottir:2022xki,Hannesdottir:2024cnn,Hannesdottir:2024hke,Caron-Huot:2024brh} to integrals in string theory. We leave this avenue of research for future work. 

\subsection{Integrating over the loop momentum}
\label{sec:IntegrationLoopMom}

From now on, we will consider only the surface integral involving massless external states. Although this doesn't correspond to actual string theory amplitudes, for open bosonic strings the loop integrals are divergent due to the presence of tachyons. Meanwhile, for the massless surface integral there are no internal tachyon states running around the loop, so we can study these objects without worrying about such divergences.

For simplicity, we start by writing the loop integrand omitting all the curves on the surface other than $Y_1,\, Y_2$. This is of course not going to change the result of integrating over the loop momentum, since all the remaining exponents are independent of $\ell^\mu$.

Using the expression for the $u$-variables derived from the bubble fatgraph, the Koba-Nielsen factor takes the following form in the corresponding positive parametrization:
\begin{equation} 
\begin{aligned} 
    \ah_2^{(1)} = \int\frac{d^d\ell}{(i\pi)^{d/2}}\int_{D}\frac{dy_{1,0}}{y_{1,0}}\frac{dy_{2,0}}{y_{2,0}}\ y_{1,0}^{Y_1}y_{2,0}^{Y_2}(1+y_{1,0})^{Y_2-Y_1}(1+y_{2,0})^{Y_1-Y_2}.
\end{aligned}
\end{equation}
where $D$ is the modified integration domain in the $y$-variables defined earlier, $\{y_{1,0},y_{2,0}\geq 0 \cap y_{1,0}y_{2,0}\leq 1\}$, where the infinitely spiraling curves are defined. 

In analogy to the usual procedure with Feynman integrals, we can now substitute $Y_1=\ell^2$ and $Y_2 = (\ell+p)^2$ and integrate out the loop momentum $\ell^\mu$, since the dependence is simply Gaussian. Defining $y_{i,0}=e^{-t_{i,0}}$ and reinstating the product over the rest of the curves on the surface, after loop-integration we arrive at:
\begin{equation}\label{eq:loop-integrated amp}
\begin{aligned} 
    \ah_2^{(1)} &= \int_{t_{1,0}+t_{2,0}\geq 0} \frac{dt_{1,0}\,dt_{2,0}}{(t_{1,0}+t_{2,0})^{d/2}}\ \Bigg(\prod_{q=0}^\infty \left(u_{1,2}^{(q)}\right)^{X_{1,2}^{(q)}}\left(u_{1,1}^{(q)}\right)^{X_{1,1}^{(q)}}\left(u_{2,2}^{(q)}\right)^{X_{2,2}^{(q)}} \left(u_{\Delta}^{(q)}\right)^{\Delta(q)} \Bigg)\\ &\times \exp \Bigg[ -\frac{\left(\log(1+e^{-t_{2,0}})-\log(1+e^{-t_{1,0}})-t_{1,0}\right)\left(\log(1+e^{-t_{1,0}})-\log(1+e^{-t_{2,0}})-t_{2,0}\right)}{t_{1,0}+t_{2,0}}p^2\Bigg],
\end{aligned}
\end{equation}
where we have left $p^2 \neq 0$ otherwise the result would be trivially zero.

As a consistency check, we can take the low-energy limit of this integral. Restoring the $\alpha'$ dependence as $X_{i,j}\to\alpha' X_{i,j}$, the low-energy regime corresponds to taking $|t_{i,j}|\to\infty$ and analyzing the tropical behavior of the integrand. In this limit the contribution from all self-intersecting curves drops out, and the resulting expression depends on what region of the integration domain we are in, explicitly, we have
\begin{equation} 
\begin{aligned} 
    \ah_2^{(1)} \to \begin{dcases} \int \frac{dt_{1,0}dt_{2,0}}{(t_{1,0}+t_{2,0})^{d/2}}\ \exp\left[ -\frac{t_{1,0}t_{2,0}}{t_{1,0}+t_{2,0}}p^2 \right], & t_{1,0},t_{2,0}>0,\\ \int \frac{dt_{1,0}dt_{2,0}}{(t_{1,0}+t_{2,0})^{d/2}}\ \exp\left[ t_{2,0}X_{1,1}^{(0)}\right], & t_{1,0}>0,\, t_{2,0}<0,\\ \int \frac{dt_{1,0}dt_{2,0}}{(t_{1,0}+t_{2,0})^{d/2}}\ \exp\left[ t_{1,0}X_{2,2}^{(0)}\right], & t_{1,0}<0,\, t_{2,0}>0.\end{dcases}
\end{aligned}
\end{equation}
The different limits of the integral precisely correspond to the Schwinger parametrizations of the bubble and tadpole diagrams, respectively. In fact, defining the piecewise linear functions $\alpha_{i,j}(t)$ and asking for their respective domains of linearity precisely gives us the 1-loop Feynman fan:
\begin{equation}
 \begin{gathered}
    \begin{tikzpicture}[line width=0.6,scale=1,baseline={([yshift=0.0ex]current bounding box.center)}]
        \coordinate (C) at (0,0);
        \coordinate (T2) at (-1.5,1.5);
        \coordinate (Y2) at (0,1.5);
        \coordinate (Y1) at (1.5,0);
        \coordinate (T1) at (1.5,-1.5);
        \draw[->] (C) -- (T1);
        \draw[->] (C) -- (Y1);
        \draw[->] (C) -- (Y2);
        \draw[->] (C) -- (T2);
        \node[scale=1,right] at (Y1) {$Y_1$};
        \node[scale=1,above] at (Y2) {$Y_2$};
        \node[scale=1,right] at (T1) {$X_{1,1}$};
        \node[scale=1,left] at (T2) {$X_{2,2}$};
        \draw[color=gray] (0.5,1) -- (0.8,1);
        \draw[color=gray] (0.95,1) circle[radius=0.15];
        \draw[color=gray] (1.1,1) -- (1.4,1);

        \draw[color=gray] (-0.9,1.2) -- (-0.4,1.2);
        \draw[color=gray] (-0.65,1.2) -- (-0.65,1.4);
         \draw[color=gray] (-0.65,1.5) circle[radius=0.1];
         
          \draw[color=gray] (1.5,-0.5) -- (1,-0.5);
        \draw[color=gray] (1.25,-0.5) -- (1.25,-0.7);
         \draw[color=gray] (1.25,-0.8) circle[radius=0.1];

         \draw[->,color=gray] (2.5,0.15) -- (3.5,0.15);
    \end{tikzpicture}
\end{gathered}
\hspace{0.3cm}
\begin{cases}
    \alpha_{1,0}(t) &= t_{1,0} - \max(0,-t_{2,0}) + \max(0,-t_{1,0}),\\
    \alpha_{2,0}(t) &= t_{2,0} - \max(0,-t_{1,0}) + \max(0,-t_{2,0}),\\
    \alpha_{1,1}(t) &= \max(0,-t_{2,0}),\\
    \alpha_{2,2}(t) &= \max(0,-t_{1,0}),\\
\end{cases} 
\label{eq:1loopFan}
\end{equation}
the tropical behavior of the integral can be written as:
\begin{equation}
    \ah_2^{(1)} \to \int \frac{dt_{1,0}dt_{2,0}}{(t_{1,0}+t_{2,0})^{d/2}}\ \exp\left[ -\frac{\alpha_{1,0}(t)\alpha_{2,0}(t)}{\alpha_{1,0}(t)+\alpha_{2,0}(t)}p^2 - \alpha_{1,1}(t)X_{1,1}^{(0)} - \alpha_{2,2}(t)X_{2,2}^{(0)} \right],
\end{equation}
where the exponent is precisely given in terms of the surface Symanzik polynomials first introduced in \cite{Arkani-Hamed:2023lbd}, which capture all the Feynman diagram contributions of the 1-loop 2-point amplitude. Thus, the curve integral defined from the surface is explicitly seen to be the $\alpha'\to0$ limit of our surface integral.

\subsection{Threshold expansion and unitarity cuts}
\label{sec:Discontinuities}

Once the loop momentum has been integrated out, there is actually not much work left to do in order to extract the unitarity cuts of our surface integral. As we know from the tree-level case, we need to localize the integrand on the region where parts of the corresponding worldsheet would degenerate into a long tube, leading to the integral developing a singularity. At that point, we implement the $i\varepsilon$ prescription and change to Lorentzian signature by deforming the $t$-contour in the imaginary direction. This procedure, just like at tree-level, can be applied in all the cones in the fan \eqref{eq:1loopFan}. 

\begin{figure}[t]
    \centering 
    \includegraphics[width=0.7\textwidth]{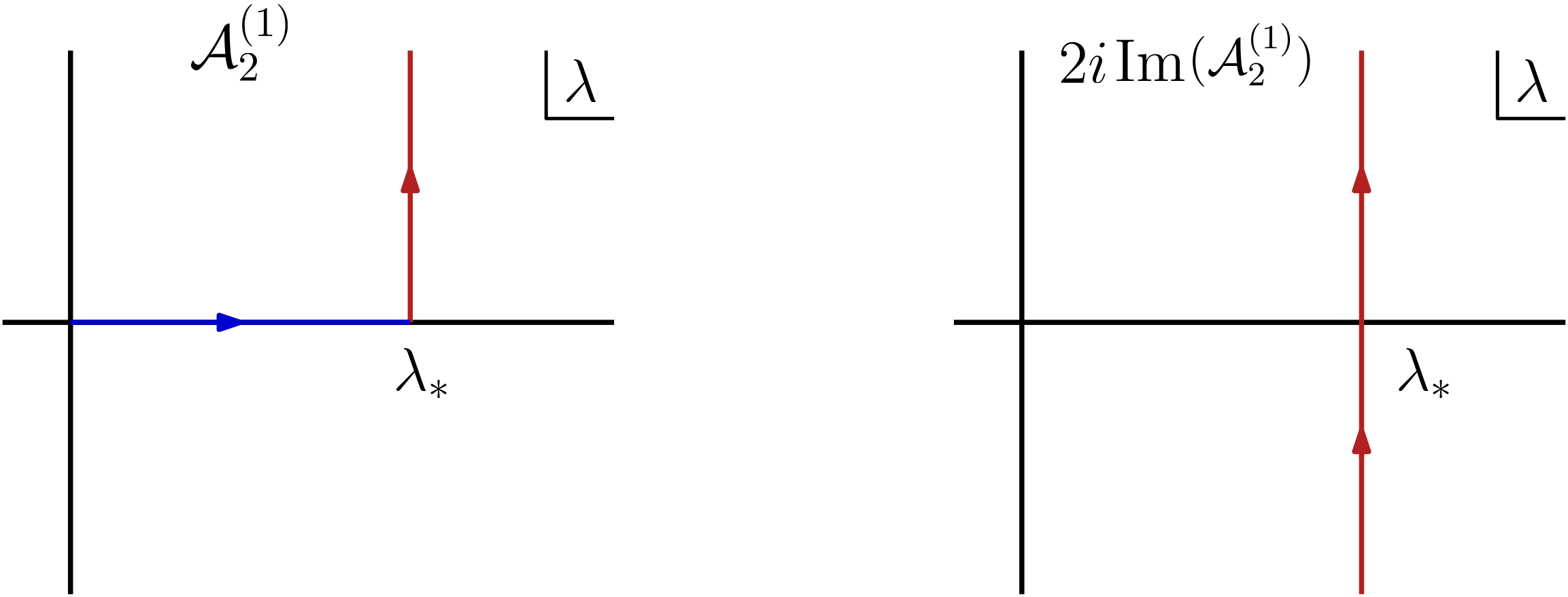}
    \caption{(Left) $i\varepsilon$ prescription on the integration contour in $\lambda$ to expose degenerations associated to the bubble diagram. (Right) Contribution to the unitarity cut of the surface integral, given by its imaginary part. The cutoff $\lambda_\star$ can be set arbitrarily large.}
    \label{fig:lambda contour}
\end{figure}

To be more concrete, we are looking for the unitarity cuts associated to the bubble diagram, where both loop propagators are on-shell. These will be localized on the bubble cone, $t_{1,0},\,t_{2,0}>0$, and thus it is convenient to introduce the following change of variables:
\begin{equation}
    t_{1,0} = \lambda\alpha,\qquad t_{2,0} = \lambda(1-\alpha),
\end{equation}
where the cone now corresponds to $0<\lambda<\infty,\  0\leq\alpha\leq1$. Moreover, now $\lambda$ is the single coordinate that parametrizes the degeneration of the worldsheet as we let it get arbitrarily large. The $i\varepsilon$ prescription is then implemented in the usual way by integrating on the real line up to some finite $\lambda_\star$ and then deforming the contour in the imaginary direction as $\lambda\to\lambda_\star+i\tau$, (see figure \ref{fig:lambda contour}, left).

Since at the moment we are only interested in the unitarity cuts of the amplitude, we can further simplify the computation. Indeed, by the optical theorem the unitarity cut is given by the imaginary part of the integral. At the level of the worldsheet integration, this amounts to subtracting a contour where the $i\varepsilon$ prescription has been applied in the opposite way (i.e. deforming downwards instead of upwards). The contribution over the real line then cancels out, and we are left with an integral over $-\infty<\tau<\infty$, where the cutoff $\lambda_\star$ can be formally set as large as we want. In $y$-space, this is equivalent to tracing a never-ending circle around $y_{\lambda}=y_{1,0}y_{2,0}=0$, where the radius can be taken to be arbitrarily small. This contour is depicted in figure \ref{fig:lambda contour}.

Strictly speaking to extract the full imaginary part of the amplitude, we should follow this procedure in all cones in the fan -- including the tadpole cones which in $(\lambda,\alpha)$-space correspond to $\alpha \in (-\infty,0)$ and $\alpha \in (1,+\infty)$. However, as long as we choose our tadpole kinematics $X_{1,1},\,X_{2,2}$ to be non-negative, the integrand decays exponentially with $\lambda_\star$ in those regions, and therefore this regions won't contribute to the imaginary part. 

Expanding the different terms of the integral \eqref{eq:loop-integrated amp} in the tropical limit $\lambda_\star\to\infty$ yields an infinite sum over normal thresholds. Schematically:
\begin{equation}\label{eq:threshold expansion bubble}
    \text{Im}\,\ah_2^{(1)} \sim \frac{1}{2}\sum_{n_1,n_2\geq0} \int_{\uparrow}\frac{d\lambda}{\lambda^{1-d/2}}\int_0^1 d\alpha\ f_{n_1,n_2}\left(X^{(q)},p^2,\alpha\right)\exp[-\lambda(\alpha(1-\alpha)p^2+\alpha n_1+(1-\alpha)n_2)],
\end{equation}
where $f_{n_1,n_2}$ is the coefficient in the expansion containing $n_1$ powers of $y_{1,0} = e^{-t_{1,0}}$ and $n_2$ powers of $y_{2,0} = e^{-t_{2,0}}$. Therefore, we have that from the product over self-intersecting curves outside the exponential in \eqref{eq:loop-integrated amp}, only curves with self intersection $q\leq \min(n_1,n_2)$ contribute to $f_{n_1,n_2}$. Here, the upwards arrow $\uparrow$ in the integration domain for $\lambda$ denotes the contour over $-\infty<\tau<\infty$ with $\lambda_\star\to\infty$. This form makes it clear that the contribution of the bubble cone to the imaginary part of the amplitude is simply a sum over Schwinger parametrized bubble integrals, each with a different set of integer $m^2$ for the internal loop propagators.

Since we are taking $\lambda$ to have an arbitrarily large real part in our integration contour, a certain term in the sum only has a non-zero contribution if $p^2\leq -(\sqrt{n_1}+\sqrt{n_2})^2$, as otherwise the exponent in \eqref{eq:threshold expansion bubble} is uniformly negative. Of course, this is just the familiar statement that the external kinematics have to be above the corresponding mass threshold to appear in the unitarity cut.

To illustrate this more concretely, we will consider the imaginary part of $\ah_2^{(1)}$ for $-4< p^2<-1$, which includes the contribution from mass levels $(0,0)$, $(0,1)$ and $(1,0)$. Thus, in the tropical expansion we can discard all curves with $q\geq1$, leaving us only with:
\begin{equation}
\begin{aligned} 
    \text{Im}\,\ah_2^{(1)} &\to \int_{\uparrow}\frac{d\lambda}{\lambda^{d/2-1}}\int_0^1 d\alpha\ \frac{(1+e^{-\lambda\alpha}+e^{-\lambda})^{X_{1,1}^{(0)}}(1+e^{-\lambda(1-\alpha)}+e^{-\lambda})^{X_{2,2}^{(0)}}}{\left[(1+e^{-\lambda\alpha})(1+e^{-\lambda(1-\alpha)})\right]^{X_{1,1}^{(0)}+X_{2,2}^{(0)}}}\\ &\times \exp \Bigg[ \frac{\left(\log\left(\frac{1+e^{-\lambda(1-\alpha)}}{1+e^{-\lambda\alpha}}\right)-\lambda\alpha\right)\left(\log\left(\frac{1+e^{-\lambda(1-\alpha)}}{1+e^{-\lambda\alpha}}\right)+\lambda(1-\alpha)\right)}{\lambda}p^2\Bigg].
\end{aligned}
\end{equation}
We can now perform the expansion in the limit $\lambda_\star\to\infty$, keeping only contributions of order $y_{1,0}=e^{-t_{1,0}}=e^{-\lambda\alpha}$ and $y_{2,0}=e^{-t_{2,0}}=e^{-\lambda(1-\alpha)}$. The result is:
\begin{equation}
\begin{aligned}
    \text{Im}\,\ah_2^{(1)} = \int_{\uparrow}\frac{d\lambda}{\lambda^{d/2-1}}\int_0^1 &d\alpha\ e^{-\lambda \alpha(1-\alpha)p^2} + \left[ (2\alpha-1)p^2 - X_{2,2}^{(0)} \right]e^{-\lambda(\alpha(1-\alpha)p^2 + \alpha)} \\ &+ \left[ (1-2\alpha)p^2 - X_{1,1}^{(0)} \right]e^{-\lambda(\alpha(1-\alpha)p^2 + (1-\alpha))} + \mathcal{O}(y_{1,0}y_{2,0}).
\end{aligned}
\end{equation}
We can actually evaluate the $\alpha$-integral analytically by noticing that all the terms in the expansion have a negative exponent outside of the region $0\leq\alpha\leq1$ and thus, in the tropical limit, all of these vanish for $\alpha \notin [0,1]$. As a result, we can formally extend the $\alpha$-integration domain to the whole real line, while leaves us with a simple Gaussian:
\begin{equation}
    \text{Im}\,\ah_2^{(1)} = \frac{1}{2}\int_{\uparrow}\frac{d\lambda}{\lambda^{(d-1)/2}}\ \sqrt{\frac{\pi}{-p^2}}\left[ e^{-\frac{\lambda}{4}p^2} + \left(2-X_{1,1}^{(0)} - X_{2,2}^{(0)}\right)e^{-\frac{\lambda(1+p^2)^2}{4p^2}} \right]
\end{equation}
Lastly, the integral over $\lambda$ can be performed by using the Hankel representation of the Gamma function:
\begin{equation}
    \frac{1}{\Gamma(z)} = \frac{1}{2\pi i}\int_\mathcal{C} ds\ s^{-z} e^s,
\end{equation}
where $\mathcal{C}$ is any contour wrapped around the negative real axis anti-clockwise. Since our contour can be deformed in that way, we can use this expression to finally obtain:
\begin{equation}\label{eq:bubble unitarity cuts}
    \text{Im}\, \ah_2^{(1)} = \frac{i\pi^{3/2}}{\sqrt{-p^2}\Gamma\left(\frac{d-1}{2}\right)}\left[ \left(\frac{-p^2}{4}\right)^{(d-3)/2} + \left(2-X_{1,1}^{(0)} - X_{2,2}^{(0)}\right)\left(\frac{(1+p^2)^2}{4(-p^2)}\right)^{(d-3)/2} \right].
\end{equation}
We can now make the connection with the leading singularities we calculated in the last section at the level of the integrand. Indeed, these expressions should coincide up to an integration of the remaining Lorentz invariant phase space of the loop momentum. To see this, recall that our leading singularities for these mass levels were given by:
\begin{equation}
\begin{aligned}
    \text{LS}_{0,0}&=1,\\
    \text{LS}_{0,1}&=1-X_{1,1},\\
    \text{LS}_{1,0}&=1-X_{2,2}.
\end{aligned}
\end{equation}
Meanwhile, the integral over a Lorentz invariant phase space for the cut bubble Feynman integral with masses $m_1^2,\, m_2^2$ in $d$ dimensions is well known (see e.g. \cite{Hannesdottir:2022bmo}):
\begin{equation}
    \text{Cut}\,\ih_{bub}(m_1^2,m_2^2) = \frac{i\pi^{3/2}}{\sqrt{-p^2}\Gamma\left(\frac{d-1}{2}\right)}\left(\frac{\Delta}{4(-p^2)}\right)^{(d-3)/2}\Theta(-p^2-(m_1+m_2)^2),
\end{equation}
where $\Delta=(p^2+(m_1+m_2)^2)(p^2+(m_1-m_2)^2)$ is the discriminant of the Schwinger potential. By comparison to \eqref{eq:bubble unitarity cuts}, we can thus write:
\begin{equation}
    \text{Im}\, \ah_2^{(1)} = \text{Cut}\,\ih_{bub}(0,0) + \left(1-X_{2,2}^{(0)}\right)\,\text{Cut}\,\ih_{bub}(1,0)+ \left(1-X_{1,1}^{(0)}\right)\,\text{Cut}\,\ih_{bub}(0,1).
\end{equation}
This agrees perfectly with our findings at the integrand level. We stress again that the ability to extract the leading singularities without performing the full loop integration is a non-trivial consequence of the fact that surfaceology directly hands us the \emph{integrand} rather than the loop-integrated object. Nonetheless, even after loop integration, the discontinuities of the amplitude are organized in a very hierarchical manner in terms of the self-intersecting curves. At a certain mass level, we only need to keep a finite number of them to compute the discontinuity.

\subsection{1-loop integration contour}

So far, we have seen that it is straightforward to implement the $i\varepsilon$ prescription in our surface integral to extract the imaginary part of the amplitude. One important aspect in this case is that the finite integration over the real $\lambda$-line cancels out and we can directly work in the tropical limit $\lambda_\star\to\infty$. This is analogous to the procedure followed in the standard worldsheet integration picture. For the superstring, the finite contributions of the planar annulus and the M\"obius strip cancel among each other up to a residue \cite{Eberhardt:2022zay,Eberhardt:2023xck}. Of course, in the case of open bosonic string theory, this residue contains the usual tachyon divergences, which means that the result is not really well defined \footnote{In \cite{Manschot:2024prc}, this is dealt with by formally consider the remaining contribution to the integral, for which it is possible to take the tropical limit directly.}.

In our formalism, we are studying a stringy integral with no tachyons, but we don't have the analogue of the cancellation between the disk and the M\"obius strip. Our integrand is defined on a single surface (the punctured disk), and there are no extra contributions that cancel the finite integration over the real line (blue part of the left contour in figure \ref{fig:lambda contour}). However, the advantage of our description is that the $t$-integrand is already written in a form that manifestly tells us how to deform the different contributions to guarantee the convergence of the integral.

\subsubsection*{Bubble Cone}

As we have stressed already, a divergence only arises when the external kinematics are above the threshold to produce the internal particles corresponding to the degeneration of the worldsheet. Therefore, if we perform the threshold expansion within the bubble cone as in \eqref{eq:threshold expansion bubble}, it is clear that only a finite number of terms will potentially cause the integrand to blow up. For all other contributions where $-(\sqrt{n_1}+\sqrt{n_2})^2<p^2$, the integrand will decay exponentially as $\lambda\to\infty$. Moreover, the finite number of terms that diverge take the simple form of Schwinger parametrized bubble integrals in field theory, for which we can apply the $i\varepsilon$ to regularize them. This can be done in a straightforward way, by deforming the contour in $\lambda$ upwards in the imaginary direction and replacing $p^2\to p^2-i\varepsilon$, or by using more sophisticated methods like the ones developed in \cite{Hannesdottir:2022bmo}. After subtracting the divergent terms, the remainder can be integrated simply over the real $\lambda$-line, since it will be perfectly finite.

With all of this in mind, we propose a systematic way to compute the 2-point integral without any divergences. For a fixed value of the external kinematics $p^2$, perform the tropical expansion of the integrand in the bubble cone up to the thresholds with $\lceil p^2\rceil<-(\sqrt{n_1}+\sqrt{n_2})^2$:
\begin{equation}\label{eq: bubble finite expansion}
    \ah_2^{(1)} \sim \int_0^\infty\frac{d\lambda}{\lambda^{1-d/2}}\int_{0}^1 d\alpha\ \left(\sum_{\substack{n_1,n_2\geq0\\\lceil p^2\rceil<-(\sqrt{n_1}+\sqrt{n_2})^2}} f_{n_1,n_2}\exp\left[-\lambda\vh_{n_1,n_2}(\alpha)\right]\right)(u_\Delta)^\Delta + \mathcal{R}^{(n_1,n_2)},
\end{equation}
where $\vh_{n_1,n_2}(\alpha) = \alpha(1-\alpha)p^2 + n_1\alpha+n_2(1-\alpha)$ and the tropical expansion of the remainder $\mathcal{R}^{(p^2)}$ only contains thresholds that are inaccessible for the chosen kinematics. Here, we have also explicitly included the contribution from the closed curve $\Delta$, which becomes important to regulate the UV divergence at $\lambda\to 0$. As we have seen before, by choosing the exponents $\Delta(q)=q\Delta$, the infinite product becomes:
\begin{equation}
    (u_\Delta)^\Delta := \prod_{q=0}^\infty u_{\Delta} ^{\Delta(q)}=\left(1-e^{-\lambda}\right)^\Delta.
\end{equation}
This has a trivial contribution to the limit $\lambda\to\infty$, but for $\lambda\to0$ it will scale as $\lambda^{\Delta}$, so we can choose $\Delta$ appropriately to soften the divergence coming from the measure after loop integration. 

For each term in the finite sum, we can integrate up to some fixed value $\lambda_\star$ and then deform the contour only in the bubble cone $0\leq\alpha\leq1$ as:
\begin{equation}
    \lambda \to \lambda_\star \left[\delta+ i\alpha(1-\alpha)\right]\tau,\quad 0\leq\tau<\infty,
\end{equation}
where $\delta>0$ is small enough so that the imaginary part of the deformation dominates over the real one. As opposed to the tree-level case, the contour deformation introduced in this case does not allow for a repackaging of the $\tau$ integral into a phase factor times a finite integral. Therefore, in order to make the integrals convergent in the bubble cone we must add a regulator $p^2\to p^2-i\varepsilon$. Meanwhile, for the remainder contribution $\mathcal{R}_{n_1,n_2}$ we can simply integrate over the real $\lambda$-line, since it won't generate any divergence. Note that the factor $\alpha(1-\alpha)$ ensures that the deformation into the imaginary direction is a continuous one, as it becomes zero at the boundaries of the bubble cone.

Note that this implementation of the $i\varepsilon$ prescription is simple enough that it is not strictly necessary to decompose the integrand into the finite sum of accessible thresholds and the remainder function. Instead, one could deform the contour uniformly for the whole integrand and regulate the divergence with the $i\varepsilon$ kinematic deformation. However, separating the different contributions is not only more natural but also numerically more stable. On the one hand, it explicitly reflects the fact that the inaccessible thresholds (encoded in the remainder function $\rh_{n_1,n_2}$) don't contribute to the imaginary part of the amplitude, in accordance to the integration contour remaining purely real. On the other hand, when it comes to numerically evaluating the integral along the contour, if the deformation is performed in all the terms then the overall object is significantly more oscillatory, which compromises the accuracy of the result. 

\subsubsection*{Tadpole Cones}

Finally, to obtain the full answer, we also need to include the contributions from the tadpole cones. Of course, in both string theory and field theory, tadpoles are rather singular contributions with well-known interpretations. However, in our context, since the tadpole propagators $X_{1,1}$ and $X_{2,2}$ are treated as free kinematic variables, it is sensible to examine the contributions from these cones and study their dependence on the $X_{i,i}$. This serves as a template for what happens at higher points in cones containing both tree and loop propagators. 

In these cones, the tropical limit corresponds to the Schwinger parametrization of the tadpole Feynman diagrams, and as such the presence of divergences depends on the kinematics $X_{1,1}$ and $X_{2,2}$. Not surprisingly, for these regions the $i\varepsilon$ prescription is much simpler to implement, since the degenerating modulus doesn't involve any complicated dependence from the loop integration.

To see this, we do the same trick as in the tree-level case and map the tadpole cones into the positive quadrant using \eqref{eq:cone_map}, with the $g$-vectors as depicted in \eqref{eq:1loopFan}. For instance, for the tadpole cone with propagators, $\{X_{1,1},Y_1\}$, we obtain the map:
\begin{equation}
    t_{1,0} = \hat{t}_{1,0}+\hat{t}_{1,1},\quad t_{2,0} = -\hat{t}_{1,1}, \quad \text{ with } \hat{t}_{1,0},\,\hat{t}_{1,1} \in \mathbb{R}^+,
\end{equation}
which results in the following expression for the 2-point function after loop integration:
\begin{equation}
\begin{aligned}
    \ah_2^{(1)} &= \int_{\mathbb{R}^+} \frac{d\hat{t}_{1,0}\, d\hat{t}_{1,1}}{\hat{t}_{1,0}^{d/2}}\ \exp\left[ -\hat{t}_{1,1}X_{1,1}^{(0)}+\frac{1}{\hat{t}_{1,0}}\log^2\left( \frac{1+e^{-\hat{t}_{1,1}}}{1+e^{-\hat{t}_{1,0}-\hat{t}_{1,1}}}\right) p^2 \right]\\ &\times \left( \frac{1+e^{-\hat{t}_{1,0}}+e^{-\hat{t}_{1,0}-\hat{t}_{1,1}}}{(1+e^{-\hat{t}_{1,0}-\hat{t}_{1,1}})(1+e^{-\hat{t}_{1,1}})} \right)^{X_{1,1}^{(0)}} \left( \frac{1+e^{-\hat{t}_{1,1}}+e^{-\hat{t}_{1,0}-\hat{t}_{1,1}}}{(1+e^{-\hat{t}_{1,0}-\hat{t}_{1,1}})(1+e^{-\hat{t}_{1,1}})} \right)^{X_{1,1}^{(0)}}.
\end{aligned}
\end{equation}
Here, we have omitted the $u$-variables of the rest of the curves in the surface for brevity. It is then clear that the divergence of the integral is determined by the term $-\hat{t}_{1,1}X_{1,1}^{(0)}$ in the exponential in the first line. To regulate the infinity that arises as $\hat{t}_{1,1}\to\infty$ when $X_{1,1}^{(0)}<0$, we take $X_{1,1}^{(0)}\to X_{1,1}^{(0)} - i\varepsilon$ and deform the contour in $\hat{t}_{1,1}$ towards the imaginary axis from some finite real value $R_\star$:
\begin{equation}
    \hat{t}_{1,1}\to R_\star + i \tau,\quad 0\leq\tau<\infty.
\end{equation}
As expected, the contour deformation is identical to the tree-level case, since the kinematics associated to the moduli that produce this degeneration are independent of the loop momenta. In particular, the integrand remains quasi-periodic under $\hat{t}_{1,1} \to \hat{t}_{1,1} + 2\pi i$, and therefore we can once again repackage the $\tau$ integration into a phase factor times a finite integral from $[0,2\pi]$. By doing this, we are also able to extend the validity of the contour for any value of $X_{1,1}$. 
Note also that the contour deformation is still continuous when considering the whole Feynman fan, since the boundary with the bubble cone corresponds to $\hat{t}_{1,1}=0$ and we only start adding an imaginary part for some non-zero, positive $R_\star$.

The procedure outline above provides a consistent way to regularize the 2-point 1-loop surface integral, but it's not without flaws. It is by now well known in the field theory context that implementing the $i\varepsilon$ prescription via a kinematic deformation $p^2\to p^2-i\varepsilon$ is not always consistent with causality, and one has to be careful to land on the physical branch of the amplitude. The deformation might also be different from diagram to diagram, something that becomes even more important for our surface amplitudes, as the Feynman fan spans over the complete set of topologies. Works like \cite{Hannesdottir:2022bmo} have provided better alternatives to the deformation in the context of Feynman integrals that are realized purely at the level of the integration contour, and which ensure causality and analyticity in a much more systematic way.

Nevertheless, the principles we have brought up in this section seem to indicate that an extension of such techniques to our surface integrals is likely feasible. The key observation is that for a fixed value of the kinematics, one only needs to worry about the divergences caused by a finite number of terms in the sum \eqref{eq: bubble finite expansion}. These contribution take the form of simple Feynman integrals, which then should be treated as in field theory. Of course, there are several additional subtleties, such as how to connect the different cones into a single contour deformation, or how to treat the UV regularization from the closed $\Delta$ curve contribution. We leave a systematic treatment of these issues to future works.

\section{Outlook}\label{sec:outlook}

In this work, we have used the surface formulation of scattering amplitudes to expose a number of different properties of stringy integrals in a transparent way. We take advantage of the way in which this $u$-variables and their positive parametrizations manifest all the singularities to provide various contour prescriptions, as well as to compute unitarity cuts of the amplitude. We now conclude by commenting on various avenues of further exploration suggested by our results.

An important question to understand is how the methods described here can be applied to known string theories, such as Type I supertrings or regular open/closed bosonic strings. In this work, we have mainly studied the massless \emph{surface} integrals that arise naturally as UV regularizations of $\Tr(\Phi^3)$ theory. However, as pointed out in Section \ref{sec:intro}, there are various reasons for which these objects are interesting. One of them is the fact that at tree-level, they form a basis for all possible open string integrals -- more formally, a basis for the twisted cohomology of $\mathcal{M}_{0,n}$. It remains to be seen if the same holds at loop level. If it does, then our methods can be directly imported to the integrals describing different types of string amplitudes.

Another important feature of the surface formulation of stringy integrals is the ability to extract residues or leading singularities at the level of the integrand directly, without explicitly having to perform the loop integration. As we saw, one can use the constraints obtained at tree level to fix the form of loop-level unitarity cuts, which provides a powerful tool to test the consistency of these families of integrals with unitarity. A simple analysis at $n=2$ for low mass levels was enough to rule out some of the simplest stringy integrals. It would be interesting to explore whether imposing these constraints at higher points will lead us to discard even broader classes of integrals on unitarity grounds, and answer the question: is the usual tachyonic string amplitude the \textit{only} stringy integral consistent with unitarity at all mass levels? 

It is also natural ask whether our definition of the contours at loop-level extends to higher multiplicity and higher loops. For instance, the decomposition of the 1-loop surface integrand into finitely many terms resembling standard Feynman integrals hints at the possibility of implementing the $i\varepsilon$ prescription in exactly the same fashion as in field theory. If this approach can be extended to higher points and/or loops, one could study analyticity and causality properties of string amplitudes in a much more systematic manner. A new challenge arises beyond planar one-loop amplitudes, due to the presence of the Mapping Class Group (MCG), which gives rise to infinitely many redundant copies of the cones in the fan. In \cite{Arkani-Hamed:2023lbd}, a method to mod out by the action of the MCG has been proposed. It would be interesting to explore whether the use of the ``Mirzakhani trick'' for modding out by the MCG can be naturally combined with our ideas for defining the integration contour.

Finally, our analysis has made clear that for large values of the kinematics, the amplitude that results from integrating along our contours exhibits huge cancellations between different regions in parameter space. This stands as a practical obstacle to numerically computing the value of the amplitude in extreme kinematic regimes, and is only slightly remedied when using our more sophisticated generalized Pochhammer contour. At 4-points, it is clear that one can engineer a closed contour that completely avoids the large oscillations by following a path of stationary phase. However, understanding whether there is always a unique contour that completely avoids the cancellations for any number of particles could lead to a better understanding of the asymptotic behavior of the amplitudes.

\acknowledgments 
We thank Nima Arkani-Hamed, Lorenz Eberhardt, Hadleigh Frost, Song He, Sebastian Mizera, Andrzej Pokraka and Giulio Salvatori for useful discussions. C.F. is supported  by FCT  (2023.01221.BD and DOI  https://doi.org/10.54499/2023.01221.BD). M.S. is supported by the US Department of Energy under contract DE-SC0010010 Task F.
\appendix

\section{Numerical checks}\label{app:numerical checks}

In this appendix, we collect a series of numerical computations to show the validity of our contour prescriptions, as well as to observe various aspects of the stringy integrals in different regimes. We focus on the tree-level case, since at loop-level we don't have an exact answer to compare to. We explicitly test the accuracy of our contour at large values of the kinematics, for $n=5$ and $n=6$, where the cancellation features highlighted in the main text come into play and also where the integrals become highly oscillatory. We have also added a \texttt{Mathematica} notebook as an ancillary file to this paper which contains the code used to produce the plots.

\subsection{A single $X_{i,j}$ large at 5-points}

\begin{figure}[t]
    \centering
    \includegraphics[width=\textwidth,valign=c]{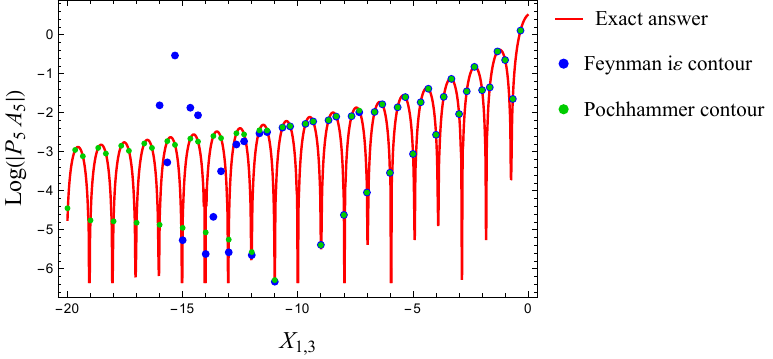} 
    \caption{Logarithmic plot of the normalized 5-point amplitude in the regime where only $X_{1,3}$ (varying in the $x$-axis) is set negative and large. We represent both the results from the standard $i\varepsilon$ prescription (blue dots) and the generalized Pochhammer contour (green dots), the latter being clearly more accurate for large values of $X_{1,3}$.}
    \label{fig:one X large} 
\end{figure}

To begin with, we consider the regime in the 5-point tree-level amplitude where one of the planar Mandelstam invariants, say $X_{1,3}$, is taken to be negative and large. The rest of the variables $\{X_{1,4},X_{2,4},X_{2,5},X_{3,5}\}$ are fixed positive and (relatively) small. This of course does not correspond to a physical scattering process, but we can use it to see how negative we can take one of the invariants before the prescription starts losing accuracy. The result is displayed in figure \ref{fig:one X large}, where we have normalized the amplitude by a factor of $\sin(\pi X_{1,3})$ to remove the poles corresponding to resonances. Here, we can see that our generalized Pochhammer contour described in section \ref{sec:Pochhammer} agrees with the exact answer for very large values of $X_{1,3}$. Meanwhile, the contour implementing the $i\varepsilon$ prescription in a more standard way as described in section \ref{sec:TreeContour} breaks down once the kinematics become too negative.
\subsection{Hard scattering limit at 5-points}

\begin{figure}[t]
    \centering
    \noindent\hspace{-1cm}\includegraphics[width=0.75\textwidth,valign=c]{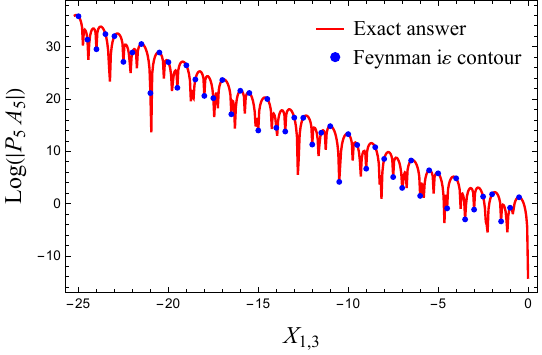} 
    \caption{Logarithmic plot of the normalized 5-point amplitude in the regime of hard scattering at fixed angle. Interestingly, the standard $i\varepsilon$ contour is enough to reach very large values of the kinematics, and the exponential growth of the amplitude is manifest.}
    \label{fig:hard scattering 5pt} 
\end{figure}

We now consider an actual physical regime for the kinematics, namely a fixed angle scattering process at high energies. We set $p_i = \sqrt{x}\,\hat p_i$ with
\begin{equation}
\begin{aligned}
    \hat p_1 &= (1,0,0,1),\quad \hat p_2 = (1,0,0,-1),\\
    \hat p_3 = \left(-\frac{2}{3},0,\frac{\sqrt{2}}{3},-\frac{\sqrt{2}}{3}\right),\  &\hat p_4 = \left(-\frac{2}{3},\frac{\sqrt{2}}{3},-\frac{\sqrt{2}}{3},0\right),\  \hat p_3 = \left(-\frac{2}{3},-\frac{\sqrt{2}}{3},0,\frac{\sqrt{2}}{3}\right).
\end{aligned}
\end{equation}
The result is displayed in figure \ref{fig:hard scattering 5pt}, where again we have removed the resonance poles by normalizing with the appropriate sine factors. In this case, the standard $i\varepsilon$ contour is enough to reach large values of the kinematics. This is clearly sufficient to show the exponential growth of the amplitude with the center-of-mass energy in this regime.

\subsection{Regge-like limit at 5-points}

\begin{figure}[t]
    \centering
    \noindent\hspace{-1cm}\includegraphics[width=0.73\textwidth,valign=c]{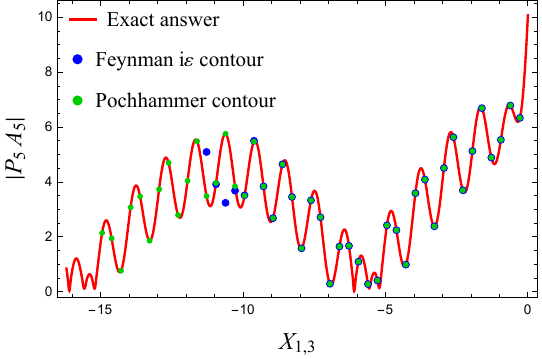} 
    \caption{Numerical plot of the normalized 5-point amplitude in the Regge-like regime where two compatible invariants are set negative and large, while keeping the rest fixed. As in the first example, the Pochhammer contour manages to accurately reproduce the amplitude at larger values of the kinematics. Still, the power-law behavior of the amplitude is relatively hard to discern.}
    \label{fig:regge 5-pt} 
\end{figure}

As a last case study for the 5-point amplitude, we will look at a more physical case of a Regge limit, which corresponds to setting two compatible variables $\{X_{1,3},\,X_{1,4}\}$ negative and large and leaving the rest fixed. For this regime, we again find that the Pochhammer contour performs significantly better than the standard $i\varepsilon$ prescription, although we are able to reach respectable values of the kinematics for both of them, see figure \ref{fig:regge 5-pt}. However, in this case it is still not enough to clearly discern the power-law behavior of the amplitude, for which we would need to probe much larger values of $|X_{1,3}|$ and $|X_{1,4}|$. This again displays the difficulties that the huge cancellations pose when numerically evaluating string integrals for asymptotic kinematic regimes. 

\subsection{6-point tree-level amplitude}

\begin{figure}[t]
    \centering
    \noindent\hspace{-1cm}\includegraphics[width=0.75\textwidth,valign=c]{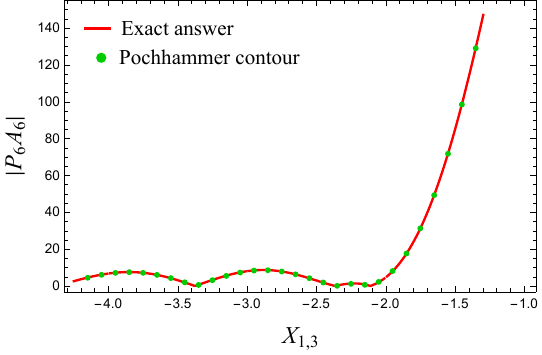}
    \caption{Numerical plot of the normalized 6-point amplitude, evaluated with the generalized Pochhammer contour (green dots). The exact answer (in red) is obtained via the series representation of the $6$-point amplitude given in \cite{Arkani-Hamed:2024nzc}.}
    \label{fig:6-pt numerics} 
\end{figure}

For the last tree-level check, we compute the 6-point tree-level amplitude. As mentioned in the main text, at higher points it becomes increasingly difficult to evaluate the integrals at large values of the kinematics, due to the large cancellations that happen across the integration domain. Thus, for this case we choose a more conservative range of kinematics in which three of the invariants $\{X_{1,3},\, X_{1,4}\, X_{1,5}\}$ are set negative while the rest are positive, and we only let $X_{1,3}$ grow towards slightly more negative values. Still, as shown in figure \ref{fig:6-pt numerics}, the generalized Pochhammer contour provides remarkably accurate results even for the largest values of $X_{1,3}$.
\section{Matching the one-loop surface integral to the bosonic string}\label{app:GSW matching}
In this appendix, we review how a simple choice of exponents for the $u$-variables of all curves on punctured disk (including both closed and open self-intersecting curves) allows the precise matching of the one-loop surface integral to the one-loop tachyon amplitude in open bosonic string theory. The original proof can be found in~\cite{GiuliosTalk,NimasTalk,combinatString}. 

According to Green-Schwartz-Witten (GSW) \cite{Green:2012pqa}, the tachyon $2$-point 1-loop planar amplitude is given by
\begin{equation}
    \int_{0}^1 \frac{\text{d}\omega}{\omega^2} \int_{\omega}^1\frac{\text{d}\rho}{\rho} \times [f(\omega)]^{2-d} \left(\frac{-2\pi}{\ln{\omega}} \right)^{d/2} (\Psi_{1,2})^{-2p_1^2},
\end{equation}
where $p_1$ is the external momenta satisfying the on-shell condition $p_1^2=-m^2$ (for the tachyon $m^2=-1$), $f(\omega)$ is defined by 
\begin{equation}
   [f(\omega)]^2=\prod_{n=1}^\infty (1-w^n)^2,
\end{equation}
which is -- up to a factor -- the familiar Dedekind-eta function, and 
\begin{equation}
   \Psi_{1,2}=\frac{1-\omega/\rho}{\sqrt{\omega/\rho}} \exp{\frac{\ln^2{\omega/\rho}}{2\ln{\omega}}}\prod_{n=1}^\infty \frac{(1-\omega^{n+1}/\rho)(1-\omega^{n-1} \rho) }{(1-\omega^n)^2},
   \label{eq:JacobiTheta}
\end{equation}
which is in turn closely related to the Jacobi-Theta function. In summary, for $p_1^2=1$, we get the following integral 
\begin{equation}
    \int \frac{\text{d}\omega\text{d}\rho}{\omega^2 \rho} \frac{\omega/\rho}{(1-\omega/\rho)^2}  \left[f(\omega)\right]^{2-d}  \left(\frac{-2\pi}{\ln{\omega}} \right)^{d/2} \left[\prod_{n=1}^\infty \frac{(1-\omega^{n+1}/\rho)(1-\omega^{n-1} \rho) }{(1-\omega^n)^2}\right]^{-2}e^{-\frac{\ln^2{\omega/\rho}}{\ln{\omega}}},
    \label{eq:GSW_Final}
\end{equation}

From surfaceology, the surface integral we associate with the punctured disk is
\begin{equation}
\begin{aligned}
    \int_{y_{1,0} y_{2,0} \leq 1} \frac{\text{d}y_{1,0} \text{d}y_{2,0}}{y_{1,0} y_{2,0}}  u_{1,0}^{Y_1+m^2}u_{2,0}^{Y_2+m^2} &\left[\prod_{q=0}^\infty u_{1,1}^{(q)}\right]^{m^2}\left[\prod_{q=0}^\infty u_{2,2}^{(q)}\right]^{m^2} \left[\prod_{q=1}^\infty u_{1,2}^{(q)} u_{2,1}^{(q)}\right]^{X_{1,2} + m^2}\\
    \times &\left[\prod_{i=1}^\infty u_{\Delta(i)}^i\right]^{\Delta_1} \left[\prod_{i=1}^\infty u_{\Delta(i)}^{i^2}\right]^{\Delta_2},
\end{aligned}
\label{eq:SurfInt1Loop}
\end{equation}
where the exponents are read directly by homology: the tadpole curves $X_{1,1}$ and $X_{2,2}$ are fixed to zero, and $X_{1,2}=p_1^2$, which we will set to $-m^2$ to impose the on-shell condition momentarily. Finally, $u_{\Delta_{(i)}}$ is the $u$-variable associated with the curve that loops $i$ times around the puncture. Note that we have two infinite products over the closed curves, one where each $u_{\Delta_{(i)}}$ is raised to $i \Delta_1$, and the second with exponents $i^2 \Delta_2$. Matching this result with GSW will fix both these exponents, as well as the mass $m^2$. 

To do this matching, we start by performing the loop-integrals in \eqref{eq:SurfInt1Loop}. This is simple because the loop-dependence is gaussian: concretely, we have $Y_1=\ell^2$ and $Y_2=(\ell+p_1)^2$, so the loop integral is simply
\begin{equation}
    \int \frac{\text{d}^d \ell}{\pi^{d/2}}\ \exp{\log{u_{1,0}}\ell^2 +\log{u_{2,0}}(\ell+p_1)^2},
\end{equation}
from which we get 
\begin{equation}
   \int_{y_{1,0} y_{2,0}\leq 1 }  \frac{\text{d}y_{1,0} \text{d}y_{2,0}}{y_{1,0} y_{2,0}} (u_{1,0}u_{2,0})^{m^2} \left[\frac{-2\pi}{\log{(u_{1,0} u_{2,0})}}\right]^{d/2} \exp{\frac{\log{(u_{1,0})}\log{(u_{2,0})}}{\log{(u_{1,0}u_{2,0})}}p_1^2} \times \cdots 
   \label{eq:A1}
\end{equation}

\begin{figure}[t]
    \centering 
    \includegraphics[width=\textwidth]{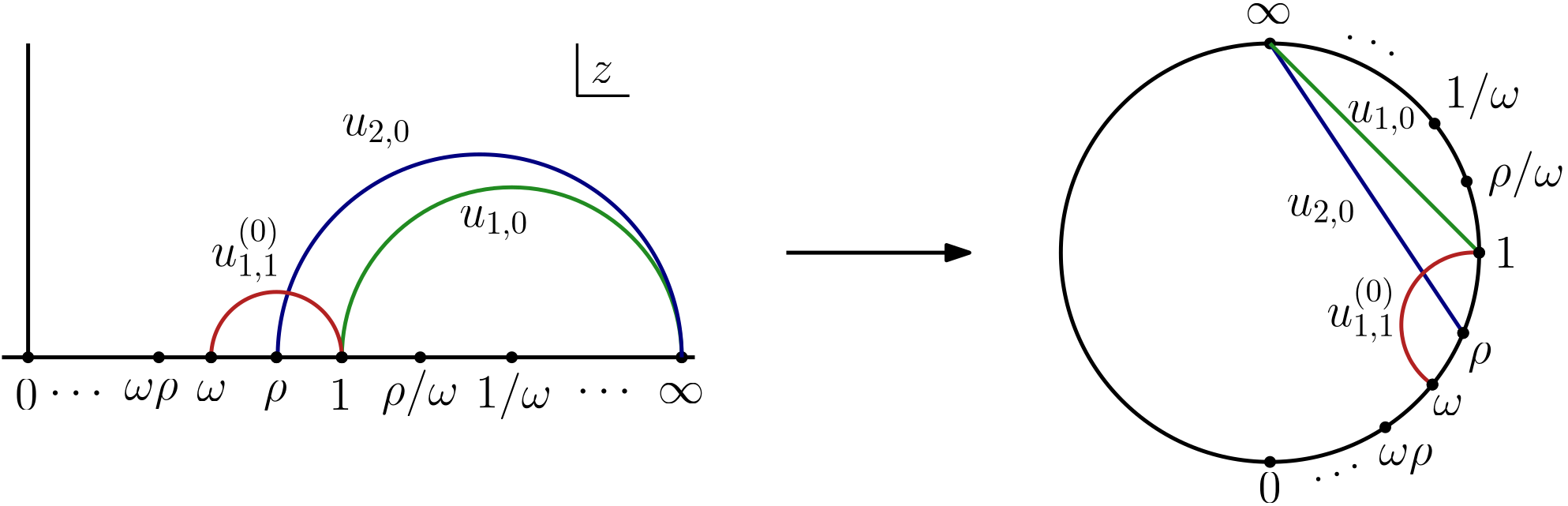}
    \caption{Upper-half plane picture of the 2-point 1-loop amplitude in terms of worldsheet coordinates. On the right, we have compactified the space into a closed circle.}
    \label{fig:upper half plane}
\end{figure}

Now we can derive the map between the $u$'s and the respective $y$'s to the GSW coordinates $\rho$ and $\omega$, by using the upper half plane picture. In standard worldsheet coordinates, we use the $\text{GL}(1)$ invariance of the integrand to fix $z_1=1$ and define $z_2=\rho$, $\omega=z_1\,z_2$. In the complex $z$-plane, we then have to identify an infinite number of pairs $(z_1,\,z_2)$ up to rescaling under $\omega$. Of course, in our curve language this just corresponds to the fact that we can have infinitely many copies of a curve joining the same marked points of the punctured disk, with a different number of windings around the puncture. In figure \ref{fig:upper half plane}, we have represented the worldsheet marked points and some examples of the curves in the upper-half $z$-plane.

The $u$-variables in terms of these coordinates are then given by the usual cross-ratios:
\begin{equation}
    u_{i,j} = \frac{z_{i-1,j}\,z_{i,j-1}}{z_{i,j}z_{i-1,j-1}},\quad z_{i,j}\equiv z_j-z_i.
\end{equation}
Thus, we can write the spiraling $u_{i,0}$'s as follows:
\begin{equation}
    u_{1,0} \equiv u_{1,\infty} = \frac{\omega}{\rho}, \quad \quad  u_{2,0} \equiv u_{\rho,\infty} = \rho,
\end{equation}
and using this we can easily derive the map between $(\omega,\rho)$ and the positive coordinates $(y_{1,0},y_{2,0})$:
\begin{equation}
    u_{1,0} u_{2,0} = y_{1,0} y_{2,0} = \omega, \quad \quad  u_{2,0} =\rho = \frac{y_{2,0}(1+y_{1,0})}{(1+y_{2,0})}.
\end{equation}
We indeed see that the integration domain in $y_{1,0},\,y_{2,0}$ precisely maps to the integration domain in $(\omega,\rho)$-space corresponding to $\omega \in [0,1]$ and $\rho \in [\omega,1]$. Plugging this into \eqref{eq:A1}, we find
\begin{equation}
   \int  \frac{\text{d}\omega \text{d}\rho}{\omega \rho} \frac{1-\omega}{(1- \omega/\rho)(1-\rho)} \omega^{m^2}\left[\frac{-2\pi}{\log{(\omega)}}\right]^{d/2} \left[\frac{1}{\omega/\rho} \right]^{-p_1^2}\exp{-\frac{\log{(\omega/\rho)^2}}{\log{(\omega)}}p_1^2}  \times \cdots 
   \label{eq:A2}
\end{equation}

The next step is to see what the result for the infinite products curves in \eqref{eq:SurfInt1Loop} is. Let's start with the closed curves. For these, we find:
\begin{equation}
    \prod_{i=1}^\infty u_{\Delta(i)}^i  = 1-\omega, \quad \quad  \prod_{i=1}^\infty u_{\Delta(i)}^{i^2} = (1-\omega)  \prod_{n=2}^\infty (1-\omega^n)^2.
\end{equation}
As for the tadpole curves, using the upper half-plane picture we obtain:
\begin{equation}
\begin{aligned}
    &u_{1,1}^{(0)} \equiv u_{1,\omega}= \frac{(1-\rho)(1-\omega^2/\rho)}{(1-\omega)^2}, \quad u_{1,1}^{(1)} \equiv u_{1,\omega^2}= \frac{(1-\omega \rho)(1-\omega^3/\rho)}{(1-\omega^2)^2}, \quad \cdots \\
    &u_{2,2}^{(0)} \equiv u_{\rho,\omega \rho}= \frac{(1-\omega \rho)(1-\omega/\rho)}{(1-\omega)^2}, \quad u_{2,2}^{(1)} \equiv u_{\rho,\omega^2\rho}= \frac{(1-\omega^2 \rho)(1-\omega^2/\rho)}{(1-\omega^2)^2}, \quad \cdots
\end{aligned}
\end{equation}
Hence, we can write the infinite products over the tadpole curves as:
\begin{align}
    &\prod^\infty_{q=0} u_{1,1}^{(q)} = \prod_{n=1}^{\infty} \frac{(1-\omega^{n+1}/\rho)(1-\omega^{n-1}\rho)}{(1-\omega^n)^2},\\ 
    &\prod^\infty_{q=0} u_{2,2}^{(q)} = \frac{(1-\omega/\rho)}{(1-\rho)}\prod_{n=1}^{\infty} \frac{(1-\omega^{n+1}/\rho)(1-\omega^{n-1}\rho)}{(1-\omega^n)^2}, 
\end{align}
which are precisely the infinite products appearing in $\Psi_{1,2}$ in \eqref{eq:JacobiTheta}!

Finally, for the self-intersecting curves $X_{1,2}^{(q)}$ we find the general expression: 
\begin{equation}
    u_{1,2}^{(q)}\equiv u_{1,\omega^q \rho} = \frac{(1 - \omega^q)(1 -\omega^{q+1})}{(1 - \omega^{q+1}/\rho) ( 1-\omega^q\rho)},
\end{equation}
so that when we take the infinite product we find 
\begin{equation}
   \prod^\infty_{q=1} u_{1,2}^{(q)} = \prod^\infty_{q=1}\frac{(1 - \omega^q)(1 -\omega^{q+1})}{(1 - \omega^{q+1}/\rho) ( 1-\omega^q\rho)} =  \frac{(1-\rho)}{(1-\omega)}\prod^\infty_{q=1}\frac{(1 - \omega^q)^2}{(1 - \omega^{q+1}/\rho) ( 1-\omega^{q-1}\rho)}.
\end{equation}
The infinite product of $u_{2,1}^{(q)}$ is identical, as these two curves have the same $u$-variables. Note, however, that neither of these curves will contribute to the surface integral once we go on-shell, as their exponents are zero when $X_{1,2} = p_1^{2}=-m^2$. 

Putting all the pieces that survive the on-shell limit together, we land on the following integral:
\begin{equation}
\begin{aligned}
\int  \frac{\text{d}\omega \text{d}\rho}{\omega \rho} &\frac{(1-\omega)\,  \omega^{m^2}}{(1- \omega/\rho)(1-\rho)}\left(\frac{-2\pi}{\log{(\omega)}}\right)^{d/2} \left[\frac{1}{\omega/\rho} \right]^{-p_1^2}e^{-\frac{\log{(\omega/\rho)^2}}{\log{(\omega)}}p_1^2}\left(\frac{1-\omega/\rho}{1-\rho}\right)^{m^2} \\
&\times  \left[\prod_{n=1}^{\infty} \frac{(1-\omega^{n+1}/\rho)(1-\omega^{n-1}\rho)}{(1-\omega^n)^2}\right]^{2m^2} \times (1-\omega)^{\Delta_1-\Delta_2} \left[\prod_{n=1}^\infty (1-\omega^n)^2\right]^{\Delta_2},
\end{aligned}
\end{equation}
which precisely agrees with the result in \eqref{eq:GSW_Final} when we set $p_1^2=1$, and choose $\Delta_2=1-d/2$ and $\Delta_1=-1+\Delta_2=-d/2$.

\bibliographystyle{JHEP}\bibliography{Refs}

@article{Witten:2013pra,
    author = "Witten, Edward",
    title = "{The Feynman $i \epsilon$ in String Theory}",
    eprint = "1307.5124",
    archivePrefix = "arXiv",
    primaryClass = "hep-th",
    doi = "10.1007/JHEP04(2015)055",
    journal = "JHEP",
    volume = "04",
    pages = "055",
    year = "2015"
}

@article{Arkani-Hamed:2023lbd,
    author = "Arkani-Hamed, N. and Frost, H. and Salvatori, G. and Plamondon, P-G. and Thomas, H.",
    title = "{All Loop Scattering as a Counting Problem}",
    eprint = "2309.15913",
    archivePrefix = "arXiv",
    primaryClass = "hep-th",
    month = "9",
    year = "2023"
}

@article{Arkani-Hamed:2023mvg,
    author = "Arkani-Hamed, N. and Frost, H. and Salvatori, G. and Plamondon, P-G. and Thomas, H.",
    title = "{All Loop Scattering For All Multiplicity}",
    eprint = "2311.09284",
    archivePrefix = "arXiv",
    primaryClass = "hep-th",
    month = "11",
    year = "2023"
}

@article{Arkani-Hamed:2023jry,
    author = "Arkani-Hamed, Nima and Cao, Qu and Dong, Jin and Figueiredo, Carolina and He, Song",
    title = "{Scalar-Scaffolded Gluons and the Combinatorial Origins of Yang-Mills Theory}",
    eprint = "2401.00041",
    archivePrefix = "arXiv",
    primaryClass = "hep-th",
    month = "12",
    year = "2023"
}

@article{Arkani-Hamed:2024vna,
    author = "Arkani-Hamed, Nima and Figueiredo, Carolina and Frost, Hadleigh and Salvatori, Giulio",
    title = "{Tropical Amplitudes For Colored Lagrangians}",
    eprint = "2402.06719",
    archivePrefix = "arXiv",
    primaryClass = "hep-th",
    month = "2",
    year = "2024"
}

@article{Arkani-Hamed:2023swr,
    author = "Arkani-Hamed, Nima and Cao, Qu and Dong, Jin and Figueiredo, Carolina and He, Song",
    title = "{Hidden zeros for particle/string amplitudes and the unity of colored scalars, pions and gluons}",
    eprint = "2312.16282",
    archivePrefix = "arXiv",
    primaryClass = "hep-th",
    month = "12",
    year = "2023"
}

@article{Arkani-Hamed:2024fyd,
    author = "Arkani-Hamed, Nima and Figueiredo, Carolina",
    title = "{All-order splits and multi-soft limits for particle and string amplitudes}",
    eprint = "2405.09608",
    archivePrefix = "arXiv",
    primaryClass = "hep-th",
    month = "5",
    year = "2024"
}

@article{Schlotterer:2012zz,
    author = "Schlotterer, O.",
    title = "{Scattering amplitudes in open superstring theory}",
    doi = "10.1002/prop.201100084",
    journal = "Fortsch. Phys.",
    volume = "60",
    pages = "373--691",
    year = "2012"
}

@article{Eberhardt:2024twy,
    author = "Eberhardt, Lorenz and Mizera, Sebastian",
    title = "{Lorentzian contours for tree-level string amplitudes}",
    eprint = "2403.07051",
    archivePrefix = "arXiv",
    primaryClass = "hep-th",
    doi = "10.21468/SciPostPhys.17.3.078",
    journal = "SciPost Phys.",
    volume = "17",
    pages = "078",
    year = "2024"
}

@article{Banerjee:2024ibt,
    author = "Banerjee, Pinaki and Eberhardt, Lorenz and Mizera, Sebastian",
    title = "{Regge Limit of One-Loop String Amplitudes}",
    eprint = "2403.07064",
    archivePrefix = "arXiv",
    primaryClass = "hep-th",
    month = "3",
    year = "2024"
}

@article{Hanson,
author = {Hanson, Andrew and Sha, Ji-Ping},
year = {2006},
month = {03},
pages = {2509-2537},
title = {A contour integral representation for the dual five-point function and a symmetry of the genus-4 surface in Bbb R6},
volume = {39},
journal = {Journal of Physics A-mathematical and General - J PHYS-A-MATH GEN},
doi = {10.1088/0305-4470/39/10/017}
}

@book{Hannesdottir:2022bmo,
    author = "Hannesdottir, Holmfridur Sigridar and Mizera, Sebastian",
    title = "{What is the i\ensuremath{\varepsilon} for the S-matrix?}",
    eprint = "2204.02988",
    archivePrefix = "arXiv",
    primaryClass = "hep-th",
    doi = "10.1007/978-3-031-18258-7",
    isbn = "978-3-031-18257-0, 978-3-031-18258-7",
    publisher = "Springer",
    series = "SpringerBriefs in Physics",
    month = "1",
    year = "2023"
}

@article{Pochhammer1890,
author = {Pochhammer, L.},
year = {1890},
month = {12},
pages = {1432-1807},
title = {Zur Theorie der Euler'schen Integrale},
volume = {35},
journal = {Mathematische Annalen},
doi = {10.1007/BF02122658}
}

@article{Arkani-Hamed:2017mur,
    author = "Arkani-Hamed, Nima and Bai, Yuntao and He, Song and Yan, Gongwang",
    title = "{Scattering Forms and the Positive Geometry of Kinematics, Color and the Worldsheet}",
    eprint = "1711.09102",
    archivePrefix = "arXiv",
    primaryClass = "hep-th",
    doi = "10.1007/JHEP05(2018)096",
    journal = "JHEP",
    volume = "05",
    pages = "096",
    year = "2018"
}

@article{Arkani-Hamed:2019mrd,
    author = "Arkani-Hamed, Nima and He, Song and Lam, Thomas",
    title = "{Stringy canonical forms}",
    eprint = "1912.08707",
    archivePrefix = "arXiv",
    primaryClass = "hep-th",
    doi = "10.1007/JHEP02(2021)069",
    journal = "JHEP",
    volume = "02",
    pages = "069",
    year = "2021"
}

@article{Arkani-Hamed:2024nzc,
    author = "Arkani-Hamed, Nima and Figueiredo, Carolina and Remmen, Grant N.",
    title = "{Open string amplitudes: singularities, asymptotics and new representations}",
    eprint = "2412.20639",
    archivePrefix = "arXiv",
    primaryClass = "hep-th",
    doi = "10.1007/JHEP04(2025)039",
    journal = "JHEP",
    volume = "04",
    pages = "039",
    year = "2025"
}

@article{Veneziano:1968yb,
    author = "Veneziano, G.",
    title = "{Construction of a crossing - symmetric, Regge behaved amplitude for linearly rising trajectories}",
    doi = "10.1007/BF02824451",
    journal = "Nuovo Cim. A",
    volume = "57",
    pages = "190--197",
    year = "1968"
}

@article{Arkani-Hamed:2023jwn,
    author = "Arkani-Hamed, Nima and Cheung, Clifford and Figueiredo, Carolina and Remmen, Grant N.",
    title = "{Multiparticle Factorization and the Rigidity of String Theory}",
    eprint = "2312.07652",
    archivePrefix = "arXiv",
    primaryClass = "hep-th",
    reportNumber = "CALT-TH 2023-051",
    doi = "10.1103/PhysRevLett.132.091601",
    journal = "Phys. Rev. Lett.",
    volume = "132",
    number = "9",
    pages = "091601",
    year = "2024"
}

@book{Green:2012pqa,
    author = "Green, Michael B. and Schwarz, John H. and Witten, Edward",
    title = "{Superstring Theory Vol. 2}: {25th Anniversary Edition}",
    doi = "10.1017/CBO9781139248570",
    isbn = "978-1-139-53478-9, 978-1-107-02913-2",
    publisher = "Cambridge University Press",
    series = "Cambridge Monographs on Mathematical Physics",
    month = "11",
    year = "2012"
}

@article{Eberhardt:2023xck,
    author = "Eberhardt, Lorenz and Mizera, Sebastian",
    title = "{Evaluating one-loop string amplitudes}",
    eprint = "2302.12733",
    archivePrefix = "arXiv",
    primaryClass = "hep-th",
    doi = "10.21468/SciPostPhys.15.3.119",
    journal = "SciPost Phys.",
    volume = "15",
    number = "3",
    pages = "119",
    year = "2023"
}

@article{Eberhardt:2022zay,
    author = "Eberhardt, Lorenz and Mizera, Sebastian",
    title = "{Unitarity cuts of the worldsheet}",
    eprint = "2208.12233",
    archivePrefix = "arXiv",
    primaryClass = "hep-th",
    doi = "10.21468/SciPostPhys.14.2.015",
    journal = "SciPost Phys.",
    volume = "14",
    number = "2",
    pages = "015",
    year = "2023"
}

@article{Manschot:2024prc,
    author = "Manschot, Jan and Wang, Zhi-Zhen",
    title = "{The $i\varepsilon$-Prescription for String Amplitudes and Regularized Modular Integrals}",
    eprint = "2411.02517",
    archivePrefix = "arXiv",
    primaryClass = "hep-th",
    month = "11",
    year = "2024"
}

@article{mumford1969irreducibility,
  title={The irreducibility of the space of curves of given genus},
  author={Deligne, Pierre and Mumford, David},
  journal={Publications Mathématiques de l'IHÉS},
  volume={36},
  pages={75--109},
  year={1969},
}

@article{DHoker:1988pdl,
    author = "D'Hoker, Eric and Phong, D. H.",
    title = "{The Geometry of String Perturbation Theory}",
    reportNumber = "PUPT-1039",
    doi = "10.1103/RevModPhys.60.917",
    journal = "Rev. Mod. Phys.",
    volume = "60",
    pages = "917",
    year = "1988"
}

@InProceedings{Stieberger:2016xhs,
author="Stieberger, S.",
editor="Burgos Gil, Jos{\'e} Ignacio
and Ebrahimi-Fard, Kurusch
and Gangl, Herbert",
title="Periods and Superstring Amplitudes",
booktitle="Periods in Quantum Field Theory and Arithmetic",
year="2020",
publisher="Springer International Publishing",
address="Cham",
pages="45--76",
isbn="978-3-030-37031-2",
eprint = "1605.03630",
doi = "10.1007/978-3-030-37031-2_3"
}

@article{Mafra:2022wml,
    author = "Mafra, Carlos R. and Schlotterer, Oliver",
    title = "{Tree-level amplitudes from the pure spinor superstring}",
    eprint = "2210.14241",
    archivePrefix = "arXiv",
    primaryClass = "hep-th",
    doi = "10.1016/j.physrep.2023.04.001",
    journal = "Phys. Rept.",
    volume = "1020",
    pages = "1--162",
    year = "2023"
}

@article{Koba:1969rw,
    author = "Koba, Z. and Nielsen, Holger Bech",
    title = "{Reaction amplitude for n mesons: A Generalization of the Veneziano-Bardakci-Ruegg-Virasora model}",
    doi = "10.1016/0550-3213(69)90331-9",
    journal = "Nucl. Phys. B",
    volume = "10",
    pages = "633--655",
    year = "1969",
    url = "https://doi.org/10.1016/0550-3213(69)90331-9" 
}

@article{Bardakci:1968rse,
    author = "Bardakci, K. and Ruegg, H.",
    title = "{Reggeized resonance model for the production amplitude}",
    doi = "10.1016/0370-2693(68)90127-5",
    journal = "Phys. Lett. B",
    volume = "28",
    pages = "342--347",
    year = "1968"
}

@article{Chan:1969ex,
    author = "Chan, Hong-Mo and Tsou, Sheung Tsun",
    title = "{Explicit construction of the n-point function in the generalized Veneziano model}",
    doi = "10.1016/0370-2693(69)90523-1",
    journal = "Phys. Lett. B",
    volume = "28",
    pages = "485--488",
    year = "1969"
}

@article{Gross:1969db,
    author = "Gross, D. J.",
    title = "{Factorization and the generalized Veneziano model with satellites}",
    doi = "10.1016/0550-3213(69)90248-X",
    journal = "Nucl. Phys. B",
    volume = "13",
    pages = "467--476",
    year = "1969"
}

@article{Arkani-Hamed:2019plo,
    author = "Arkani-Hamed, Nima and He, Song and Lam, Thomas and Thomas, Hugh",
    title = "{Binary geometries, generalized particles and strings, and cluster algebras}",
    eprint = "1912.11764",
    archivePrefix = "arXiv",
    primaryClass = "hep-th",
    doi = "10.1103/PhysRevD.107.066015",
    journal = "Phys. Rev. D",
    volume = "107",
    number = "6",
    pages = "066015",
    year = "2023"
}

@article{Hannesdottir:2024cnn,
    author = "Hannesdottir, Holmfridur S. and Lippstreu, Luke and McLeod, Andrew J. and Polackova, Maria",
    title = "{Minimal Cuts and Genealogical Constraints on Feynman Integrals}",
    eprint = "2406.05943",
    archivePrefix = "arXiv",
    primaryClass = "hep-th",
    month = "6",
    year = "2024"
}

@article{Caron-Huot:2023ikn,
    author = "Caron-Huot, Simon and Giroux, Mathieu and Hannesdottir, Holmfridur S. and Mizera, Sebastian",
    title = "{Crossing beyond scattering amplitudes}",
    eprint = "2310.12199",
    archivePrefix = "arXiv",
    primaryClass = "hep-th",
    doi = "10.1007/JHEP04(2024)060",
    journal = "JHEP",
    volume = "04",
    pages = "060",
    year = "2024"
}

@article{Caron-Huot:2023vxl,
    author = "Caron-Huot, Simon and Giroux, Mathieu and Hannesdottir, Holmfridur S. and Mizera, Sebastian",
    title = "{What can be measured asymptotically?}",
    eprint = "2308.02125",
    archivePrefix = "arXiv",
    primaryClass = "hep-th",
    doi = "10.1007/JHEP01(2024)139",
    journal = "JHEP",
    volume = "01",
    pages = "139",
    year = "2024"
}

@article{Hannesdottir:2022xki,
    author = "Hannesdottir, Holmfridur S. and McLeod, Andrew J. and Schwartz, Matthew D. and Vergu, Cristian",
    title = "{Constraints on sequential discontinuities from the geometry of on-shell spaces}",
    eprint = "2211.07633",
    archivePrefix = "arXiv",
    primaryClass = "hep-th",
    reportNumber = "CERN-TH-2022-189",
    doi = "10.1007/JHEP07(2023)236",
    journal = "JHEP",
    volume = "07",
    pages = "236",
    year = "2023"
}

@article{Hannesdottir:2024hke,
    author = "Hannesdottir, Holmfridur S. and McLeod, Andrew J. and Schwartz, Matthew D. and Vergu, Cristian",
    title = "{Applications of the Landau bootstrap}",
    eprint = "2410.02424",
    archivePrefix = "arXiv",
    primaryClass = "hep-ph",
    doi = "10.1103/PhysRevD.111.085003",
    journal = "Phys. Rev. D",
    volume = "111",
    number = "8",
    pages = "085003",
    year = "2025"
}

@article{Caron-Huot:2024brh,
    author = "Caron-Huot, Simon and Correia, Miguel and Giroux, Mathieu",
    title = "{Recursive Landau Analysis}",
    eprint = "2406.05241",
    archivePrefix = "arXiv",
    primaryClass = "hep-th",
    month = "6",
    year = "2024"
}

@article{Arkani-Hamed:2020tuz,
    author = "Arkani-Hamed, Nima and He, Song and Lam, Thomas",
    title = "{Cluster Configuration Spaces of Finite Type}",
    eprint = "2005.11419",
    archivePrefix = "arXiv",
    primaryClass = "math.AG",
    doi = "10.3842/SIGMA.2021.092",
    journal = "SIGMA",
    volume = "17",
    pages = "092",
    year = "2021"
}

@article{CurvyU,
    author = "Arkani-Hamed, N. and Frost, H. and Salvatori, G. and Plamondon, P-G. and Thomas, H.",
    title = "{Curvy World of Curves, to appear}",
    eprint = "25xx.xxxx",
    archivePrefix = "arXiv"
}

@article{Backus:2025hpn,
    author = "Backus, Jeffrey V. and Rodina, Laurentiu",
    title = "{Emergence of Unitarity and Locality from Hidden Zeros at One-Loop}",
    eprint = "2503.03805",
    archivePrefix = "arXiv",
    primaryClass = "hep-th",
    month = "3",
    year = "2025"
}

@phdthesis{Salvatori:2019ehb,
    author = "Salvatori, Giulio",
    title = "{Amplituhedra for Phi\textasciicircum{}3 Theory at Tree and Loop Level}",
    doi = "10.13130/salvatori-giulio_phd2019-10-25",
    school = "Milan U.",
    year = "2019"
}

@article{Arkani-Hamed:2019vag,
    author = "Arkani-Hamed, Nima and He, Song and Salvatori, Giulio and Thomas, Hugh",
    title = "{Causal diamonds, cluster polytopes and scattering amplitudes}",
    eprint = "1912.12948",
    archivePrefix = "arXiv",
    primaryClass = "hep-th",
    doi = "10.1007/JHEP11(2022)049",
    journal = "JHEP",
    volume = "11",
    pages = "049",
    year = "2022"
}

@article{Salvatori:2018aha,
    author = "Salvatori, Giulio",
    title = "{1-loop Amplitudes from the Halohedron}",
    eprint = "1806.01842",
    archivePrefix = "arXiv",
    primaryClass = "hep-th",
    doi = "10.1007/JHEP12(2019)074",
    journal = "JHEP",
    volume = "12",
    pages = "074",
    year = "2019"
}

@unpublished{GiuliosTalk,
title= "{Scattering Amplitudes, Positive Geometries and Surfaces}",
author = "Salvatori, Giulio",
year = "2022",
note= "{IAS High Energy Theory Seminar}",
url= {https://www.ias.edu/video/scattering-amplitudes-positive-geometries-and-surfaces},
}

@unpublished{NimasTalk,
title= "{All-Loop Scattering as a Counting Problem}",
author = "Arkani-Hamed, Nima",
note= "{Amplitudes 2022 Talk}",
}

@article{combinatString,
    author = "Arkani-Hamed, N. and Frost, H. and Salvatori, G. and Plamondon, P-G. and Thomas, H.",
    title = "{to appear}",
    eprint = "25xx.xxxx",
    archivePrefix = "arXiv"
}

@misc{devadoss2000tessellationsmodulispacesmosaic,
      author = "Satyan L. Devadoss",
      title = "Tessellations of Moduli Spaces and the Mosaic Operad",
      eprint = "9807010",
      year = "2000",
      archivePrefix = "arXiv",
      primaryClass = "math.AG",
}

@article{Gross:1987kza,
    author = "Gross, David J. and Mende, Paul F.",
    title = "{The High-Energy Behavior of String Scattering Amplitudes}",
    reportNumber = "PUPT-1062",
    doi = "10.1016/0370-2693(87)90355-8",
    journal = "Phys. Lett. B",
    volume = "197",
    pages = "129--134",
    year = "1987",
    eprint = "9807010"
}

@article{Huang:2016tag,
    author = "Huang, Yu-tin and Schlotterer, Oliver and Wen, Congkao",
    title = "{Universality in string interactions}",
    eprint = "1602.01674",
    archivePrefix = "arXiv",
    primaryClass = "hep-th",
    doi = "10.1007/JHEP09(2016)155",
    journal = "JHEP",
    volume = "09",
    pages = "155",
    year = "2016"
}

@article{Arkani-Hamed:2024pzc,
    author = "Arkani-Hamed, Nima and Frost, Hadleigh and Salvatori, Giulio",
    title = "{The Cut Equation}",
    eprint = "2412.21027",
    archivePrefix = "arXiv",
    primaryClass = "hep-th",
    month = "12",
    year = "2024"
}

@article{Gross:1987ar,
    author = "Gross, David J. and Mende, Paul F.",
    title = "{String Theory Beyond the Planck Scale}",
    reportNumber = "PUPT-1067",
    doi = "10.1016/0550-3213(88)90390-2",
    journal = "Nucl. Phys. B",
    volume = "303",
    pages = "407--454",
    year = "1988"
}

@article{DHoker:1989cxq,
    author = "D'Hoker, Eric and Phong, D. H.",
    title = "{Conformal Scalar Fields and Chiral Splitting on Superriemann Surfaces}",
    reportNumber = "UCLA-88-TEP-38",
    doi = "10.1007/BF01218413",
    journal = "Commun. Math. Phys.",
    volume = "125",
    pages = "469",
    year = "1989"
}

\end{document}